\documentstyle[12pt,psfig]{article}
\def\mybaselinestretch{1.0}        % normal 1.0; double spaced 1.5;
            \newif\ifdodraftheader \newif\ifdomarginnotes
\newcommand{\SKIP}[1]{}             % To skip whole passages
\def\Titleofthispaper{%
  Critical dynamics in the 2d classical XY-model: a spin dynamics study
}%
\def\Preprintnumber{}
\def\Preprintdate  {June 1996}
\def\Preprintnumber{\rule{0ex}{1ex}cond-mat/yymmddd }    %!!!
\def\Authorsofthispaper{
     {Hans Gerd Evertz$^{a,b}$ and D. P. Landau$^a$}
}%
\def\Theiraddresses{
     $^a$ Center for Simulational Physics, 
     University of Georgia,  
     Athens, GA 30602                    \\[1ex]
     $^b$ Theor.\ Physik, Univ.\ W\"urzburg, 97074 W\"urzburg, Germany \\[1ex]
     {\tt evertz@physik.uni-wuerzburg.de}       \\
     {\tt dlandau@uga.cc.uga.edu}
}%
\def\Abstracttext{%
Using spin-dynamics techniques we have performed large-scale computer
simulations
of the dynamic behavior of the 
classical three component \XY-model
(i.e.\ the anisotropic limit of an easy-plane Heisenberg ferromagnet),
on square lattices of size up to $192^2$,
for several temperatures below, at, and above \Tkt.
The temporal evolution of spin configurations was determined
numerically from coupled equations of motion for individual spins by
a fourth order predictor-corrector method, with initial spin
configurations generated by a hybrid Monte Carlo algorithm.
The neutron scattering function \Sqw\  was calculated from 
the resultant space-time displaced spin-spin correlation function.
Pronounced spin-wave peaks were found both in the in-plane and the
out-of-plane scattering function over a wide range of temperatures.
The in-plane scattering function \Sxx\ also has
a large number of clear but weak additional peaks, 
which we interpret to come from two-spin-wave scattering.
In addition, we observed a small central peak in \Sxx,
even at temperatures well below the phase transition.
We used dynamic finite size scaling theory to extract the dynamic
critical exponent $z$.
We find $z=1.00(4)$ for all $T \leq \TKT$,
in excellent agreement with theoretical predictions,
although the shape of \Sqw\  is not well described by current theory.
}%% END OF ABSTRACTTEXT
    \newcommand{\note}[1]{}
    \ifdomarginnotes  \renewcommand{\note}[1]
                          {{\bf [{ #1\/}]}\message{Note: #1}}
                      \fi
    \ifdodraftheader  
                      \fi
\newcount\hour   \newcount\minute
\hour  = \time   \divide\hour by 60
\minute= \hour   \multiply\minute by -60   \advance\minute by \time
\ifnum\minute<10 \def\Time{\number\hour:0\number\minute}
           \else \def\Time{\number\hour:\number\minute}  \fi
\newcommand{\figurebox}[2]{\fbox{\vbox to #1{\hbox to #2{\hfil}\vfil}}}
\newcounter{itemnumber}

\newcommand{\bit}{\begin{itemize}}
\newcommand{\eit}{\end{itemize}}
\newcommand{\beq}[1]{\begin{equation}\label{#1}}
\newcommand{\eeq}{\end{equation}}
\newcommand{\eq}[1]{eq.\ (\ref{#1})}       % reference an equation by its label
\newcommand{\half}{{1\over2}}		   % One half
\newcommand{\ident}{\equiv}

\newcommand{\mylt}{\!<\!}
\newcommand{\mygt}{\!>\!}

\newcommand{\lsim}{\raisebox{-3pt}{$\stackrel{<}{\sim}$}}
\def\3{\ss}

\newlength{\LLL}
\newlength{\SSS}
\newlength{\captionwidth}
\newcommand{\mycaption}[2]{\label{#1} \parbox[t]{\captionwidth}
                           {\vskip2mm Fig.\ #1. #2}}
\def\S{{\bf S}}
\def\q{{\bf q}}
\def\om{$\omega$\ }
\def\Sqw{$S(q,\omega)$}
\def\Sxxqw{$S^{xx}(q,\omega)$}
\def\Sxx{$S^{xx}$}
\def\Szz{$S^{zz}$}
\def\Tkt{$T_{KT}$}
\def\TKT{T_{KT}}
\def\XY{$XY$}
\def\ee{{\bf\hat{e}}}
\def\tmax{{t_{max}}}
\def\tcut{{t_{cutoff}}}
\def\Heff{{\bf H}_{\mbox{\protect\scriptsize eff}}}
\def\DO{\Delta\omega}
\begin{document}
\psrotatefirst
% TOP OF TITLE PAGE: "Draft", Time, Preprintnumber and -date.
  \parbox[t]{20ex}{\ifdodraftheader \fbox{\bf Draft version}   \fi}
  \hfill           
  \begin{tabular}[t]{l}
      \ifx\Preprintnumber\empty\else \rule{0ex}{1ex}\Preprintnumber \\[.5ex]\fi
      \ifx\Preprintdate  \empty\else \rule{0ex}{1ex}\Preprintdate \fi
  \end{tabular}
  \ifdomarginnotes 
               %%%\note{\today} \note{\Time} \fi
              {\marginpar{\raggedright\tiny \today}}
              {\marginpar{\raggedright\tiny \Time}}  \fi

%
% AUTHORS AND TITLE
% The following is adapted from article.sty (\maketitle)
    \renewcommand{\thefootnote}{{\protect\fnsymbol{footnote}}}
    \vskip2em
    \begin{center}
           {\Large \bf \vbox{\vspace{4ex}} \Titleofthispaper \par}\vskip1.5em
           {\large \lineskip .5em
               \begin{tabular}[t]{c} \vbox{\vspace{2em}}
                                   \Authorsofthispaper
               \end{tabular}\par}
           \vspace{1em}
           {\normalsize \begin{tabular}[t]{c} \vbox{\vspace{0em}}
                                 \Theiraddresses
                        \end{tabular}\par}  
    \vskip1em
    \end{center} \par 
    \renewcommand{\thefootnote}{{\protect\arabic{footnote}}}
    \vfill
%
% ABSTRACT
\pagebreak[3]
\begin{abstract} {\protect\normalsize \noindent \Abstracttext}
\end{abstract}
\vspace*{1em}
\vfill
\thispagestyle{empty} \setcounter{page}{0} \newpage
\renewcommand{\baselinestretch}{\mybaselinestretch} % for double spacing
\protect\small \protect\normalsize                  % do use \baselinestretch
%%%%%%%%%%%%%%%%%%%%%%%%%%%%%%%%%%%%%%%%%%%%%%%%%%%%%%%%%%%%%%%%%%%%%%%%%%%%%%%
%        BEGIN MAIN TEXT OF PAPER
%%%%%%%%%%%%%%%%%%%%%%%%%%%%%%%%%%%%%%%%%%%%%%%%%%%%%%%%%%%%%%%%%%%%%%%%%%%%%%%
%
%******************************************************************************
 \section{Introduction}
%******************************************************************************
%

The two-dimensional \XY-model 
%%(anisotropic Heisenberg model)
is one of the `special' models of magnetism.
It undergoes an unusual phase transition to a state with bound,
topological excitations (vortex pairs) but no long range order \cite{KT}.
The (three component) XY-model may be viewed as a special case 
of the anisotropic Heisenberg model 
in which the coupling between the z-components of spins vanishes.
It has static properties which are similar to those of the 
``plane rotator'' model, in which the spins have only two components.
The static properties which of both models have been determined  
by numerical simulation 
\cite{Gerling_Landau_statics,XYsimulations,Rotatorsimulations}
and found to be consistent with the predictions of the
Kosterlitz-Thouless theory.
For example, the susceptibility shows an essential singularity instead of
a power law divergence, and computer simulations show that vortex pairs 
unbind at \Tkt.
The model is critical, i.e.\ it has infinite correlation length $\xi$,
at all temperatures $T\leq \TKT$.
The spin-spin correlation function decays algebraically with distance 
for all $T\leq \TKT$, but with a power $\eta$ which varies with temperature.

The dynamic behavior of the model should be governed by the 
dynamic critical exponent $z$,
which describes the divergent behavior of the
relevant time scale \cite{HohenbergHalperin}, 
i.e.\ $\tau \propto \xi^{z}$ for all $T\leq\TKT$.
Recently, finite size scaling for critical dynamics 
in the neutron scattering function
has been developed \cite{ChenLandau},
and successfully applied to the study of a 3-dim.\ Heisenberg ferromagnet.
The \XY-model has true dynamics which can be determined by integrating 
the equations of motion for each spin;
its critical dynamics has been studied 
theoretically \cite{Villain,NelsonFisher,Menezes,Pereira}
with different predictions for the nature of the neutron
scattering function.
In contrast, the ``plane rotator'' model does not possess equations of motion;
it has only stochastic, i.e.\ purely relaxational time dependence,
which has also been examined by Monte Carlo 
simulation \cite{Rotatorsimulations}.
Note, however, that {\em different}
dynamic exponents are expected for stochastic and for
true dynamics \cite{HohenbergHalperin}.

In this paper we present the first large scale simulation study of the
true dynamic behavior of the \XY-model. 
Great care was taken to ensure that statistical as well as systematic errors
were both well understood and quite small.
An earlier, much less complete study \cite{Landau_Gerling_dynamics} 
indicated a rich
structure in the neutron scattering function which was not adequately
described by theory.
Our model is defined by the Hamiltonian
\beq{H}
{\cal H} \,=\, -J\sum_{nn} \left(            S_i^x S_j^x  \,+\,
                                             S_i^y S_j^y  
%%                         \,+\,  \lambda\,  S_i^z S_j^z         
                           \right)\,,
%%      \;\;\; \lambda\equiv 0 \,,
\eeq
where $\S_i$ is a {\em three-component} classical vector of length unity and 
the sum is over all nearest neighbor pairs.
The equation of motion of each spin is  \cite{Landau_Gerling_dynamics}
\beq{EoM}
\frac {d}{dt}\S_i \,=\, \S_i \,\times\, \Heff \,,
\eeq
where
\beq{Heff}
\Heff \,=\, -J \, \sum_{nn} 
         \left( S_j^x \ee_x  \,+\, S_j^y \ee_y \right) \,,
\eeq
and $\ee_x$ and $\ee_y$ are unit vectors in the x- and y-directions 
respectively.
Eq.\ (\ref{EoM}) represents a set of coupled equations and can be integrated
numerically.
(The plane rotator model has a hamiltonian of the same form
 as \eq{H}, but since the vectors have only {\em two components},
 an equation of motion cannot be defined in the same way as for the
 \XY-model.)

In section 2 we define the neutron scattering function,
provide dynamic finite size scaling equations, 
and summarize analytical results.
Section 3 describes the details of our simulations.
We present our data and subsequent analysis in section 4, 
and draw conclusions in section 5.

The results of our dynamic study prompted us to reexamine the 
static properties of the model.
In order to obtain more reliable values for the critical temperature
and the static critical exponent $\eta$, we carried out
static Monte Carlo simulations; our results are presented in the appendix.

%******************************************************************************
 \section{Background}
%******************************************************************************
%

%==============================================================================
 \subsection{Neutron Scattering Function \Sqw}
%==============================================================================
%
The neutron scattering function \Sqw\ 
(also called the dynamic structure factor)
is an experimental observable and is fundamental to the study of spin dynamics.
It is defined \cite{ChenLandau} 
for momentum transfer $\q$ and frequency transfer \om\ 
as the space-time fourier transform 
\beq{Sqw}
   S^{kk}(\q, \omega) \,=\, \sum_{\bf r,r'} e^{i\q \cdot ({\bf r}-{\bf r'})}
                      \int_{-\infty}^{+\infty} e^{i\omega t}\,
                    C^{kk}({\bf r}-{\bf r'},t) \,{\frac {dt} {{2\pi}}} 
\eeq
of the space-displaced, time-displaced
spin-spin correlation function 
\beq{Crt}
   C^{kk}({\bf r}-{\bf r'},t-t') \,=\, 
                \left<S_{\bf r}^k(t)S_{\bf r'}^k(t')\right> \,,
\eeq
where $k=x,y$, or $z$ is the spin component,
displacement ${\bf r}$ is in units of lattice spacings, 
and the angle brackets $\langle \,...\,\rangle$ denote the 
thermal ensemble average.
Note that in the 2-dim.\ \XY-model, 
       $\left<S_{\bf r}^k(t)\right> \ident 0$  for all components $k=x,y,z$.
The equations of motion (\ref{EoM}) are time reversal invariant,
therefore $C^{kk}({\bf r}-{\bf r'},t-t')$ is symmetric in $t$ and $t'$, 
and \Sqw\  is real-valued.

The neutron scattering function generally depends 
on the correlation length $\xi$ and 
may be written in the form \cite{HalperinHohenberg}
\beq{S5}
   S_{\xi}^{kk}(\q,\omega) \,=\, {\frac {1} {\omega_m^{kk}(\q,\xi)}} 
                              \,S_{\xi}^{kk}(\q)\; 
      f^{kk} \left( {\frac {\omega} {\omega_m^{kk}(\q,\xi)}},\q,\xi\right) \,,
\eeq
where $S_{\xi}^{kk}(\q)$ is the total intensity given by
\beq{Sq}
     S_{\xi}^{kk}(\q) \,=\,
   \int_{-\infty }^{\infty} S_{\xi}^{kk}(\q,\omega) {d\omega} \,,
\eeq
and $f^{kk}$ is a normalized shape function,
  $ \int_{-\infty}^\infty f^{kk}(x,\q,\xi) dx \,=\, 1 $.
The characteristic frequency $\omega_m^{kk}(\q,\xi)$ is a median
frequency determined by the constraint
\beq{om1} 
          {\frac{1}{2}} S_{\xi}^{kk}(\q) \,=\,
              \int_{-\omega_m^{kk}}^{\omega_m^{kk}} S_{\xi}^{kk}(\q,\omega) 
                                 {{d\omega}} \,.
\eeq
In dynamic scaling theory it is assumed that the median frequency
$\omega_m^{kk}(\q,\xi)$ is a homogeneous function of $\q$ and $\xi$, i.e.,
\beq{om2} 
    \omega_m^{kk}(\q,\xi) \,=\, q^{\normalsize z^{kk}} \, \Omega^{kk}(q\xi) \,,
\eeq
where $z^{kk}$ is the dynamic critical exponent 
and $\Omega^{kk}$ is another shape function,
and that the function $f^{kk}$ depends only on
the product of $q\xi$ but not on $\q$ and $\xi$ separately.
Therefore $S_{\xi}^{kk}(\q,\omega)$ as given in \eq{S5} 
simplifies to
\beq{sk1} 
  S_{\xi}^{kk}(\q,\omega)  \,=\,  {\frac {1}{\omega_m^{kk}(q\xi)}} \;
  S_{\xi}^{kk}(\q) \; f^{kk} \left( {\frac{\omega}{\omega_m^{kk}
                                            (q\xi)}},q\xi\right) \,.
\eeq

Note that the $z$-component of the magnetization is conserved during the time
evolution. Thus the neutron scattering function 
${S}_{\xi}^{kk}(\q,\omega)$
can be regrouped in terms of symmetry into 
the out-of-plane component \Szz\  
and the in-plane component
\beq{Sxx}
{S}_{\xi}^{xx}(\q,\omega)  \,=\, {S}_{\xi}^{yy}(\q,\omega)  \,,
\eeq
with different physical behavior.
(As reported in sections \ref{Sec:scale_o} and \ref{Sec:scale_S},
 we find that for the two different components the
 exponent $z$ is the same, but the scaling functions $\Omega$ and $f$ differ.)

%==============================================================================
 \subsection{Dynamic Finite Size Scaling}\label{DFSS}
%==============================================================================
%

At the critical temperature \Tkt\  and below, the \XY-model is
expected to be critical, with infinite correlation length $\xi$.
In this region the dynamic critical exponent $z$ can be extracted
by using the dynamic finite size scaling theory
developed by Chen and Landau \cite{ChenLandau}.

These authors also introduced a frequency resolution function
in order to smoo\-then the effects of finite length of time integration,
which was not necessary 
for most of the analyses 
in our study 
because of much longer integration times
(see section \ref{Sec:Simulations}).
Their dynamic finite size scaling relations \cite{ChenLandau} 
can then be simplified to
\beq{scal}  
   \frac {S_L^{kk}(\q,\omega)} {L^{z}\,S_L^{kk}(\q)}
                  \,=\,   
                          G^{kk}(\omega L^{\normalsize z^{kk}},qL)
\eeq
(replacing a factor $\omega$ in front of $S_L^{kk}(q,\omega)$), 
and 
\beq{ob} 
    \omega_m^{kk}(\q,L) \,\equiv\, \omega_m^{kk}(qL)   
                   \,=\,  L^{\normalsize -z^{kk}} \,\Omega^{kk}(qL) \,,
\eeq
analogous to \eq{om2} and \eq{sk1}.

We see that two different ways emerge to test dynamic scaling and 
to estimate the dynamic critical exponent $z$ :
Firstly, from \eq{ob},
$z$ is given by the slope of a graph of $\log \omega_m$ versus $\log L$ 
at fixed value of $qL$.
Secondly, \eq{scal} implies that for correctly chosen $z$
and at fixed value of $qL$, graphs of 
(${S_L^{kk}(\q,\omega)} \,/\, L^{z} \, {S_L^{kk}(\q)}$)
versus $\omega L^{z}$ 
should all fall onto the same curve for different lattice sizes.
Both procedures will only be valid for sufficiently large lattice size.

%==============================================================================
 \subsection{Analytical Results}
%==============================================================================
%
The dynamics of two-dimensional ferromagnets with easy-plane asymmetry,
specifically the \XY-model, 
were first analyzed by Villain \cite{Villain}
%for small temperatures and 
%in the harmonic approximation, 
%which is justifiable for large spin $s$
%and therefore also for our classical spins.
and by Moussa and Villain \cite{MoussaVillain}.
The in-plane scattering function 
was found to have a delta-function spin-wave peak at low temperature,
and a spin-wave peak of the form
\beq{Villain}
   S^{xx}(q,\omega) \,\sim\, \frac{1}{|\omega-\omega_q|^{1-\eta/2}}\,
\eeq
close to \Tkt. 
Here $\eta(T)$ is the critical exponent
describing the decay of the static spin-spin correlation function 
(and we now expect $\eta=\frac{1}{4}$ at \Tkt).
%
%%%%%%
%At high temperatures, Moussa and Villain predicted that $S^{xx}(q,\omega)$
%is given by the sum of two terms:
%
%\beq{Moussa_Villain_High_T}
%  S^{xx}(q,\omega) \sim 
%%%\left(\frac{2\xi}{c}\right)^{1-\eta/2} \left(
%      \frac{1-\eta/2}
%           {\left[1+\left(\frac{\hbar\omega-\hbar\omega_q)}{c/2\xi}\right)^2
%            \right]^{1-\eta/4}}
%     +\frac{\eta\pi/4}
%           {\left[1+\left(\frac{\hbar\omega-\hbar\omega_q)}{c/2\xi}\right)^2
%            \right]^{1/2-\eta/4}}
%%%     \right) 
%\;,
%\eeq
%%
%where $c$ is the spin-wave velocity, and $\xi$ the correlation length.
%%%%%%

Nelson and Fisher \cite{NelsonFisher} treated the classical XY model
in a fixed length hydrodynamic description for $T\leq \TKT$, 
without vortex contributions.
They obtained the transverse spin-spin correlation function,
\beq{NelsonC}
  C^{xx}(r,t) \,\sim\, \frac{1}{r^{\eta}} \; 
          \Phi_{\eta} \left( \frac{ct}{r} \right) \;, \mbox{~~~}
          \Phi_{\eta}(y) \,=\, \left\{ \begin{array}{ll}
                                                  1            &,\;\; y<1 \\
                                      (y+\sqrt{y^2-1})^{-\eta} &,\;\; y>1  \;,
                                   \end{array}
                               \right.
\eeq
for $r,\,ct\,\gg\,1$ and $ct \neq r$,
where $c$ is the spin-wave velocity.
The fourier transform of \eq{NelsonC} has the form
\beq{NelsonS}
  S^{xx}(q,\omega) \,\sim\, \frac{1}{q^{3-\eta}} \,
               \Psi\left(\frac{\omega}{cq}\right) \;,
\eeq
where the scaling function $\Psi$ behaves like
\beq{Psi}
      \Psi(y) \,\sim\, \frac{1}{|1-y^2|^{1-\eta}}
\eeq
around the spin-wave peak,
and 
\beq{largeomega}
  S^{xx}(q,\omega) \sim \omega^{\eta-3}
\eeq
for large values of $\omega/cq$.

Nelson and Fisher also predicted that
 the dynamic critical exponent $z$, \eq{om2}, 
which is expected \cite{HohenbergHalperin} to be $z=d/2$ for $d>2$, is
\beq{z}
  z \,=\,1  \mbox{~~~for~~} d\leq 2 \,.
\eeq
Note that the value $z=1$ and a {\em linear dispersion relation}
are also implicit in the argument $\omega/cq$ 
of the  scaling function $\Psi$ in \eq{NelsonS}.

Finally, both Villain and Nelson and Fisher predicted  a very narrow 
(delta-function) spin-wave peak
in the out-of-plane function \Szz, at $\omega=cq$.

More recently, Menezes, Pires, and Gouv\^{e}a \cite{Menezes}
have performed a low temperature calculation
which includes the contribution of out-of-plane fluctuations
to the in-plane correlation functions.
They worked in the harmonic spin-wave approximation
which is justifiable for large spin $s$
and therefore also for our classical spins,
and they used a projection operator technique.
They found a spin-wave peak similar to that of Nelson and Fisher,
\def\obar{\hat{\omega}}
\beq{MenezesSW}
   S^{xx}(q,\omega) \,\sim\, \eta^2 \frac{1}{q^3 |\obar|\;|1-\obar^2|} \;,
    \mbox{~~~if~~}  \left\{ \begin{array}{l}
                      e^{-1/\eta} \,\ll\, k  \,\ll\,\pi \mbox{~~and~}\\
                      e^{-1/\eta} \,\ll\, |1-\obar^2|  \,\ll\,\pi     \;,
                          \end{array}
                   \right.
\eeq
where $\obar=\omega/(cq)$.
In addition to the spin-wave peak, 
they found a logarithmically diverging central peak, 
i.e.\ a signal at very small $\omega$,
which diverges like
\beq{Mcentral}
  S^{xx}(q,\omega) \,\sim\, \frac{1}{q} \,\frac{1}{\log|\obar|}
   \mbox{~~~+~ (less divergent terms)}  \;.
\eeq

Of course, a central peak at small temperature can also be caused
by other mechanisms, 
for example vortex pairs diffusing like a dilute pair of 
solitons \cite{Pereira}.

The dynamic behavior of the \XY-model is different {\em above} \Tkt.
For a phase transition of Kosterlitz-Thouless
type, the spin stiffness should drop discontinuously to zero at \Tkt,
i.e.\ the spin-wave peak is predicted to
disappear \cite{NelsonFisher,NelsonKosterlitz}.
Above \Tkt, vortex-antivortex pairs unbind,
and their diffusion leads to a strong central peak in \Sqw.

Mertens et al. \cite{MertensOne,MertensTwo} calculated \Sqw\  above \Tkt,
assuming an ideal dilute gas of unbound vortices moving in the presence
of renormalized spin-waves, and screened by the remaining vortex-antivortex 
pairs.
They found a Lorentzian central peak for \Sxx, 
\beq{MertensSxx}
   S^{xx}(q,\omega) \,\sim\, \frac{\gamma^3 \xi^2}
        {\left(\omega^2 + \gamma^2[1+(\xi q)^2]\right)^2} \;,
\eeq
and a Gaussian central peak for \Szz,
\beq{MertensSzz}
   S^{zz}(q,\omega) \,\sim\, \frac{n_v \bar{u}}{q^3}
        e^{-\left(\frac{\omega}{\bar{u}q}\right)^2} \;,
\eeq
where 
$\gamma = \frac{1}{2}\sqrt{\pi}\bar{u}/\xi$, 
$\bar{u}$ is the rms vortex velocity,
$\xi$ the correlation length, and
$n_v \sim (2\xi)^{-2}$ the free vortex density;
and they compared their results to numerical simulations (see below).

%==============================================================================
 \subsection{Previous numerical work}\label{Sec:previous_numerical}
%==============================================================================
%%
Gerling and Landau \cite{Landau_Gerling_dynamics} carried out 
spin dynamics simulations on the XY model with $L\leq 204$
and found both spin-wave peaks and a central peak. The resolution 
was too limited, however, to allow quantitative comparison with theory
or to extract an estimate for the dynamic exponent.

Mertens et al.\ \cite{MertensOne,MertensTwo} 
performed spin dynamics simulations,
with fixed system size $L=100$ and very low statistics 
(3 starting configurations) at $T=0.5$ and $T=1.1$.
Below $T_c$ they observed only spin-wave peaks in both \Sxx\ and \Szz;
above $T_c$ they saw a strong central peak and a weak spin-wave peak in \Sxx,
and vice versa in \Szz.
The width and intensity of the central peaks were compatible with 
\eq{MertensSxx} and \eq{MertensSzz}.

Other earlier numerical work on dynamical behavior has been 
exploratory \cite{Kawabata,Shirakura}.

%==============================================================================
 \subsection{Experiments}\label{Sec:experiments}
%==============================================================================
%%
The closest physical realizations of the XY-model are 
materials with very large anisotropy,
more generally described by strongly anisotropic Heisenberg models.
%
%The $XY$-model is the limiting case of an anisotropic Heisenberg model.
Several experiments have studied the dynamics of 
%corresponding anisotropic 
such materials \cite{Wiesler,experiments},
like $Rb_2CrCl_4$, $K_2CuF_4$, and $CoCl_2$.

In a recent study on 
stage-2 $CoCl_2$ intercalated graphite,
Wiesler et al. \cite{Wiesler}
found four temperature regimes with different behavior.
There are indications of a Kosterlitz-Thouless transition at a temperature 
``$T_u$'',
though some properties disagree with KT predictions.
Between temperatures ``$T_l$'' and ``$T_u$'', they observed  spin-wave peaks.
It is not clear whether ta central peak is present there.
(In this temperature region the long range part of the scattering 
function shows true 2-dimensional character, 
whereas for $T<T_l$ 3-dimensional correlations develop.)
Above ``$T_u$'', the in-plane scattering function showed the expected
central peak, and the out-of-plane function exhibited
damped spin-waves.

In experimentally available materials both defects
and the effects of residual threedimensional couplings limit
the effective size of the two-dimensional KT-like system to
$L_{\exp}=O(100)$ lattice spacings \cite{Bramwell,Wiesler}.
Remarkably, this size is similar to the lattice sizes of the present
numerical study.
For further discussion and an extensive listing of relevant literature,
see the recent overview contained in \cite{Wiesler}.

%******************************************************************************
 \section{Simulations} \label{Sec:Simulations}
%******************************************************************************
%

We have studied the two-dimensional 
classical \XY-model with Hamiltonian given in \eq{H}
on $L\times L$ lattices with periodic boundary conditions
for $16\leq L\leq 192$,
at temperatures $T=0.4,\  0.6,\  0.7,\ 0.725,\ $ and $0.8$
in units of $J/k_B$.
Most of these temperatures are in the critical region
$T\lsim \TKT = 0.700(5)$ (see appendix).

Equilibrium configurations were created at each temperature using 
a Monte Carlo method which combined
cluster updates of the $x$ and $y$ spin components 
(using the Wolff embedding method \cite{SW,WolffXY})
with vectorized Metropolis and overrelaxation \cite{Overrelaxation} 
spin reorientations.
After each single-cluster update, two Metropolis and eight overrelaxation
sweeps were performed.
Use of the cluster algorithm was important,
since critical slowing down was severe
for most of our simulations;
the inclusion of cluster flipping
reduced Monte Carlo autocorrelation times at $L=192$ and $T=0.6$ 
from more than $300$ to about $3$ hybrid sweeps,
while requiring only a factor of two more CPU time per sweep.
We performed 200 hybrid sweeps between equilibrium configurations,
and discarded the first 5000 hybrid sweeps for equilibration.

Between 500 and 1200 equilibrium configurations were generated
for each lattice size and temperature.
We found this many configurations to be necessary in order to
sufficiently reduce statistical errors in the resulting
neutron scattering function.
The error bars in our figures represent the statistical errors
for averages over the equilibrium configurations,
drawn from the canonical ensemble.

Starting with each equilibrium configuration, 
the time dependence of the spins was determined from the 
coupled set of equations of motion, \eq{EoM},
and was integrated numerically
using a vectorized fourth order predictor-corrector method \cite{ChenLandau},
with a time step size of $\delta t=0.01J^{-1}$.
The maximum integration time was generally $\tmax = 400 J^{-1}$;
a few runs were also performed for lattice size $256\times 256$ with 
$\tmax= 800 J^{-1}$ and produced the same physical results.

The time-displaced, space-displaced spin-spin correlation functions 
$C({\bf r}-{\bf r'},t-t')$, \eq{Crt},
were measured for each time integration, with 
\beq{tcut}
  0 \,\leq\, t' \,\leq\, 0.1 \;\tmax                 \mbox{~~~and~~~}
  0 \,\leq\, (t-t') \,\leq\, \tcut \,\ident\, 0.9 \;\tmax  \,,
\eeq
and were then averaged.
By fourier transformation in space and in time, \eq{Sqw}, we obtained
the neutron scattering function \Sqw.
The time integration in \eq{Sqw} was performed using Simpson's rule,
with a time step of $0.1J$, 
which has been shown \cite{ChenLandau} to be sufficiently small.

To reduce memory and computer time requirements,
we restricted ourselves to momenta $\q=(q,0)$ and $(0,q)$,
with $q$ determined by the periodic boundary conditions,
\beq{q}  
     q  \,=\,  n_q \;{\frac {2\pi} {L}}, \hspace{1.5cm}
     n_q  \,=\, 1,2,\dots,L  \,,  \vspace*{2mm}
\eeq
and data from these two spatially equivalent directions were averaged together
to further enhance the statistical accuracy.
We used fast fourier transforms to increase the efficiency of the
program in calculating correlation functions.

%%
%%%============================================================================
%% \subsection{Frequency resolution}    \label{Sec:tmax}
%%%============================================================================

The frequency resolution $\DO$ of our results is determined by the
time integration cutoff 
$\tcut = 0.9\: \tmax$, see \eq{tcut},
which will introduce oscillations of period $2\pi/\tcut$
into \Sqw.
Since we observed very sharp spin-wave peaks (see section \ref{Sec:Sqw}),
we chose to integrate the equations of motion to very large times.
We used  $\tmax=200 J^{-1}$ for $L\leq 96$, 
and      $\tmax=400 J^{-1}$ for $L\geq 128$.
A theoretical delta-function in frequency will then become a widened peak
with a width at half maximum of
\beq{DO}
   \DO \,\approx\, 1.2\, \frac{\pi}{\tcut} \,=\, \left\{ \begin{array}{ll}
                                           0.021\, J, & L \leq  96 \\
                                           0.010\, J, & L \geq 128 \,.
                                                     \end{array}
                                             \right.
\eeq
in the simulation data.
To smoothen the oscillations, previous spin dynamics 
studies \cite{Landau_Gerling_dynamics,ChenLandau}
have employed a frequency resolution function,
replacing 
\beq{smoothen}
 C^{kk}(r,t)  \mbox{~~~~~by~~~~~}
      C^{kk}(r,t) \;\; \exp(-\half(t\,\delta\omega)^2)
\eeq
to compute \Sqw.
Because of the large values of $\tmax$ in our study,
we achieved a very small frequency resolution $\DO$, 
and the oscillations were not noticeable for most of our data.
We therefore did not generally use a frequency resolution function 
and could significantly simplify the analysis of our data.

The numerical integration of the equations of motion can potentially
become unstable at very large integration times.
We checked that for our calculations,
in which we integrate to much larger times than 
previous studies, 
we do not encounter this problem.
We verified that the constants of motion 
(energy and magnetization in $z$-direction)
do remain constant, 
with a relative variation of less than $3 \times 10^{-6}$.
We also verified that the neutron scattering function 
remains virtually unchanged when an additional integration 
of length $t=200J^{-1}$
is performed from each equilibrium configuration before starting
to calculate time-displaced spin-spin correlation functions.
All simulations were carried out using highly vectorized programs
on the Cray C90 at the Pittsburgh Supercomputing Center.

%******************************************************************************
 \section{Results}\label{Sec:Results}
%******************************************************************************
%
We now present our results for the dynamic structure factor
\Sqw, its dependence on temperature, frequency, momentum, and lattice size,
and we analyze its dynamic scaling behavior.
With few exceptions 
%%(figs.\ 1(b), 5(b),(c)  and fig. 6),
%%Except for section (\ref{Sec:large_omega}), 
we have analyzed the data without the use of a frequency resolution function
\eq{smoothen}.
(The effects of such a function, and of integrating to shorter maximum times
are described together with initial results in reference \cite{SDshort}.)

% FIGURE 1--------------------------------------------------------------
\setlength{\LLL}{\textwidth}
\addtolength{\LLL}{-20mm}
\setlength{\SSS}{\textheight}
\addtolength{\SSS}{-7em}
\divide\SSS by 2
\begin{figure}[htbp] 
  \begin{center}
    \vskip-10mm
      \parbox[t]{\LLL}{\psfig{file=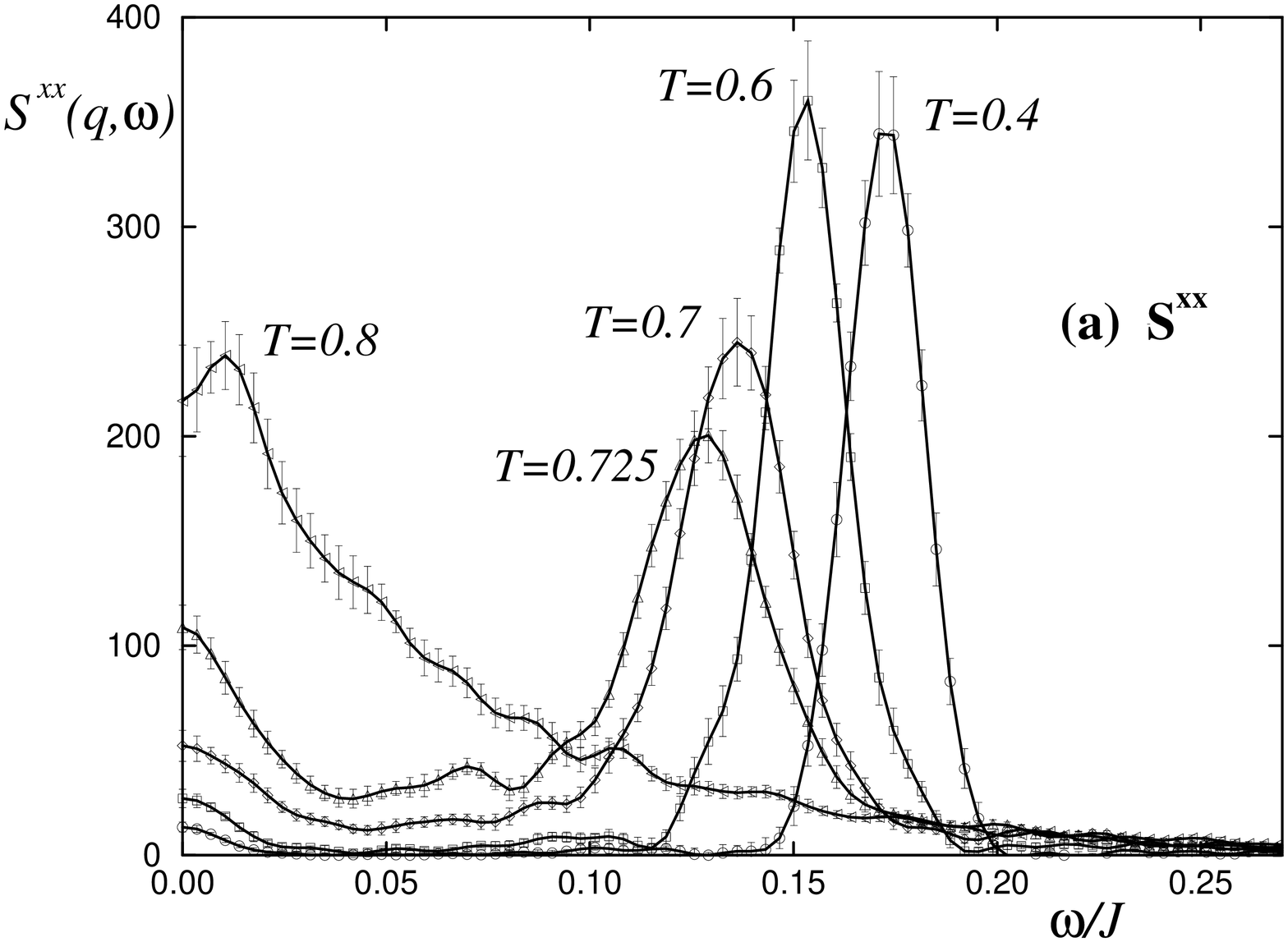,angle=-90,width=\LLL,height=\SSS} }
      \\[-1ex]
      \parbox[t]{\LLL}{\psfig{file=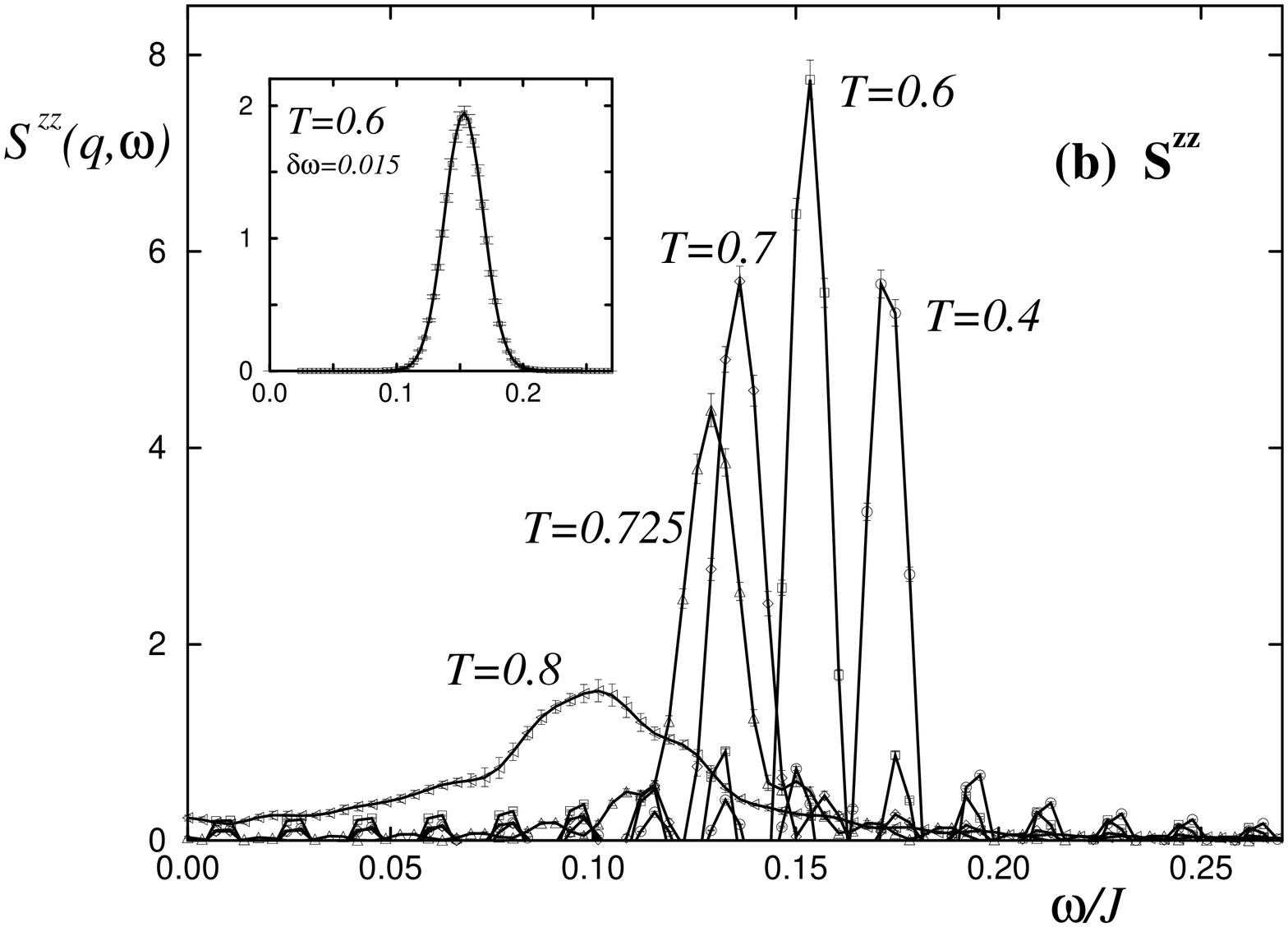,angle=-90,width=\LLL,height=\SSS} }
 \vskip-3ex
  \mycaption{1}{Temperature dependence of the
                 neutron scattering function \Sqw\,
                 as a function of frequency $\omega$.
                 %% for temperatures $T=0.4,\;T=0.6,\;0.7,\;0.725$ and $0.8$
                 %% in units of $J/k_B$.
                 The transition temperature is $\TKT\sim 0.700(5)$.
                 Lattice size $L=128$ and 
                 momentum $q\,=\,2\times\frac{2\pi}{L}\,=\,\frac{\pi}{32}$
                 in all cases. 
                 (a) $xx$-component. 
                 (b) $zz$-component;
                              the inset shows the data at $T=0.6$,
                              smoothened with a resolution function, 
                              \eq{smoothen}, with $\delta\omega=0.015$.
                }%%mycaption
%                 Note the sharp spin-wave peaks, and the additional structure
%                 in the $xx$-component at all temperatures.
%                 The intensity is much higher for the $xx$-component.
 \end{center}
\end{figure}
%--------------------------------------------------------------

In order to investigate the $XY$-model in the critical phase,
we chose several temperatures $T\leq \TKT$, including the
best previous estimate of $\TKT\approx 0.725 J/k_B$ 
\cite{Gerling_Landau_statics},
and one temperature well above the transition, $T=0.8 J/k_B$.
The results of our analysis prompted us to perform additional
static Monte Carlo studies, which are described in the appendix.
They provided an improved  estimate of 
$\TKT = 0.700(5) J/k_B$.
In order to elucidate the situation closer to the transition,
we performed additional (but less extensive)
spin dynamics simulations at $T=0.700 J/k_B$.

%==============================================================================
 \subsection{Spin-Wave Peak} \label{Sec:Sqw}
%==============================================================================
%
%% TEMPERATURE DEPENDENCE
%
Figure 1 shows the temperature dependence of \Sqw\ 
as a function of $\omega$,
for lattice size $L=128$
and fixed small momentum  $q=\pi/48$ (i.e.\ $n_q=2$ in \eq{q}).
Here, as in other results which we shall show, the error bars are determined
from the statistical variation of results obtained from different 
initial spin states.
The in-plane component \Sxx, fig.\ 1(a), exhibits a very strong and 
moderately sharp spin-wave peak at temperatures $T\lsim \TKT$. 
Even at the lowest temperature, however, the width of the peak is larger than
the minimum value \eq{DO}  due to  finite cutoff time.
The position of the peak moves towards lower \om\ as the temperature increases,
and the peak broadens slightly.
Just above the transition, at $T=0.725$, there is still 
both a strong spin-wave peak and a sizeable central peak in \Sxx.
At higher temperature, the spin-wave peak disappears completely
(for this low momentum) and only a large central peak remains.
Note that from KT-theory \cite{KT} one would expect complete disappearance
of a spin-wave peak at all $T\mygt\TKT$.

There is sizeable additional structure in \Sxx\ away from the spin-wave peak
at temperatures up to \Tkt.
We will discuss this structure in the following sub-section.

The out-of-plane component \Szz, shown in fig.\ 1(b), 
is two orders of magnitude weaker
than the in-plane component.
It exhibits a very sharp spin-wave peak for $T\leq \TKT$,
whose width is limited by our $\omega$-resolution.
The finite time cutoff \eq{DO} produces very noticeable oscillations in \Sqw.
(The magnitude of these oscillations is  minute compared
to the intensity of the spin-wave peaks in \Sxx.)
The oscillations can be smoothened by convoluting \Sqw\  
with a gaussian resolution function in frequency, as is shown in the inset.
%%%This procedure masks, however, the sharp nature of the spin-wave peak.
No central peaks are visible in \Szz\ at $T\leq \TKT$.
At $T=0.725$, the peak in \Szz\ is still present, with a larger
width similar to that in \Sxx.
In contrast to \Sxx,
there is a clear, but weak,
spin-wave peak in \Szz\, 
even at $T=0.8 > \TKT$ and small momentum.
It is of similar intensity as the peak at lower temperatures.
(See also section \ref{Sec:experiments}).

% FIGURE 2--------------------------------------------------------------
\setlength{\LLL}{\textwidth}
\setlength{\SSS}{2mm}
\divide \LLL by 2
\addtolength{\LLL}{-\SSS}
\begin{figure}[tbp] 
  \begin{center}
    \vskip-5mm
    \mbox{
      \parbox[t]{\LLL}{\psfig{file=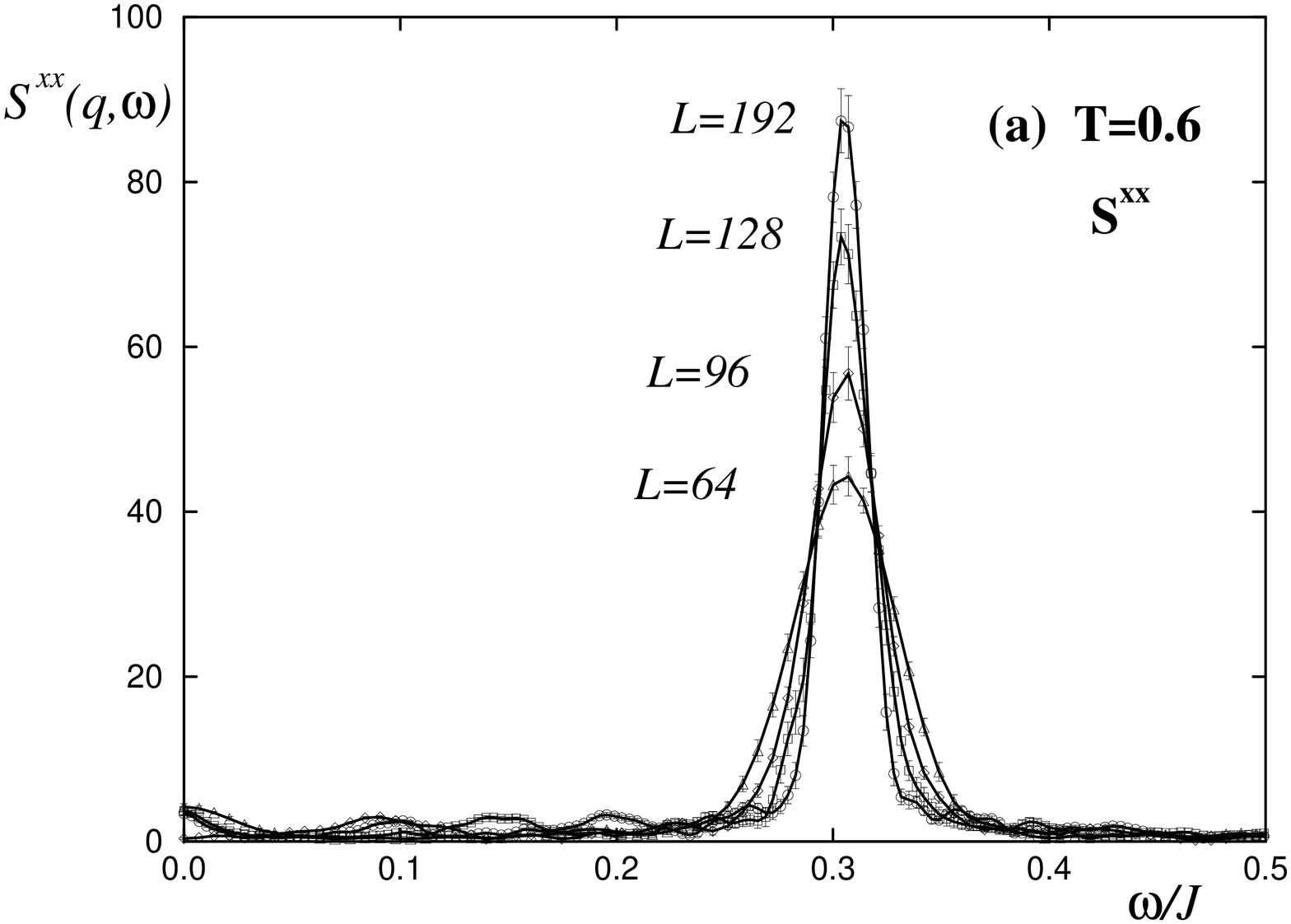,angle=-90,width=\LLL} }
%%      \hskip\SSS
      \parbox[t]{\LLL}{\psfig{file=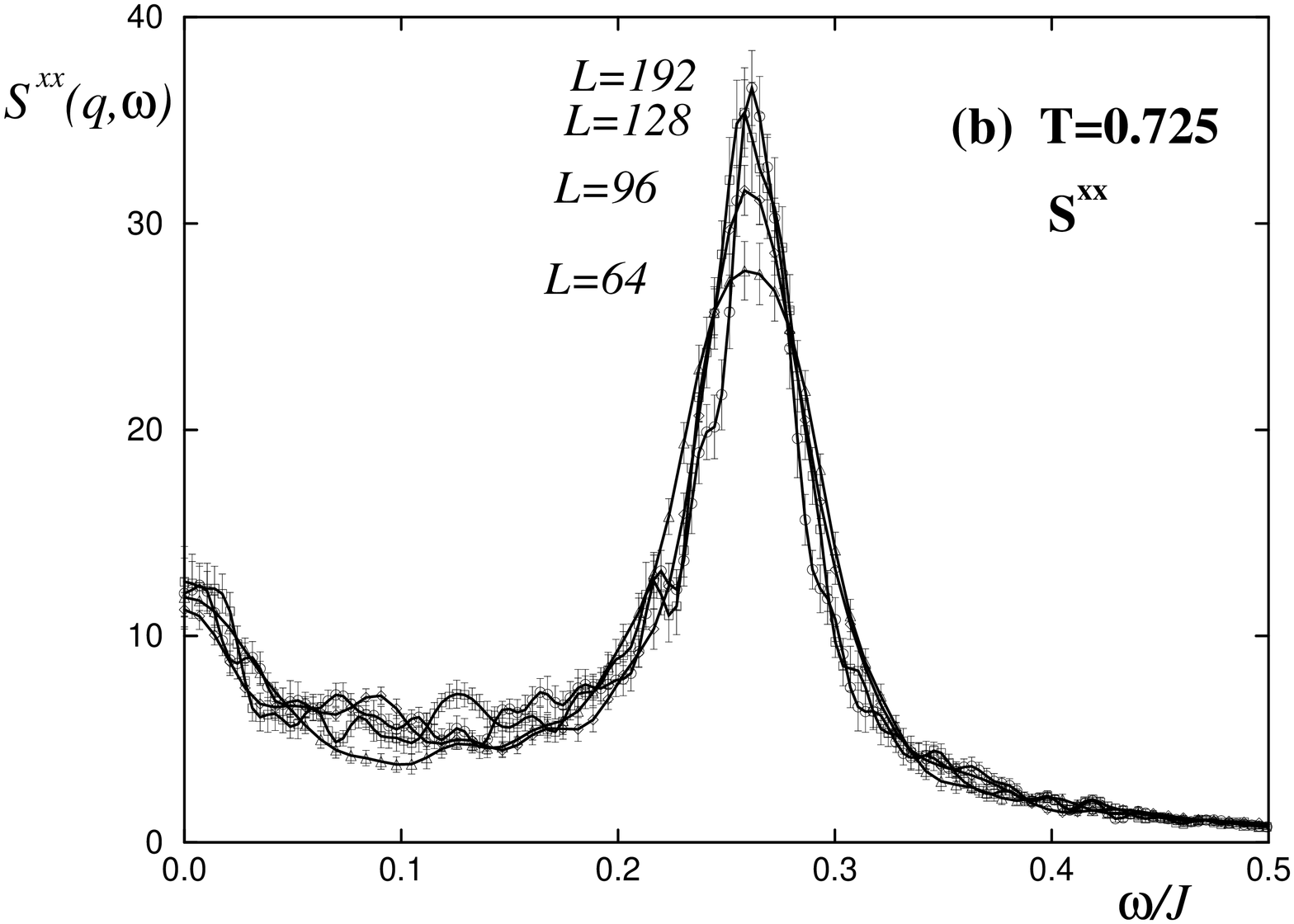,angle=-90,width=\LLL} }
    }%%mbox
    \\
    \mbox{
      \parbox[t]{\LLL}{\psfig{file=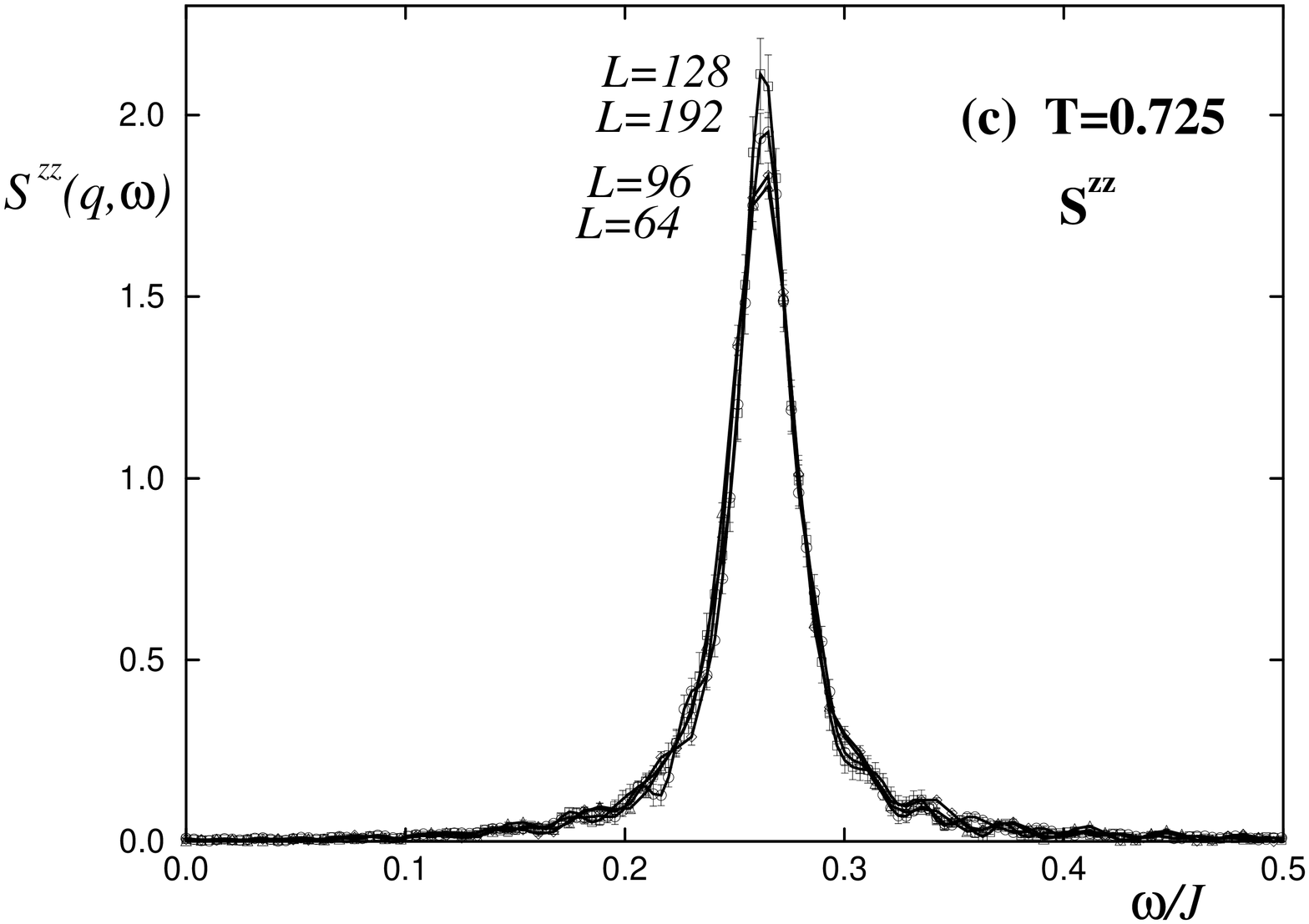,angle=-90,width=\LLL} }
%%      \hskip\SSS
      \parbox[t]{\LLL}{\psfig{file=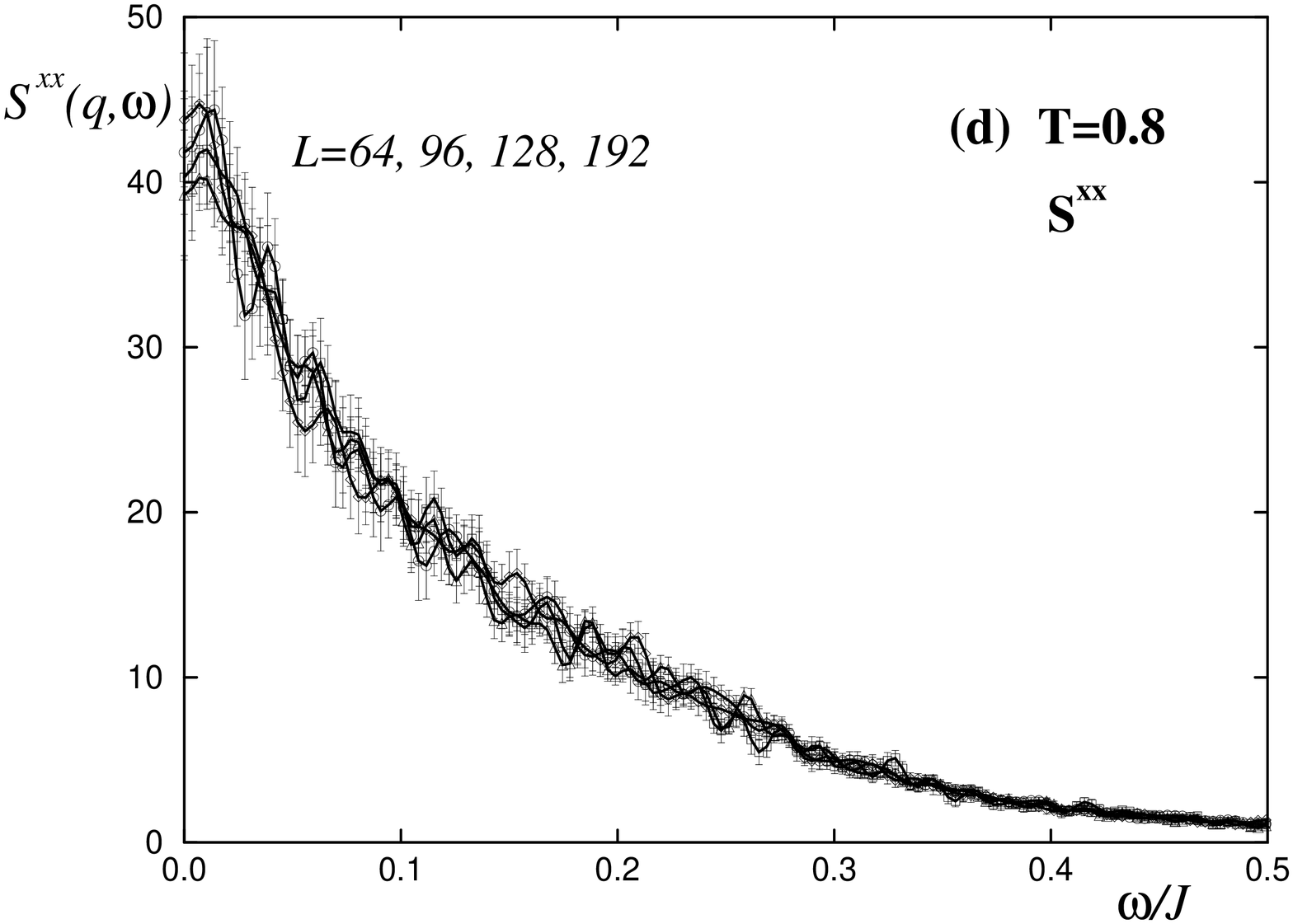,angle=-90,width=\LLL} }
    }%%mbox
    \\
    \mbox{
      \parbox[t]{\LLL}{\psfig{file=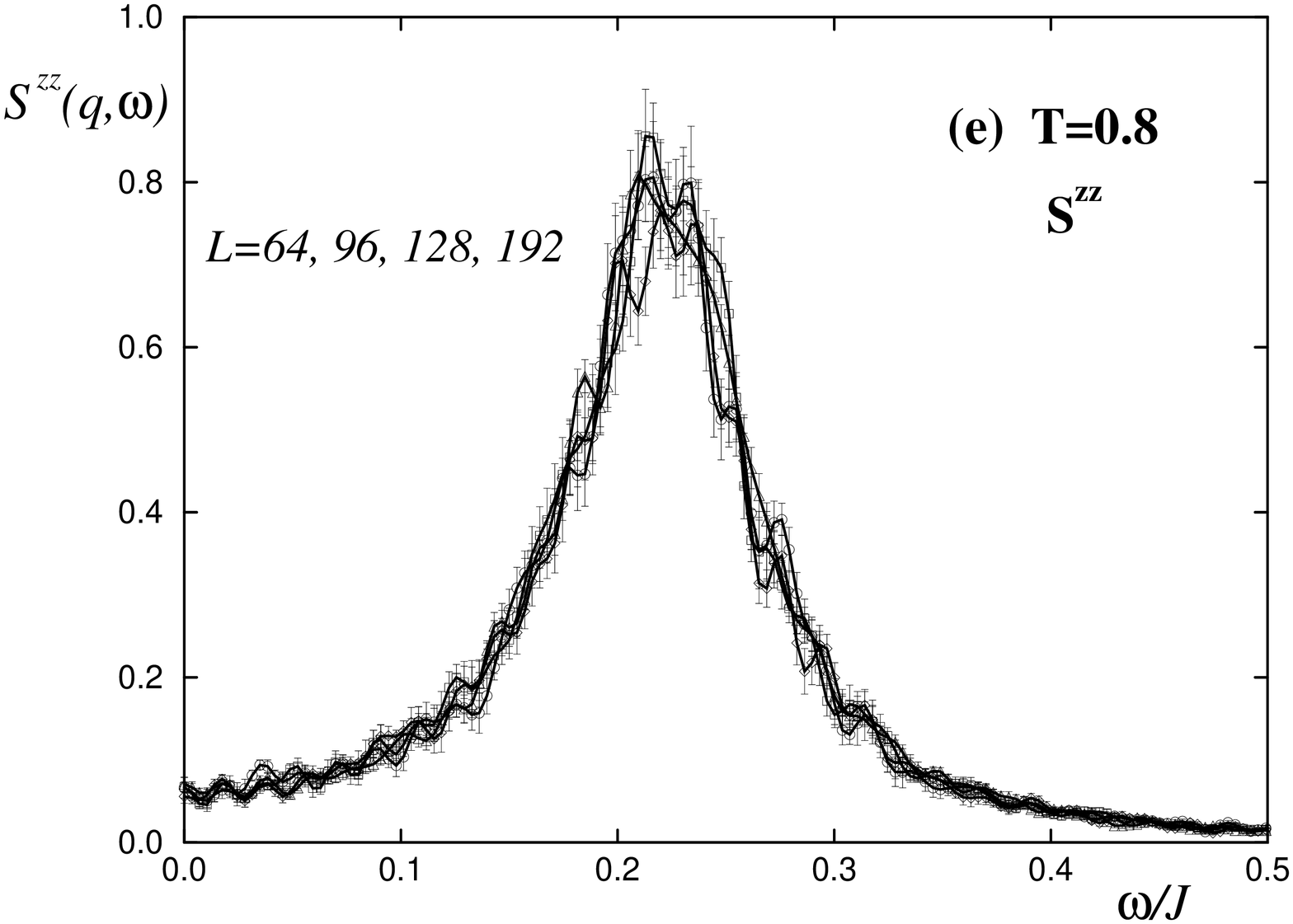,angle=-90,width=\LLL} }
    }%%mbox
    \mycaption{2}{Lattice size dependence of \Sqw, 
                   at fixed momentum $q=\pi/16$. 
                   (a)-(d): $xx$-component; (e): $zz$-component
                 }%%mycaption
  \end{center}
\end{figure}
%--------------------------------------------------------------

%% FINITE SIZE DEPENDENCE AT FIXED Q
%
Figure 2 shows the lattice size dependence of \Sqw, 
at fixed momentum $q=\pi/16$.
Below the transition, Fig. 2(a), the intensity of the spin-wave peak
depends strongly on lattice size, whereas its position is constant.
(The out-of-plane component \Szz\ is dominated by finite time cutoff effects 
for $T\mylt\TKT$, and we do not show it here).
Just above the transition, at $T=0.725$, the spin-wave peak in \Sxx\ 
appears to gain intensity slightly as $L$ increases,
whereas neither its central peak 
nor the spin-wave peak in \Szz\ 
show any finite size effects.
At higher temperature, fig.\ 2(d) and 2(e), 
there is no visible lattice size dependence 
in either \Sxx\  or \Szz.
Notice the two different vertical scales 
for \Sxx\ and \Szz.
Data taken for $L=16$ and $L=32$ exhibit such strong finite size rounding
that we have chosen not to show the data here.
%

% FIGURE 3--------------------------------------------------------------
\setlength{\LLL}{\textwidth}
\setlength{\SSS}{6mm}
\divide \LLL by 2
\addtolength{\LLL}{-\SSS}
\begin{figure}[tbp] 
  \begin{center}
    \vskip-5mm
    \mbox{
      \parbox[t]{\LLL}{\psfig{file=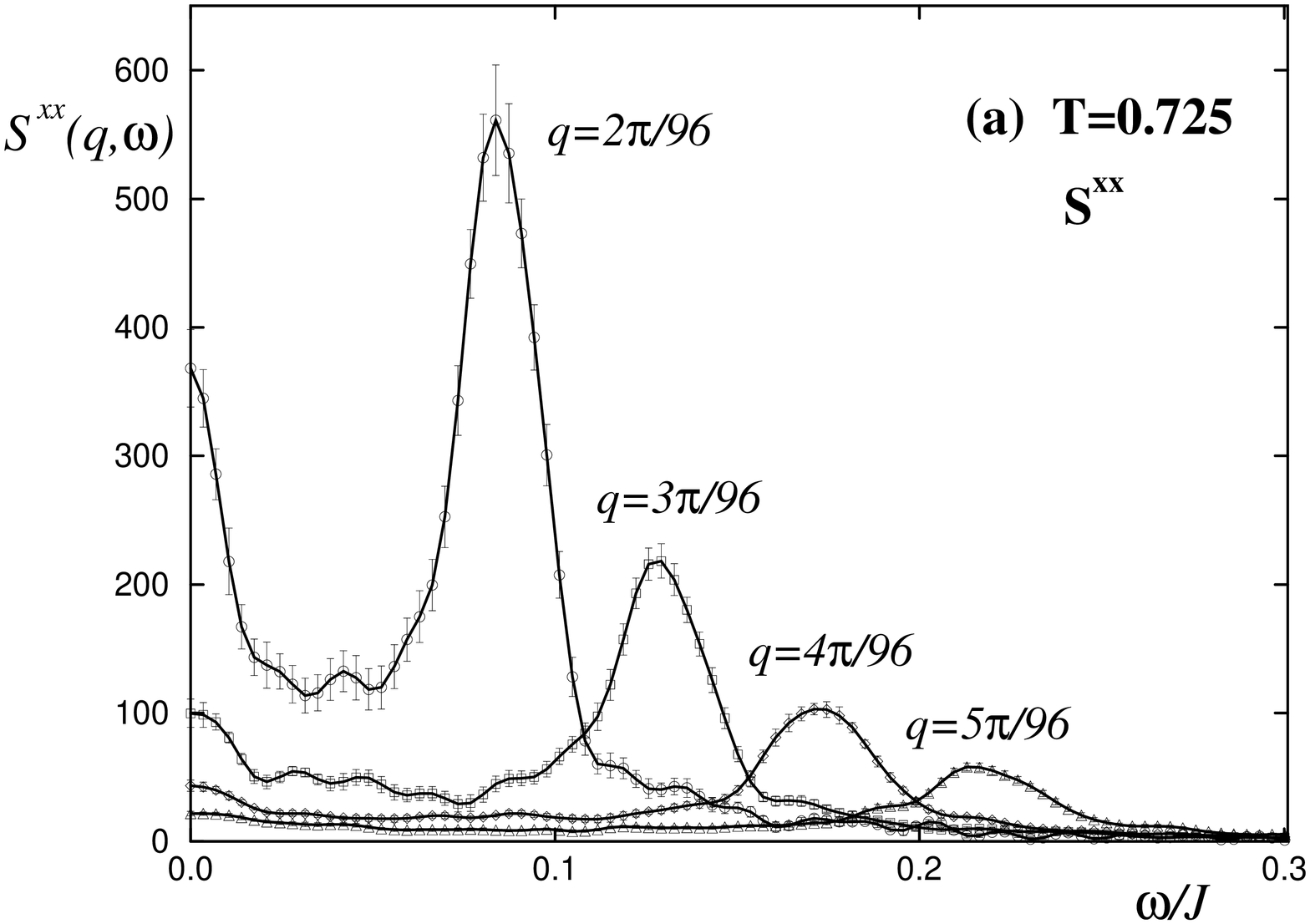,angle=-90,width=\LLL} }
%%      \hskip\SSS
      \parbox[t]{\LLL}{\psfig{file=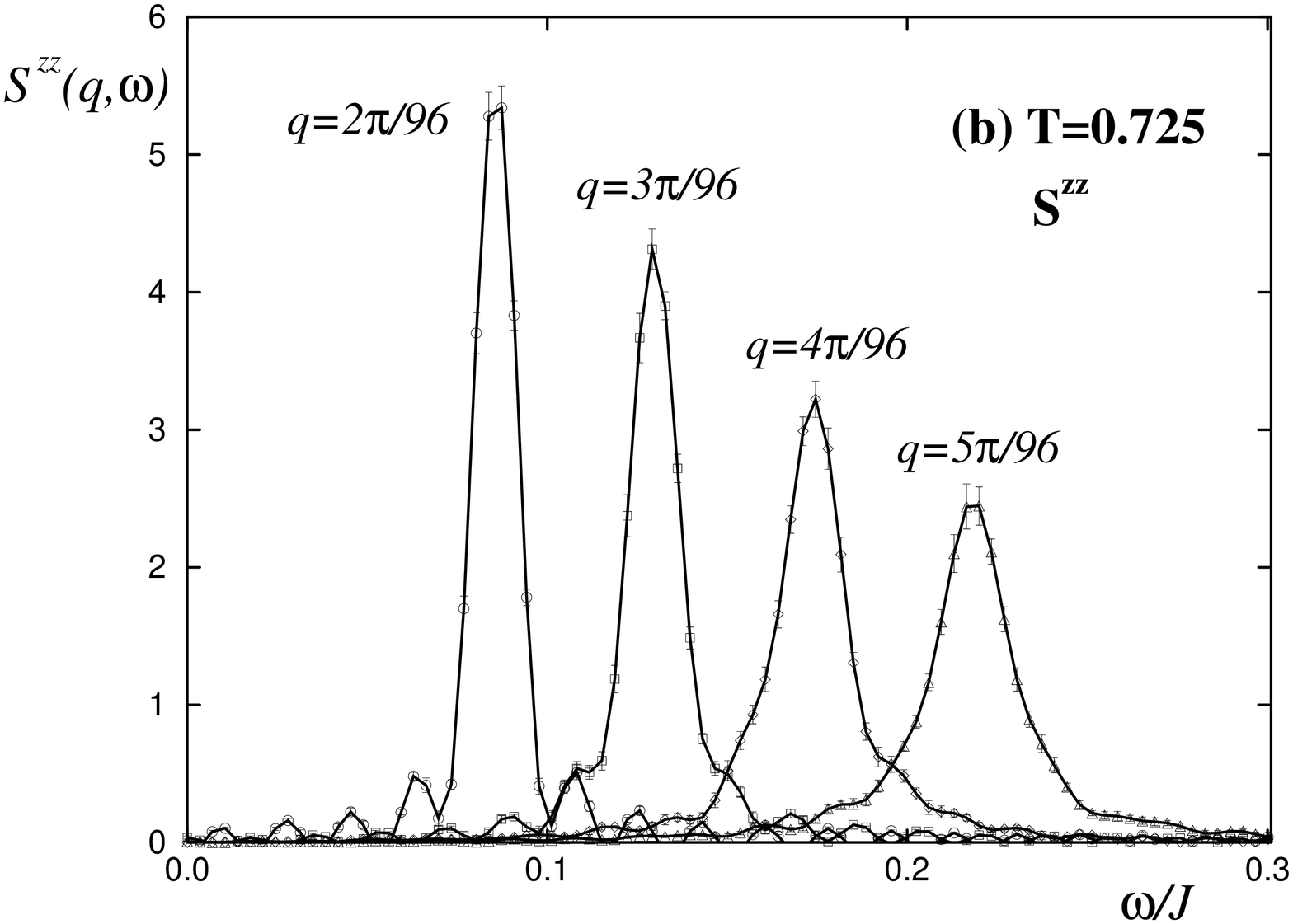,angle=-90,width=\LLL} }
    }%%mbox
    \\
    \mbox{
      \parbox[t]{\LLL}{\psfig{file=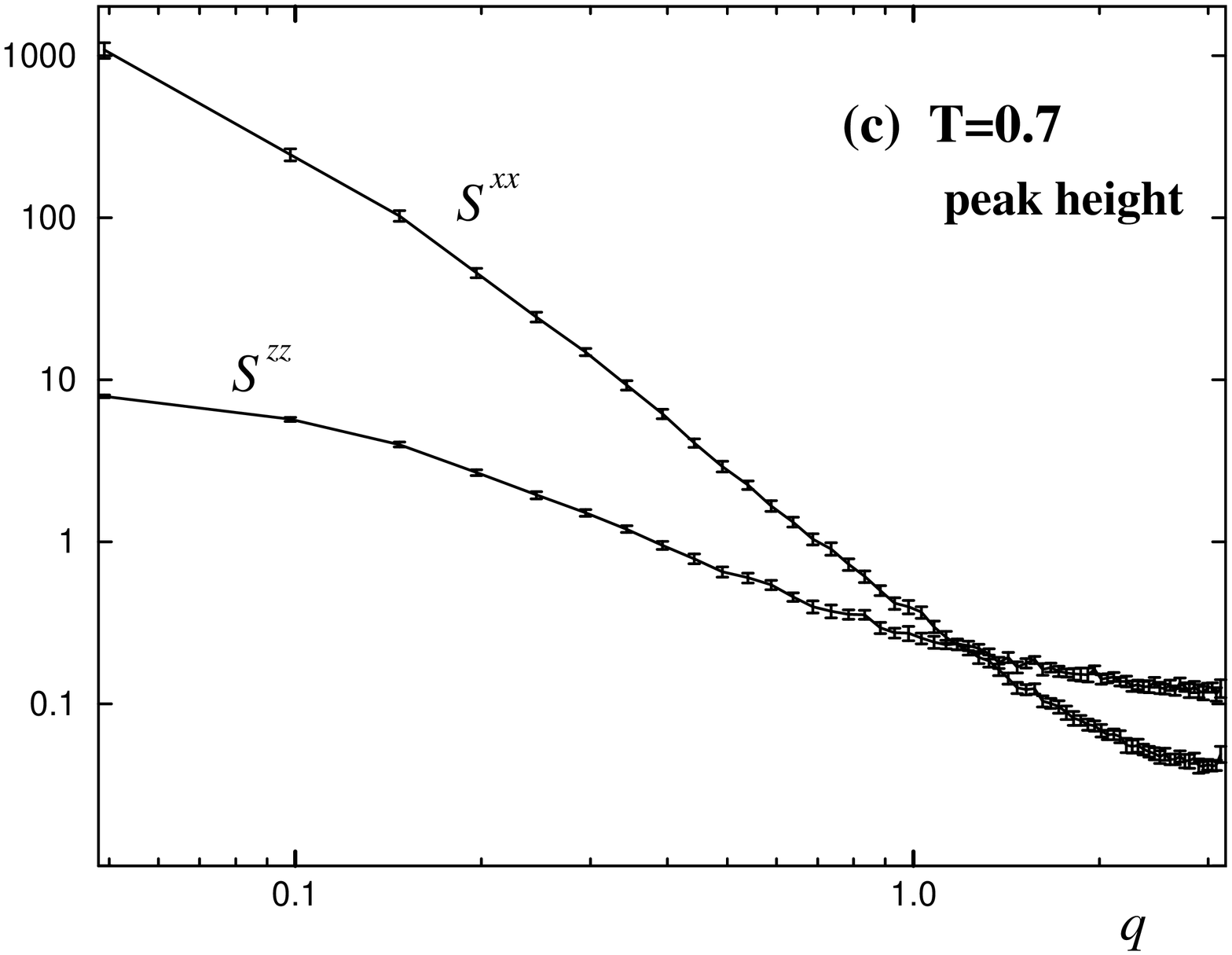,angle=-90,width=\LLL} }
%%      \hskip\SSS
      \parbox[t]{\LLL}{\psfig{file=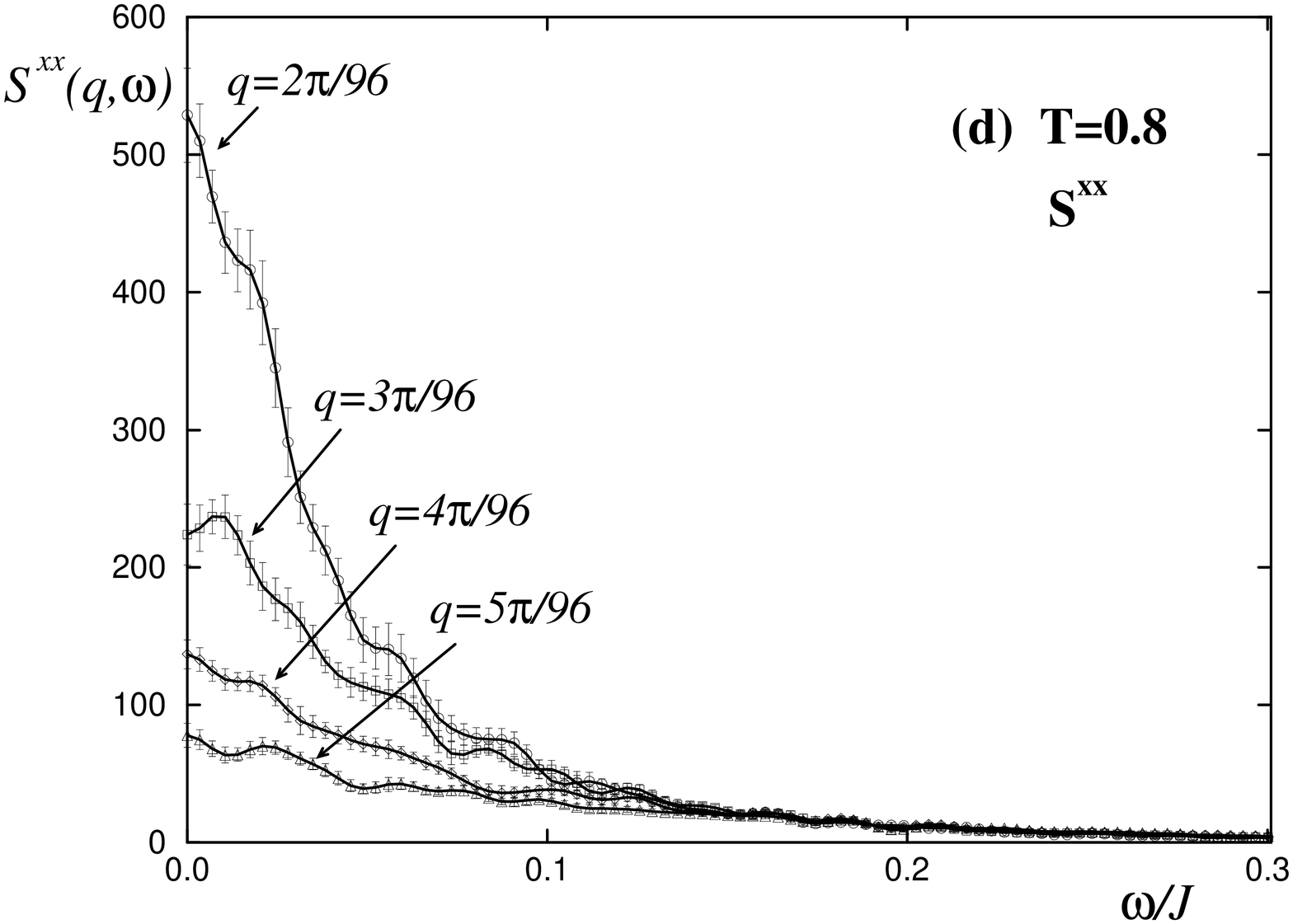,angle=-90,width=\LLL} }
    }%%mbox
    \\
    \mbox{
      \parbox[t]{\LLL}{\psfig{file=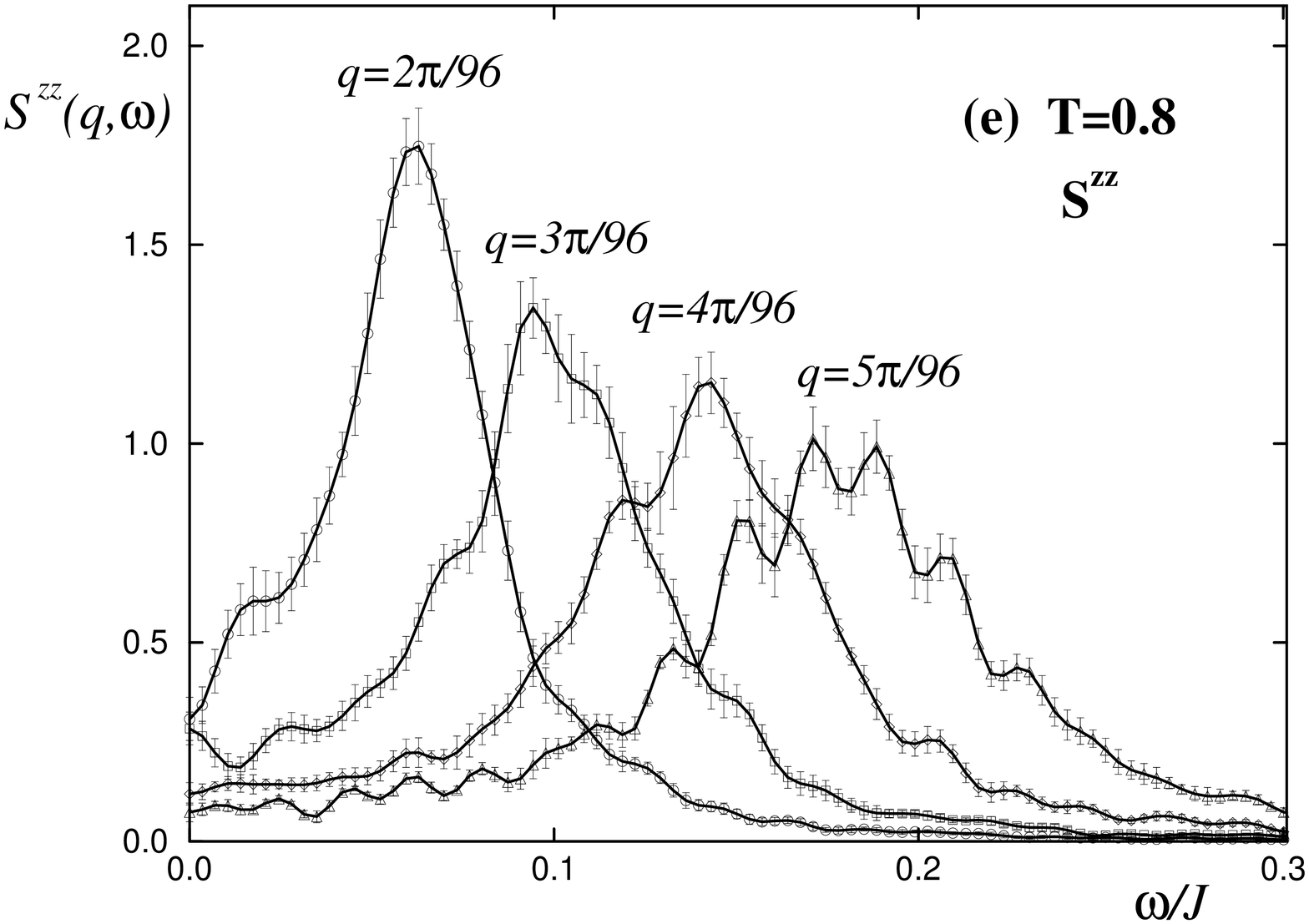,angle=-90,width=\LLL} }
%%      \hskip\SSS
      \parbox[t]{\LLL}{\psfig{file=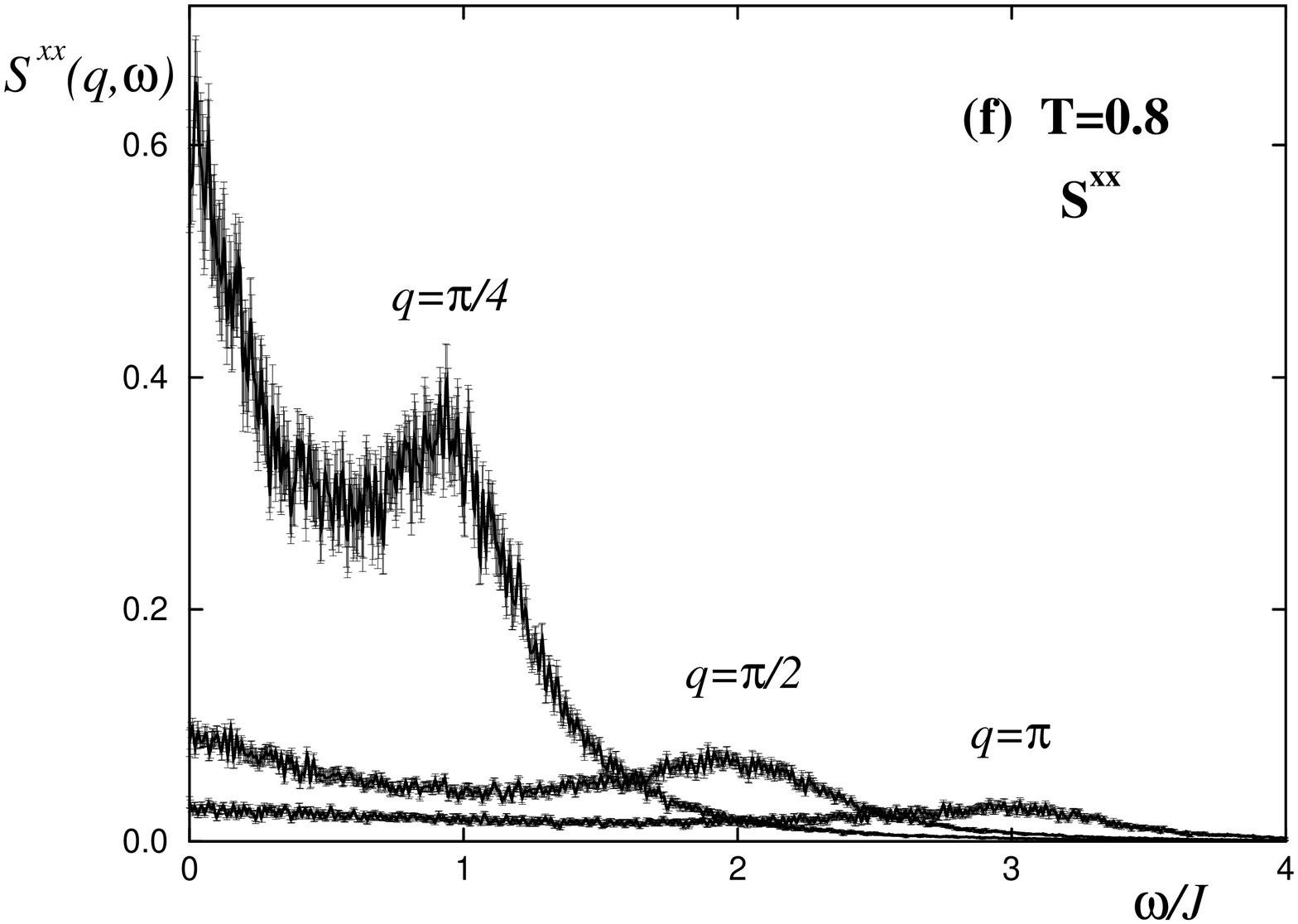,angle=-90,width=\LLL} }
    }%%mbox
    \mycaption{3}{Momentum dependence of \Sqw\, 
                 at fixed lattice size $L=192$.
                   (a) T=0.725, $xx$-component;
                   (b) T=0.725, $zz$-component;
                   (c) T=0.7, height of spin-wave peak as function of $q$
                               (at $L=128$);
                   (d) T=0.8, $xx$-component;
                   (e) T=0.8, $zz$-component;
                   (f) T=0.8, $xx$-component at large $q$.
                 }%%mycaption
  \end{center}
\end{figure}
%--------------------------------------------------------------

%% MOMENTUM DEPENDENCE AT FIXED L
%
In figure 3 we show the momentum dependence of \Sqw.
Fig.\ 3(a) and (b) display the behavior at $T=0.725$, 
which is qualitatively similar to that at lower temperatures.
The position of the spin-wave peak 
is the same for \Sxx\ and \Szz\ and is proportional to momentum
for small q.
As $q$ increases, the peak broadens, and becomes less intense,
yet it remains  well defined.
The additional structure in \Sxx\ is strongly momentum-dependent,
as will be discussed below.

For the $zz$-component, 
both the total intensity and the relative loss of intensity 
with increasing momentum are much smaller.
Our $\omega$-resolution  dominates the width of the spin-wave peak
in \Szz\  only at the smallest $q$ (which also appears in figure 1);
it is not dominant 
at higher momenta or for \Sxx.
We conclude that \Szz\ has the expected delta-function form only for
very small momentum.
(Higher order perturbation theory 
also predicts a finite width \cite{Menezes_Szz}.)
As shown in fig.\ 3(c),
the intensity of the spin-wave peak 
decreases much more rapidly for \Sxx\ than for \Szz\ with increasing $q$;
the intensities cross each other well before the zone edge is reached.
This behavior is similar at other temperatures.
We also note that at all temperatures the total intensity $S^{zz}(q)$ (not shown) 
is constant with $q$,
whereas $S^{xx}(q)$ decreases (as $q^{\eta-2}$) 
and crosses $S^{zz}(q)$ 
at a slightly larger momentum $q\approx 1.5-2$.

Well above the transition, at $T=0.8$, fig.\ 3(d),
\Sxx\  has no noticeable spin-wave peak at small momentum.
The strong central peak 
rapidly loses intensity with increasing momentum.
In marked contrast, the behavior of \Szz\  (fig.\ 3(e))
is very similar to that at
lower temperatures, with clear but broadened spin-wave peaks.
Notice that there is now non-zero intensity at small \om\ in \Szz.
Remarkably, at very large momenta spin-waves appear in \Sxx\ 
even for $T=0.8$ (fig.\ 3(f),
so that both a central peak and a spin-wave peak are present.
Note that the vertical scale in fig.\ 3(f) is $100$ times smaller 
than in fig.\ 3(d).
There is no noticeable lattice size dependence here.

% FIGURE 4--------------------------------------------------------------
\setlength{\LLL}{\textwidth}
%%\divide \LLL by 2
\addtolength{\LLL}{-40mm}
\begin{figure}[htbp] 
  \begin{center}
      \parbox[t]{\LLL}{\psfig{file=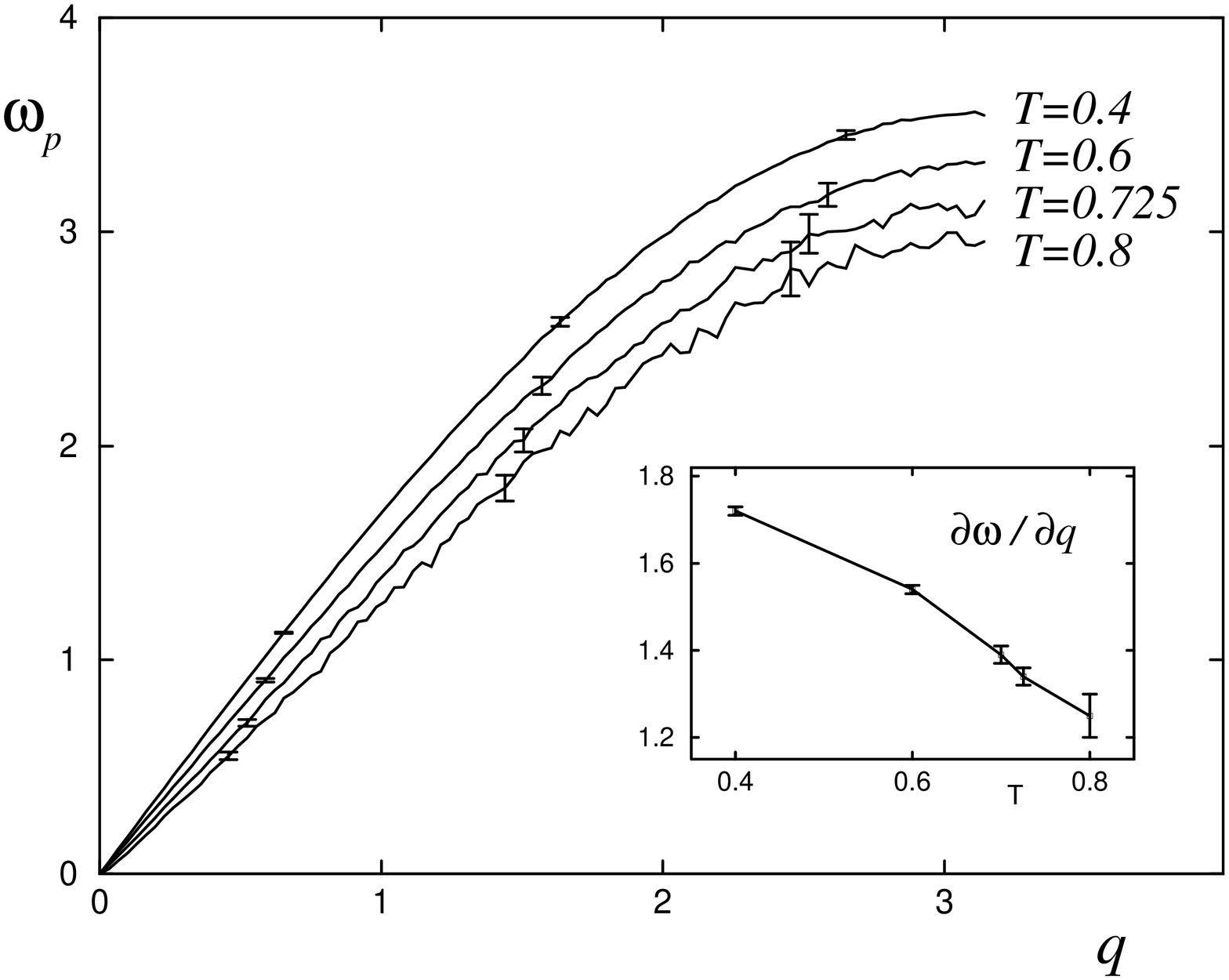,angle=-90,width=\LLL} }
    \mycaption{4}{Dispersion relation: 
                  spin-wave frequency $\omega_p^{xx}$
                  against momentum,
                  at $L=192$ and four different temperatures.
                  Note that for $T=0.8$ only \Szz\  has
                  a noticeable spin-wave peak, \Sxx\  does not.
                  The inset shows the spin-wave velocity 
                  $\frac{\partial\omega}{\partial q}$ 
                  as a function of temperature.
                 }%%mycaption
  \end{center}
\end{figure}
%--------------------------------------------------------------

%%%% Dispersion Curves %%%
%
Figure 4 shows the position $\omega_p$ 
of the spin-wave peak as a function of momentum.
The expected linear portion of the 
dispersion curve extends to rather large momenta.
With increasing temperature, the spin-wave  velocity 
$\partial\omega_p/\partial q$,
which is proportional to the spin-wave stiffness,
decreases slowly and approximately linearly, as shown in the inset,
and theoretically expected 
for small $T$ \cite{NelsonKosterlitz,Menezes_spinstiffness}. 
At $T\leq \TKT$, $\omega_p$ is the same for \Sxx\  and for \Szz,
as expected by theory \cite{Villain,NelsonFisher}.
At $T=0.8$ on the other hand, 
we can only plot the position of the residual peak in \Szz,
because \Sxx\  has dropped sharply to zero here, as expected
for a KT transition.
\pagebreak[4]

%==============================================================================
 \subsection{Additional Structure in \Sxxqw} \label{Sec:strangepeaks} 
%==============================================================================
%

If we expand the vertical scale in plots of $S^{xx}(q,\omega)$, 
we find that the in-plane component
$S^{xx}(q,\omega)$ shows  rich structure in addition to the spin-wave peak.
Note that the intensity of this structure is typically $10^{-2}$ 
of the maximum.
It is visible at all temperatures $T\lsim \TKT$.
At the lowest temperature, the absolute intensity  of this structure is
low, but the relative intensity is quite high (see also fig.\ 11(a)).
At higher temperature, the structure becomes rather smeared.
No such structure can be found in $S^{zz}(q,\omega)$.
%
%
% FIGURE 5--------------------------------------------------------------
\setlength{\LLL}{\textwidth}
\setlength{\SSS}{5mm}
\divide \LLL by 2
\addtolength{\LLL}{-\SSS}
\begin{figure}[htbp] 
  \begin{center}
    \vskip-5mm
    \mbox{
      \parbox[t]{\LLL}{\psfig{file=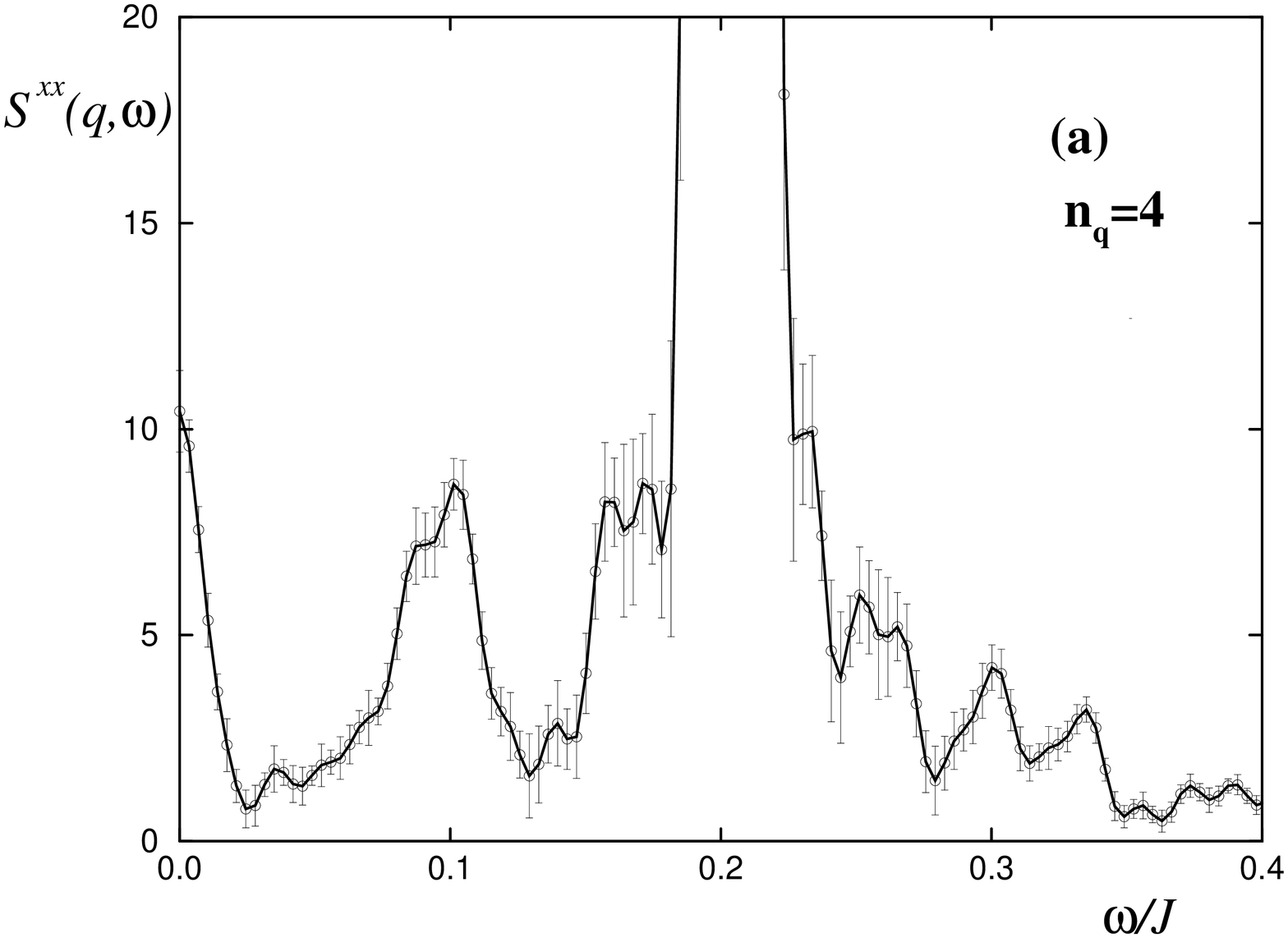,angle=-90,width=\LLL} }
%%      \hskip\SSS
      \parbox[t]{\LLL}{\psfig{file=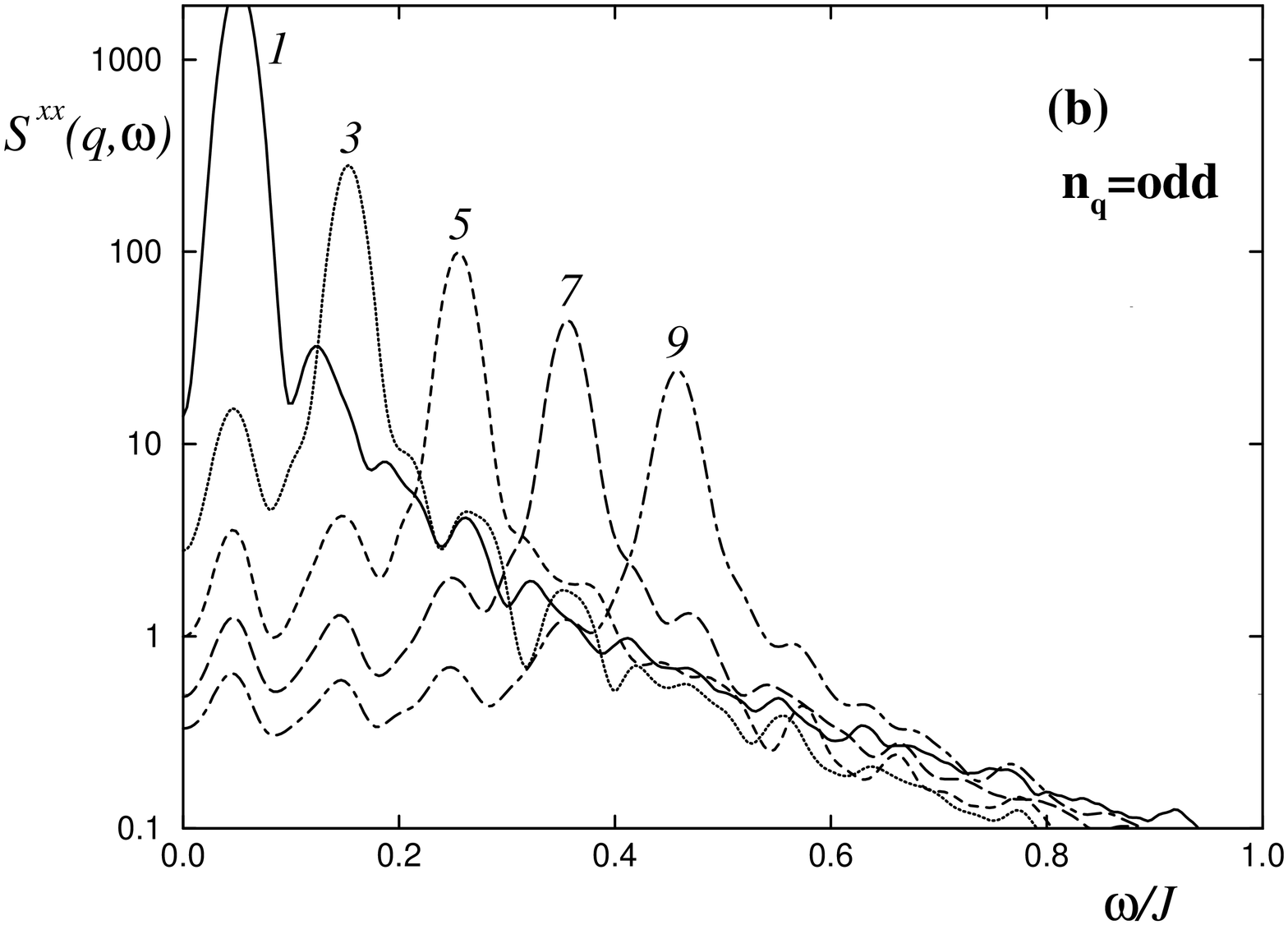,angle=-90,width=\LLL} }
    }%%mbox
    \\
    \mbox{
      \parbox[t]{\LLL}{\psfig{file=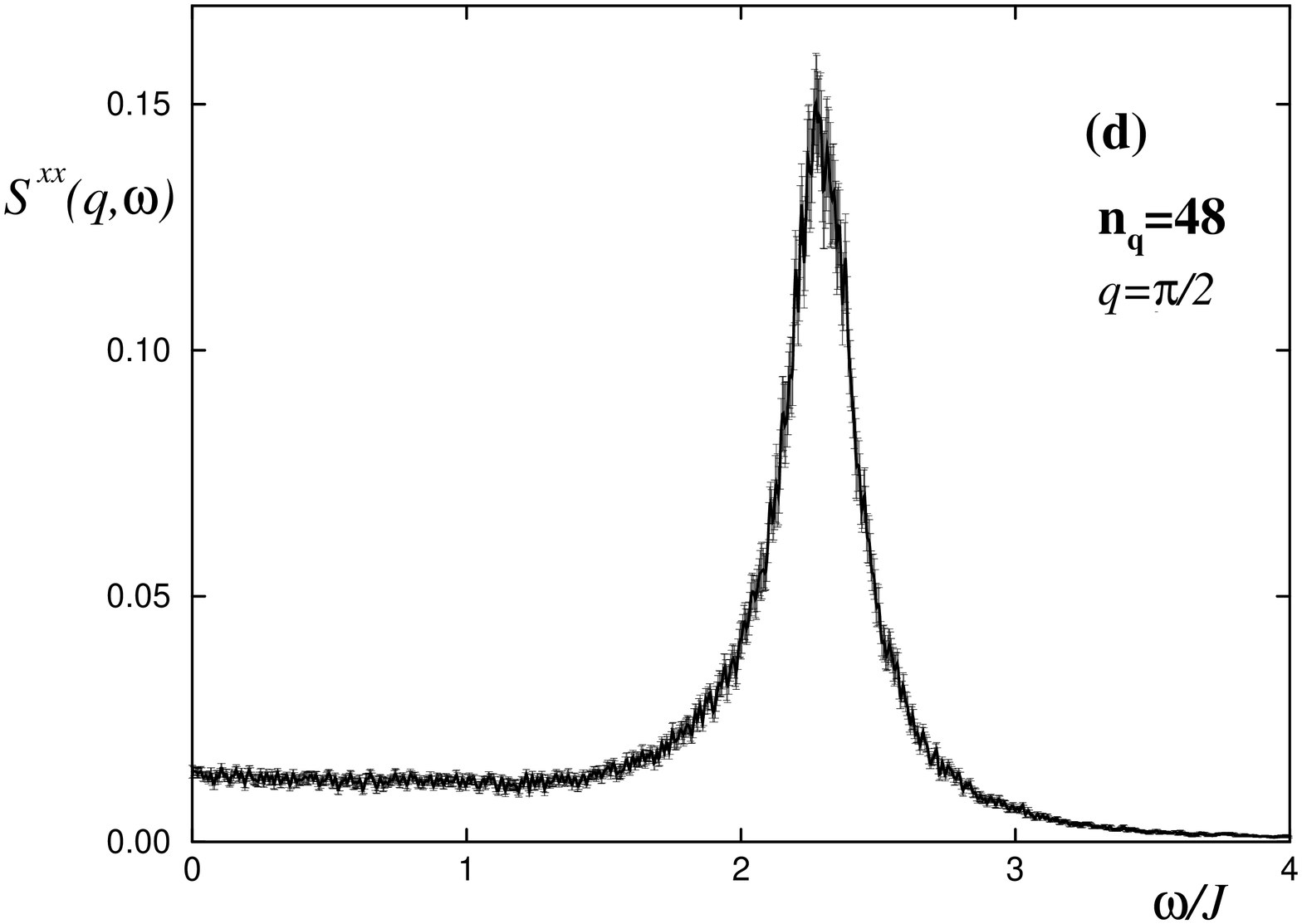,angle=-90,width=\LLL} }
%%      \hskip\SSS
      \parbox[t]{\LLL}{\psfig{file=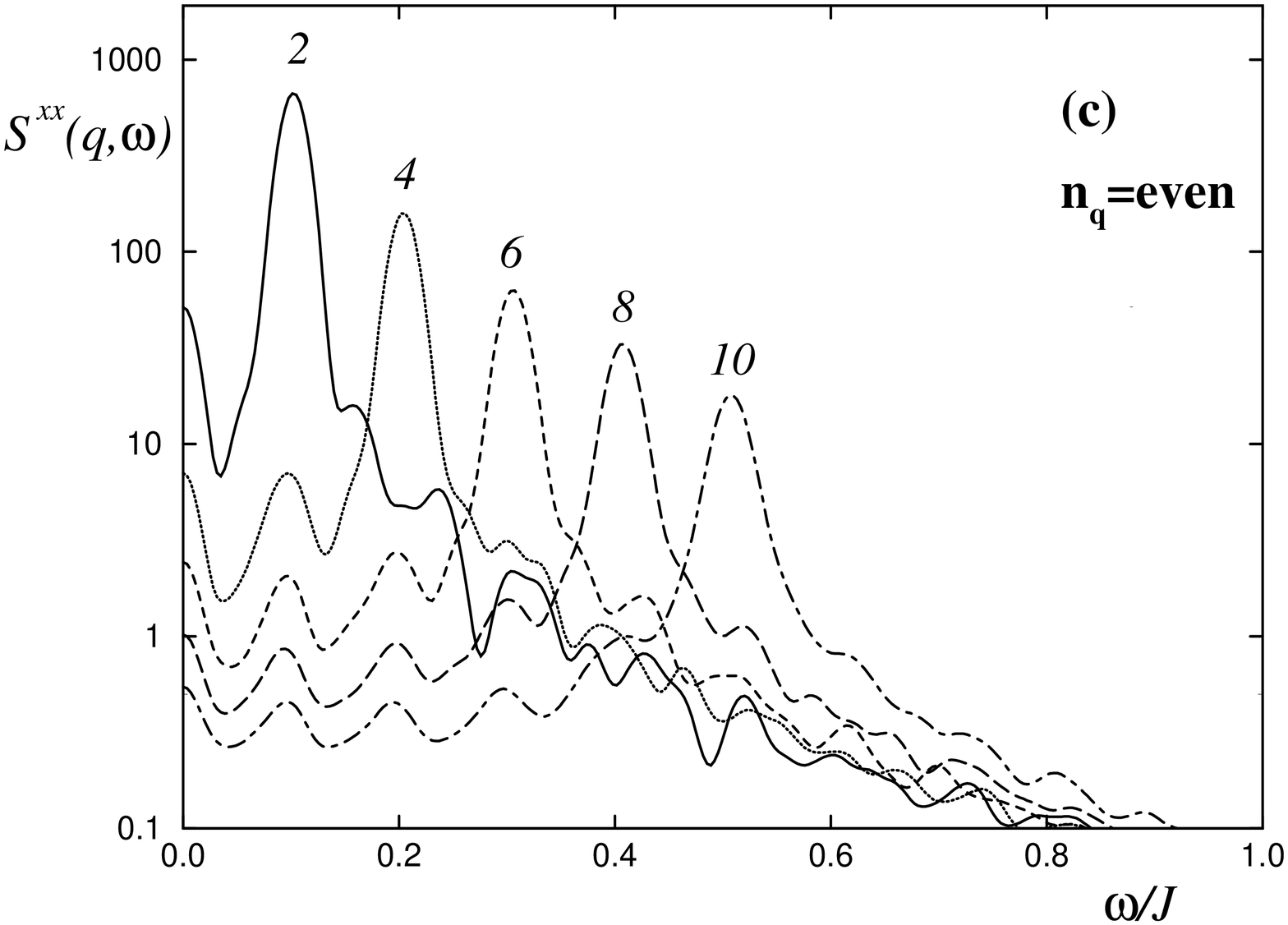,angle=-90,width=\LLL} }
    }%%mbox
    \\
    \mbox{
      \parbox[t]{\LLL}{\psfig{file=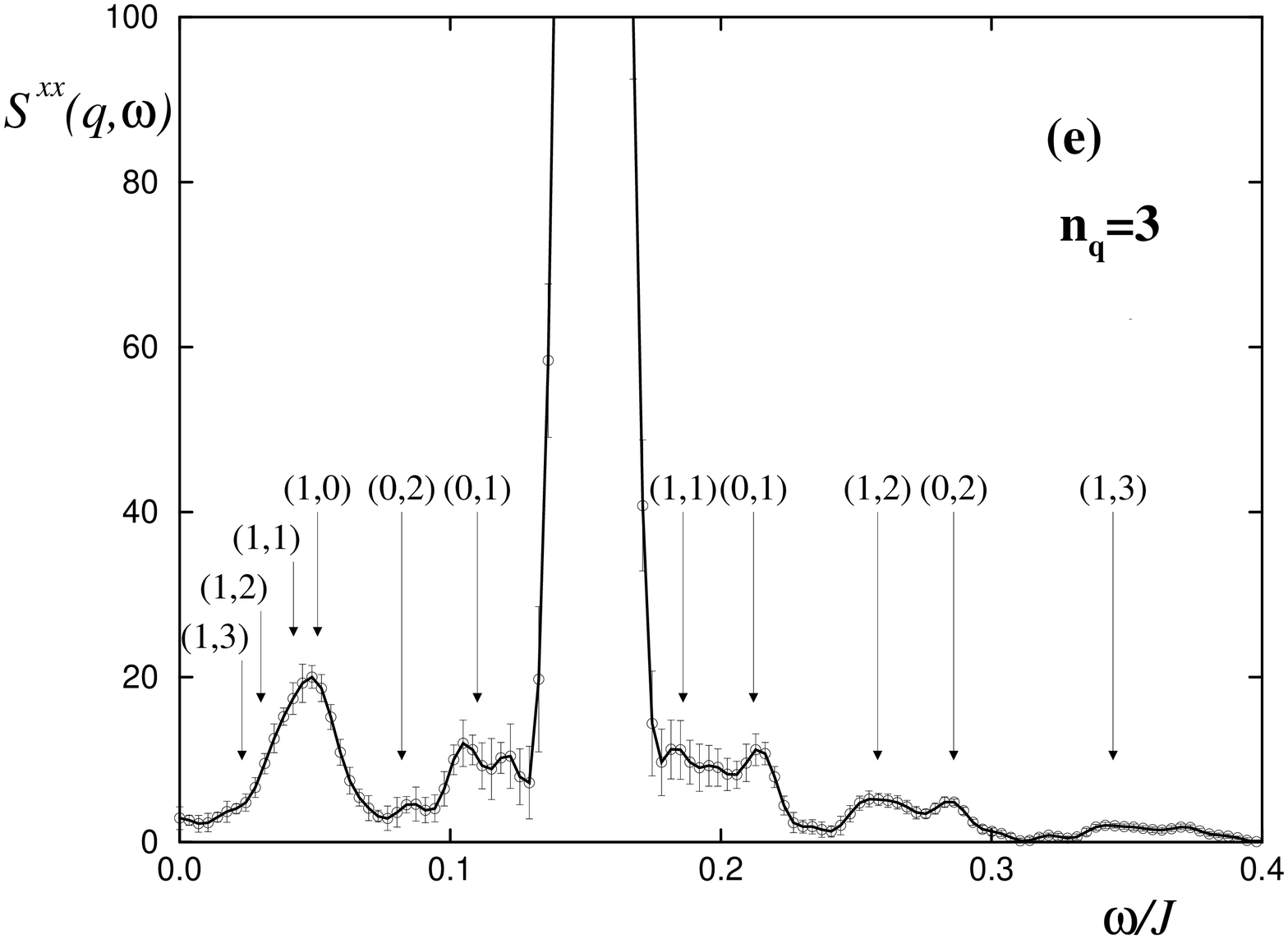,angle=-90,width=\LLL} }
    }%%mbox
    \mycaption{5}{High resolution study of the ``fine structure'' in 
                 $S^{xx}(q,\omega)$ for $T=0.6$, $L=192$.
                 (a) $n_q=4$; note that the maximum value of the
                            spin-wave peak is $\approx 160$.
                 (b),(c) log plot of $S^{xx}(q,\omega)$ 
                            at small values of $n_q$; 
                            the data are
                            smoothened, with $\delta \omega=0.01$;
                 (d) $S^{xx}(q,\omega)$ at $q=\pi/2$;
                 (e) $n_q=3$: vertical arrows show the location of
                            two-spin-wave peaks formed by spin-waves of
                            small momentum $|\vec{q}| < 4 \,\frac{2\pi}{L}$. 
                 }%%mycaption
  \end{center}
\end{figure}
%--------------------------------------------------------------
%
The locations of additional maxima in \Sxx\ 
are essentially unchanged with L when $n_q=qL/(2\pi)$ is held fixed.
Figures 5(b) and (c) show, on a logarithmic scale,
that for odd values of $n_q$ 
there are strikingly regular pronounced  peaks at 
$\omega = \omega_p/n_q,\, 3\omega_p/n_q,\,...$,
and for even $n_q$ such peaks appear at
$\omega = 0,\, 2\omega_p/n_q,\, 4\omega_p/n_q,\,...$.
At large $n_q$, fig.\ 5(d), individual peaks cannot be distinguished;
instead \Sxx\ is nearly constant
below the spin-wave peak there.
(In figs.\ 5(b) and (c), the data have been smoothened slightly
with a resolution function \eq{smoothen} with $\delta\omega=0.01$,
in order to reduce the wiggles and allow general features to be identified.)
In addition to the regularly spaced pronounced peaks,
there is further ``fine structure'' in \Sxx, 
clearly visible in fig.\ 5(a).
In the course of our study, the additional structure
became clearer as the statistical quality of the data improved;
it is apparent that the structure is not statistical noise.
Very close to the spin-wave peak, part  of the additional structure may
be due to  
the finite-time cutoff in our time-integration;
but most of the observed structure
must be due to different reasons.

One simple explanation for the observed rich structure,
which is consistent with the data
but for which we have no rigorous theory, 
is that of multi-spin-wave effects.
Of these, two-spin-wave processes are likely to be the most important.
Thus, at a given total momentum $\vec{q}$ 
we can have either a single spin-wave excitation of momentum $\vec{q}$,
or two spin-waves for which the sum or difference of momenta equals $\vec{q}$.
The result will then be both a single-spin-wave peak at a 
characteristic frequency $\omega_p(q)$ 
as well as additional sum and difference
peaks due to the two-spin-wave processes.

Of particular interest
is the case when the two spin-waves have momenta $\vec{q}_1$ and $\vec{q}_2$ 
that are collinear, so that $q=q_1+q_2$ is a scalar equation. 
Since the momenta are discrete on a finite lattice, 
$q_i=n_{q_i} \frac{2\pi}{L}$, 
this implies $n_q = n_{q_1} + n_{q_2}$.
With a linear dispersion relation $\omega=cq$,
the difference of the two spin-wave frequencies is then
$\omega = (2n_{q_1} - n_q) \, c\frac{2\pi}{L}$, 
i.e.\  just the series of additional peaks
that are visible in figures 5(b) and (c).

Using the measured dispersion relation (fig.\ 4),
we have calculated the frequencies of two-spin-wave excitations
consisting of the most likely individual spin-waves,
i.e.\ those with smallest individual momenta.
For the case of $n_q=3$ and $T=0.6$, 
these locations are marked in fig.\ 5(e).
They are identified by the coordinates $\vec{n}_{\vec{q}_1}$ 
of one of the spin-waves in reciprocal space;
the sum of the two spin-wave momenta 
must equal $\vec{q}=(3\times\frac{2\pi}{L},0)$.
The locations of the resultant excitations agree extremely well with the
positions of the small peaks in \Sqw, but we have no way of comparing 
intensities.
%Considering the low intensity of the small peaks,
%some of the structure in 
%the figure may still contain some remainder of cutoff wiggles.

The presence of distinct small peaks at
$\omega = 0,\, 2\omega_p/n_q,\, 4\omega_p/n_q,\,...$.
for even $n_q$ at $T<\TKT$ 
complicates the identification of a possible central peak.
Interpolating the intensities for odd values of $n_q$ 
(which {\em do not} show peaks at $\omega=0$)
to obtain estimates for even $n_q$, 
we conclude that there is indeed extra intensity at
$\omega=0$ which is not attributable to two-spin-wave processes.

%==============================================================================
 \subsection{Finite size scaling of characteristic frequency $\omega_m$}
 \label{Sec:scale_o}
%==============================================================================
%
           
Equation (\ref{om1}) defines the characteristic frequency $\omega_m$ of the
whole spectrum of \Sqw.
When there is only a single spin-wave peak, then $\omega_m$ coincides with the
spin-wave frequency $\omega_p$.
This is the case at $T=0.4$, where all frequencies coincide 
(within error bars),
$\omega_p^{xx} = \omega_m^{xx} = \omega_p^{zz} = \omega_m^{zz}$.
Closer to the transition, intensity between
$\omega=0$ and the spin-wave peak grows;
therefore the characteristic frequency $\omega_m^{xx}$ becomes smaller
than the spin-wave frequency $\omega_p$.
Their difference 
%between $\omega_m$ and $\omega_p$ 
is thus a measure of
the relative weight of non-single-spin-wave contributions to \Sqw.
Figure 6(a) shows the situation at \Tkt, where \Sxx\ exhibits large
non-single-spin-wave contributions and $\omega_m^{xx} < \omega_p^{xx}$.
%
% FIGURE 6--------------------------------------------------------------
\setlength{\LLL}{\textwidth}
\setlength{\SSS}{2mm}
\divide \LLL by 2
\addtolength{\LLL}{-\SSS}
\begin{figure}[htbp] 
  \begin{center}
    \vskip-5mm
    \mbox{
      \parbox[t]{\LLL}{\psfig{file=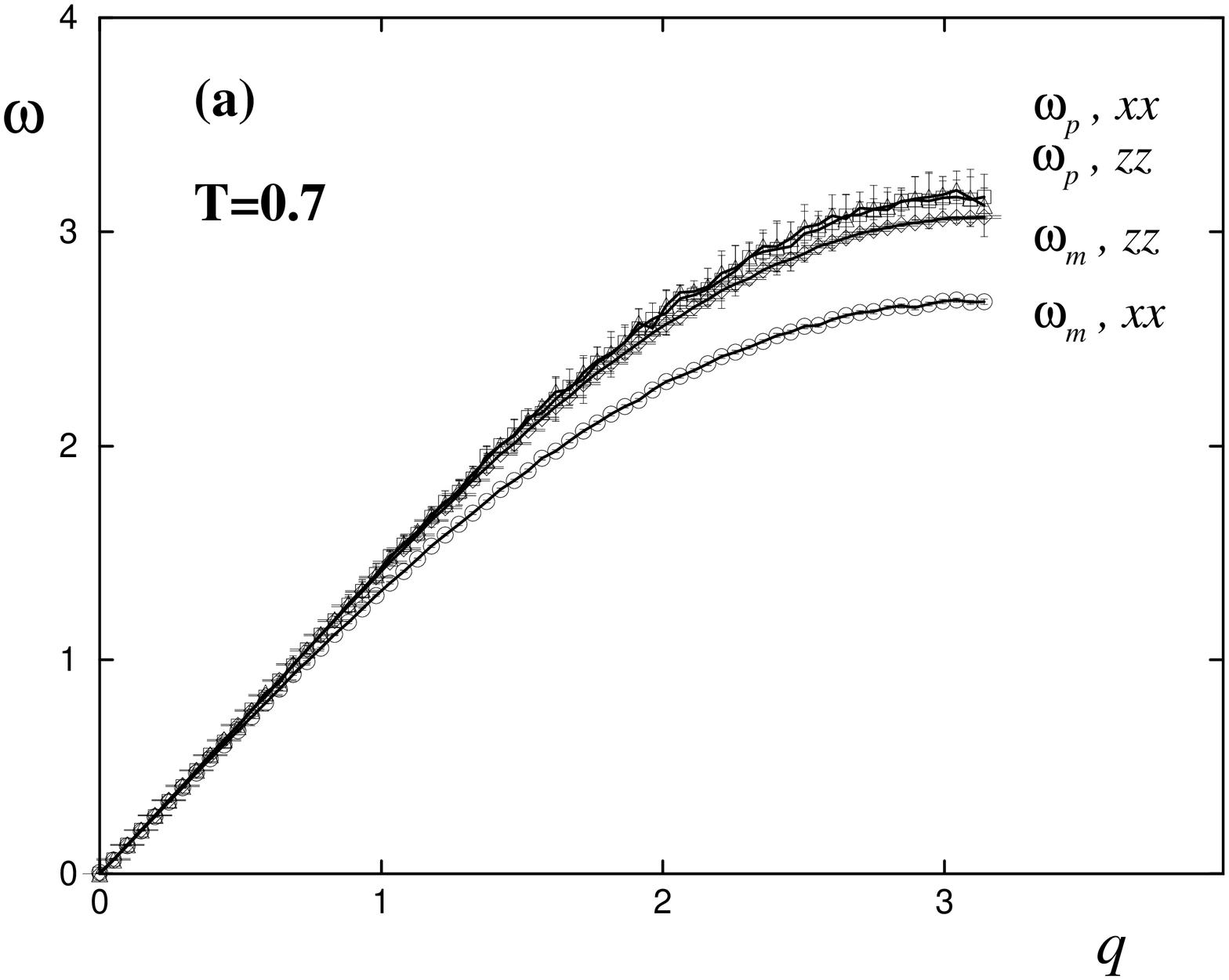,angle=-90,width=\LLL} }
%%      \hskip\SSS
      \parbox[t]{\LLL}{\psfig{file=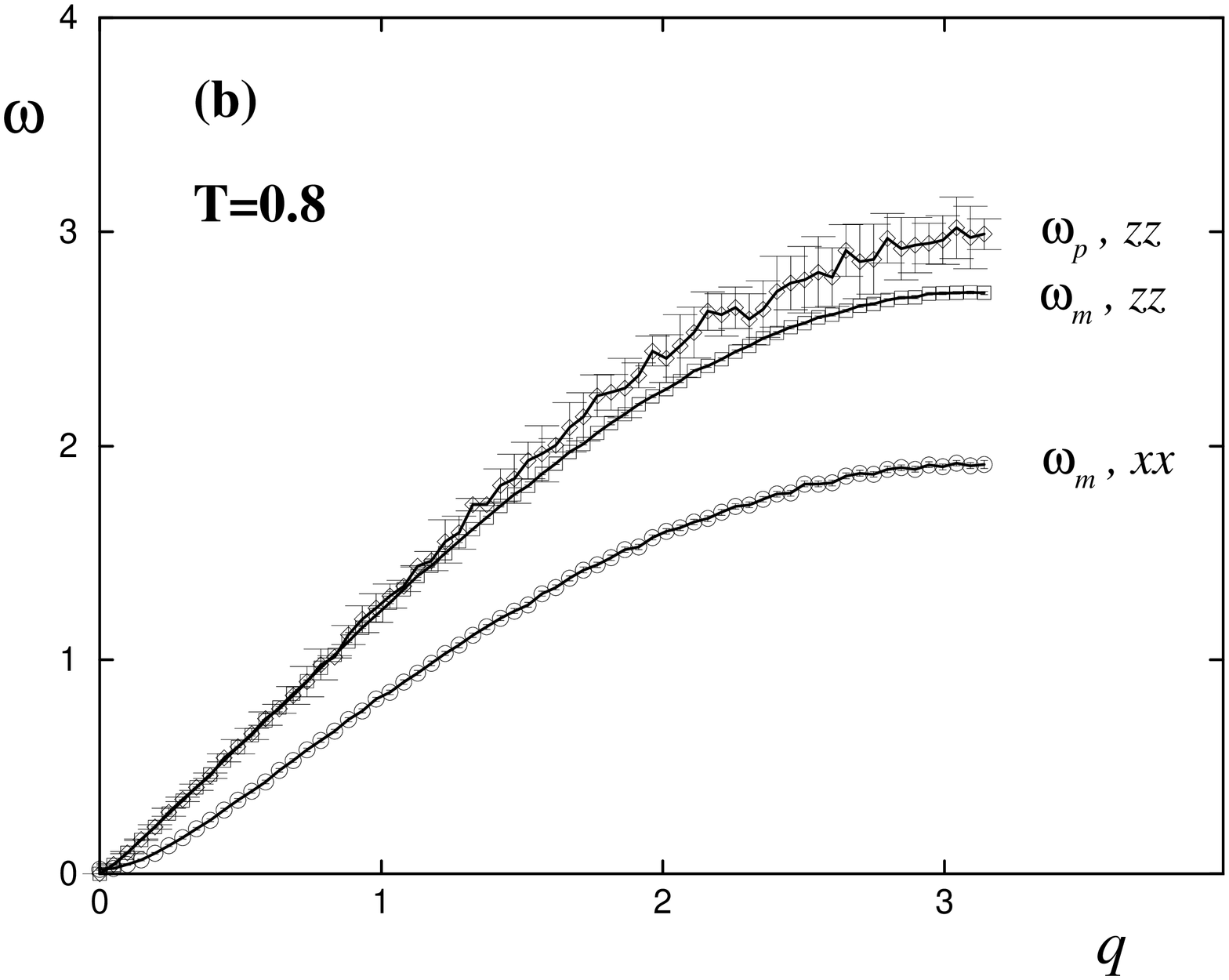,angle=-90,width=\LLL} }
    }%%mbox
    \mycaption{6}{Characteristic frequency $\omega_m$ and
                 spin-wave frequency $\omega_p$ 
                 for \Sxx\  and \Szz, at $L=128$.
                 (At $T=0.8$ there is no $xx$-spin-wave frequency 
                                $\omega_p^{xx}$).
                 }%%mycaption
  \end{center}
\end{figure}
%--------------------------------------------------------------

Above the transition, $\omega_m^{xx}$ is no longer linear in momentum
for small $q$,
as shown in fig.\ 6(b), and differs strongly from the $zz$-component.
The latter still has both a spin-wave peak that is linear in momentum, as
well as intensity at small $\omega$, so that $\omega_m^{zz}$ is 
smaller than $\omega_p^{zz}$.

The central question of {\em critical} dynamics is that of scaling,
i.e.\ whether data from lattices of different size match when properly scaled.
As mentioned in section (\ref{DFSS}), we can test scaling and extract
the dynamic critical exponent $z$ in two ways, by analyzing
the characteristic frequency $\omega_m$, or by looking at \Sqw\ itself.

We concentrate on the dynamic critical behavior of our model at $T\leq \TKT$,
the critical region,
in which the correlation length in an infinite system is divergent.
The relevant length scale on a finite lattice is therefore the lattice size $L$, 
and we expect scaling for suitable functions of $qL$, 
as described in section (\ref{DFSS}).
From the analytical results we expect the dynamic critical exponent to be
$z=1$ (\eq{z}).

In figure 7, (a),(b) 
we show  $\omega_m^{xx} \,L^z$ as a function of $qL$, for $z=1.00$ and 
at temperatures $T\leq \TKT$.
From \eq{ob} we expect the data to fall on a single curve if scaling holds.
This is indeed exactly the observed behavior at all temperatures $T\leq \TKT$.
The asymptotic behavior for large $L$ is strictly linear,
$\omega_m\,L^z \sim \,qL$; i.e.\ for $z=1$, $\omega_m \sim q$.
%
%
% FIGURE 7--------------------------------------------------------------
\setlength{\LLL}{\textwidth}
\setlength{\SSS}{2mm}
\divide \LLL by 2
\addtolength{\LLL}{-\SSS}
\begin{figure}[htbp] 
  \begin{center}
    \vskip-5mm
    \mbox{
      \parbox[t]{\LLL}{\psfig{file=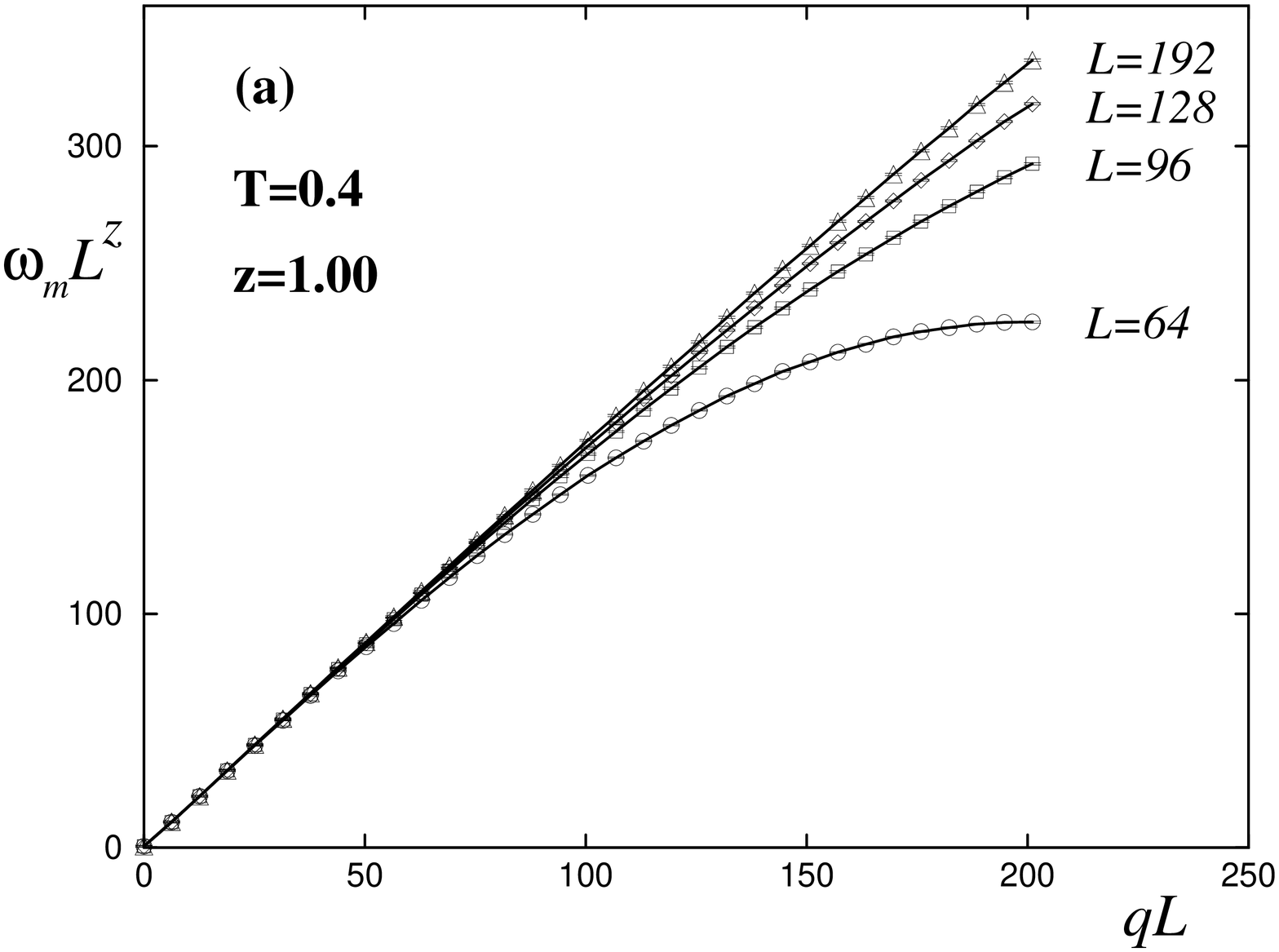,angle=-90,width=\LLL} }
%%      \hskip\SSS
      \parbox[t]{\LLL}{\psfig{file=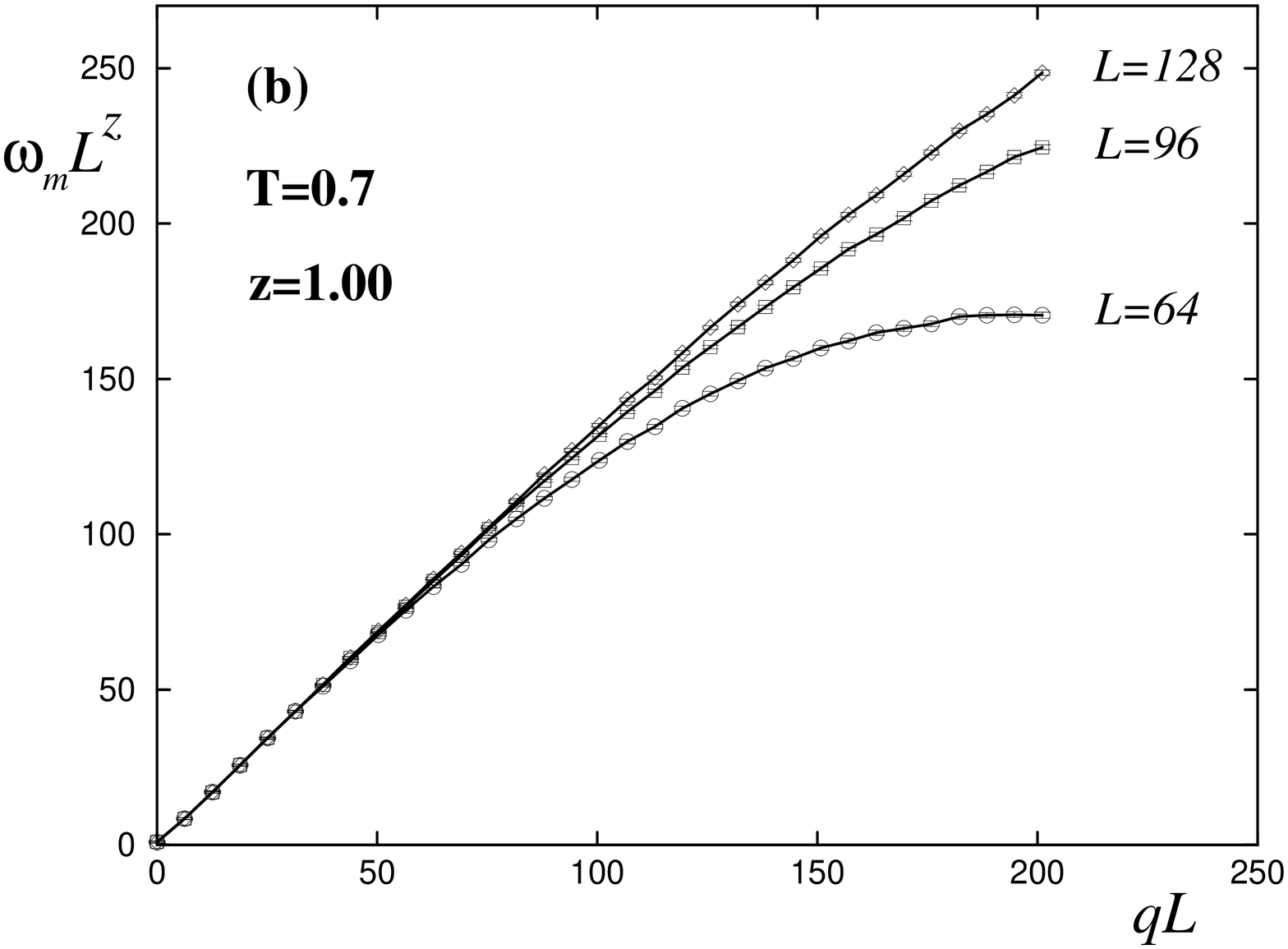,angle=-90,width=\LLL} }
    }%%mbox
    \mycaption{7}{Finite size scaling of the characteristic frequency;
                  $\omega_m^{xx} \,L^z$ is plotted against $q L$,
                  for $z=1.00$. 
                  %The  curves correspond
                  %to different lattice sizes, from $L=16$ to $192$.
                 }%%mycaption
  \end{center}
\end{figure}
%--------------------------------------------------------------
%
For each finite lattice size the dispersion curve flattens when $q$ becomes 
large.
Therefore as $L$ increases, the data in figure 7 
start to move away from the asymptotic behavior
at progressively larger values of $qL$.
We have also analyzed the scaling behavior for different values of $z$, 
e.g.\ $z=1.10$
(not shown in the figures). In that case the data for different $L$
diverge from each other immediately; 
they do not fall onto a common line even at the smallest momenta.

The scaling curves for $\omega_m^{xx}$ 
at all three temperatures $T\leq \TKT$ are very similar,
with variation only in their slope.
In contrast, we do not observe similar scaling behavior in $\omega_m^{xx}$
at $T=0.8$ above the transition (not shown in the figures). 

Analyzing the out-of-plane characteristic frequency $\omega_m^{zz}$, 
we found that (for $q\neq 0$) it has
the same scaling behavior as the in-plane component.
At $T=0.4$ the data for $\omega_m^{zz}$ and $\omega_m^{xx}$ 
are indistinguishable.
When intensity below the spin-wave peak grows in \Sxx\ at larger $T$,
the scaling curve for 
$\omega_m^{xx}$ has a smaller slope than $\omega_m^{zz}$.
Interestingly, at $T=0.8$, above the transition, 
not only are there spin-wave peaks  present in \Szz,
but $\omega_m^{zz}$ also shows the same  scaling
behavior as below the transition, with $z=1.0$.

%==============================================================================
 \subsection{Finite size scaling of \Sqw} \label{Sec:scale_S}
%==============================================================================
%
If dynamic finite size scaling holds, 
then the scaled neutron scattering function itself
should fall onto a single curve for sufficiently large lattices.
Corresponding to \eq{scal}, figure 8 shows 
$S^{xx}(q,\omega)/(L^z S(q))$ versus $\omega\,L^z$.

% FIGURE 8--------------------------------------------------------------
\setlength{\LLL}{\textwidth}
\setlength{\SSS}{6mm}
\divide \LLL by 2
\addtolength{\LLL}{-\SSS}
\begin{figure}[htbp] 
  \begin{center}
    \vskip-5mm
    \mbox{
      \parbox[t]{\LLL}{\psfig{file=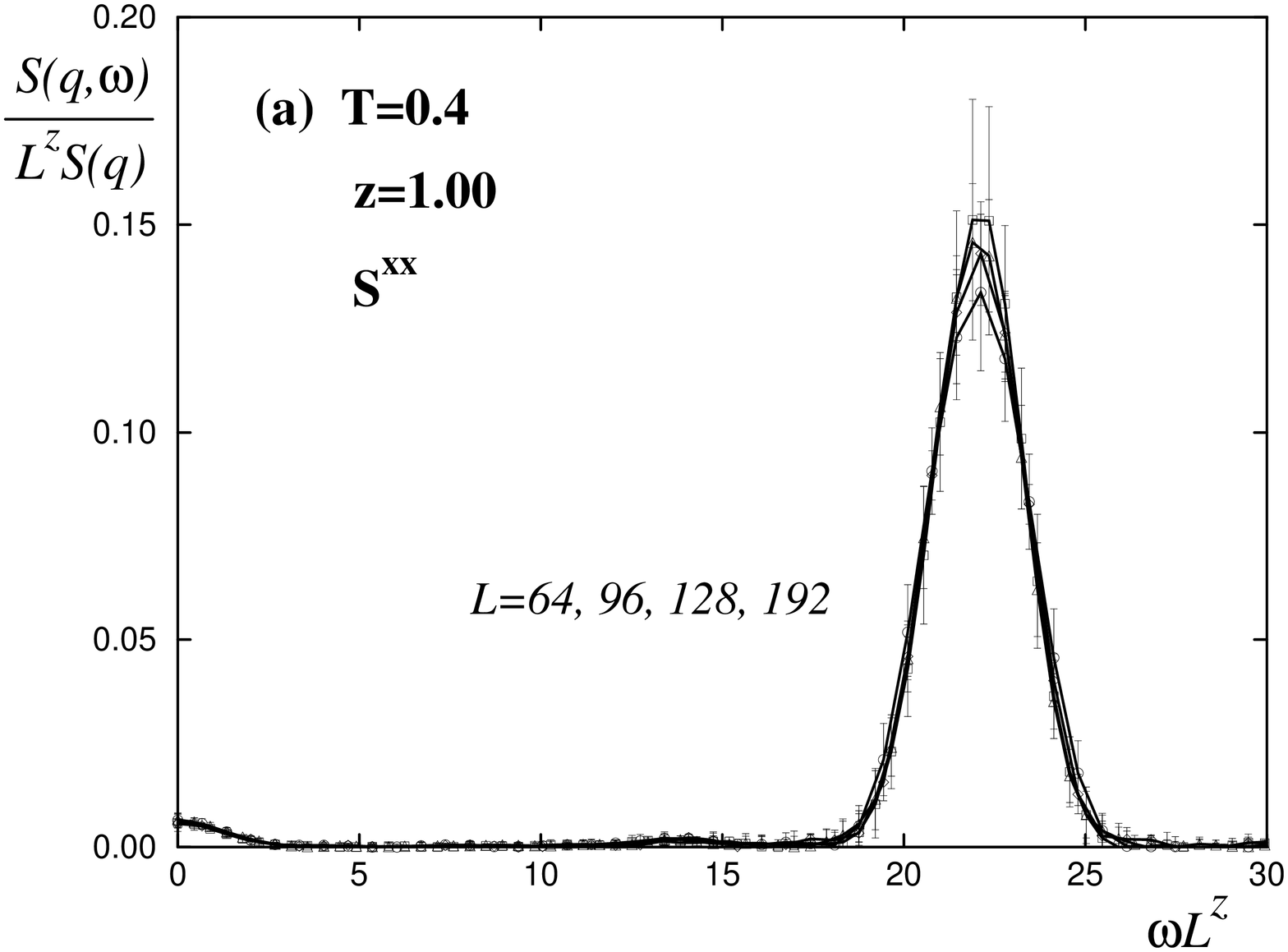,angle=-90,width=\LLL} }
%%      \hskip\SSS
      \parbox[t]{\LLL}{\psfig{file=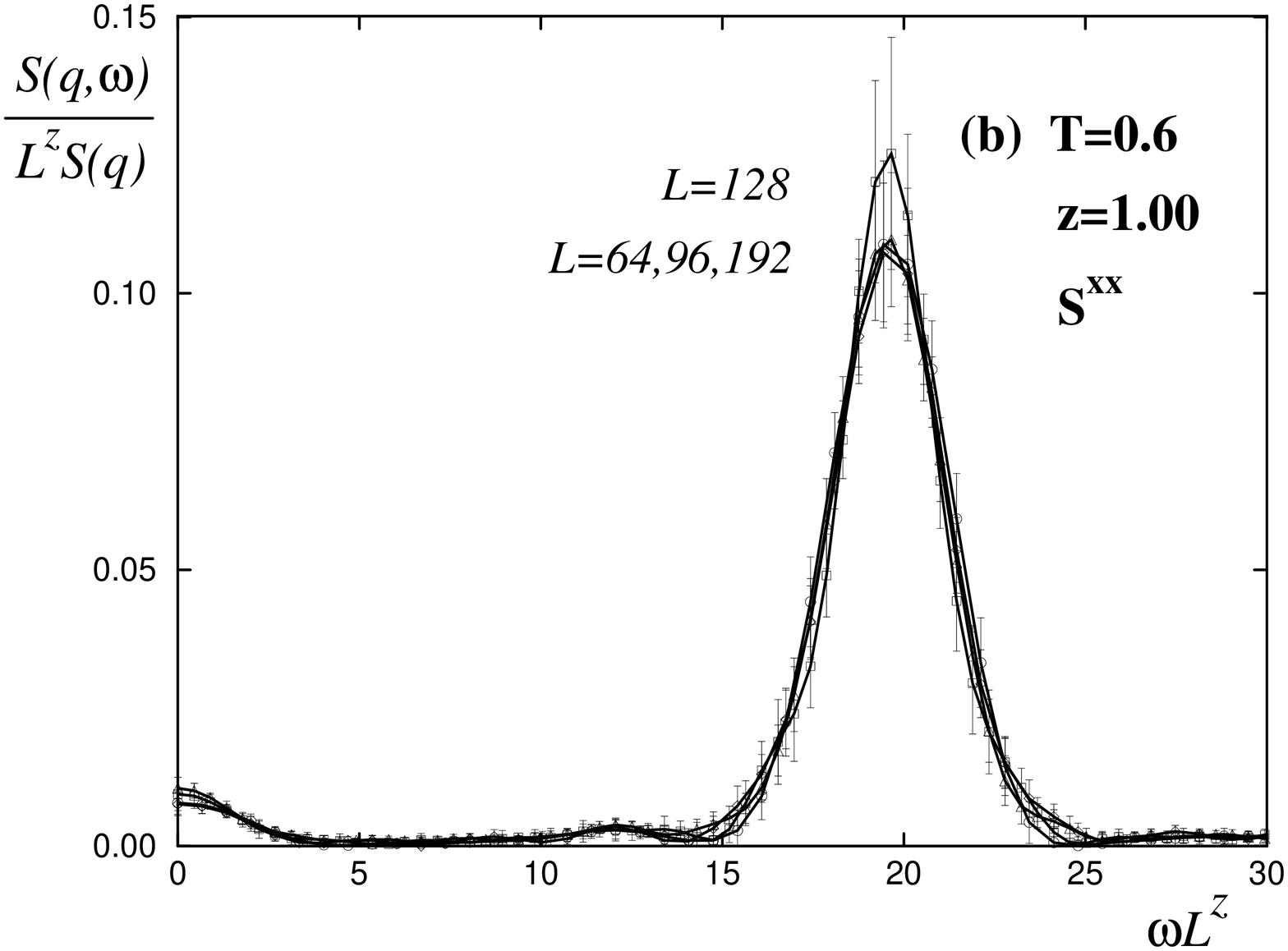,angle=-90,width=\LLL} }
    }%%mbox
    \\
    \mbox{
      \parbox[t]{\LLL}{\psfig{file=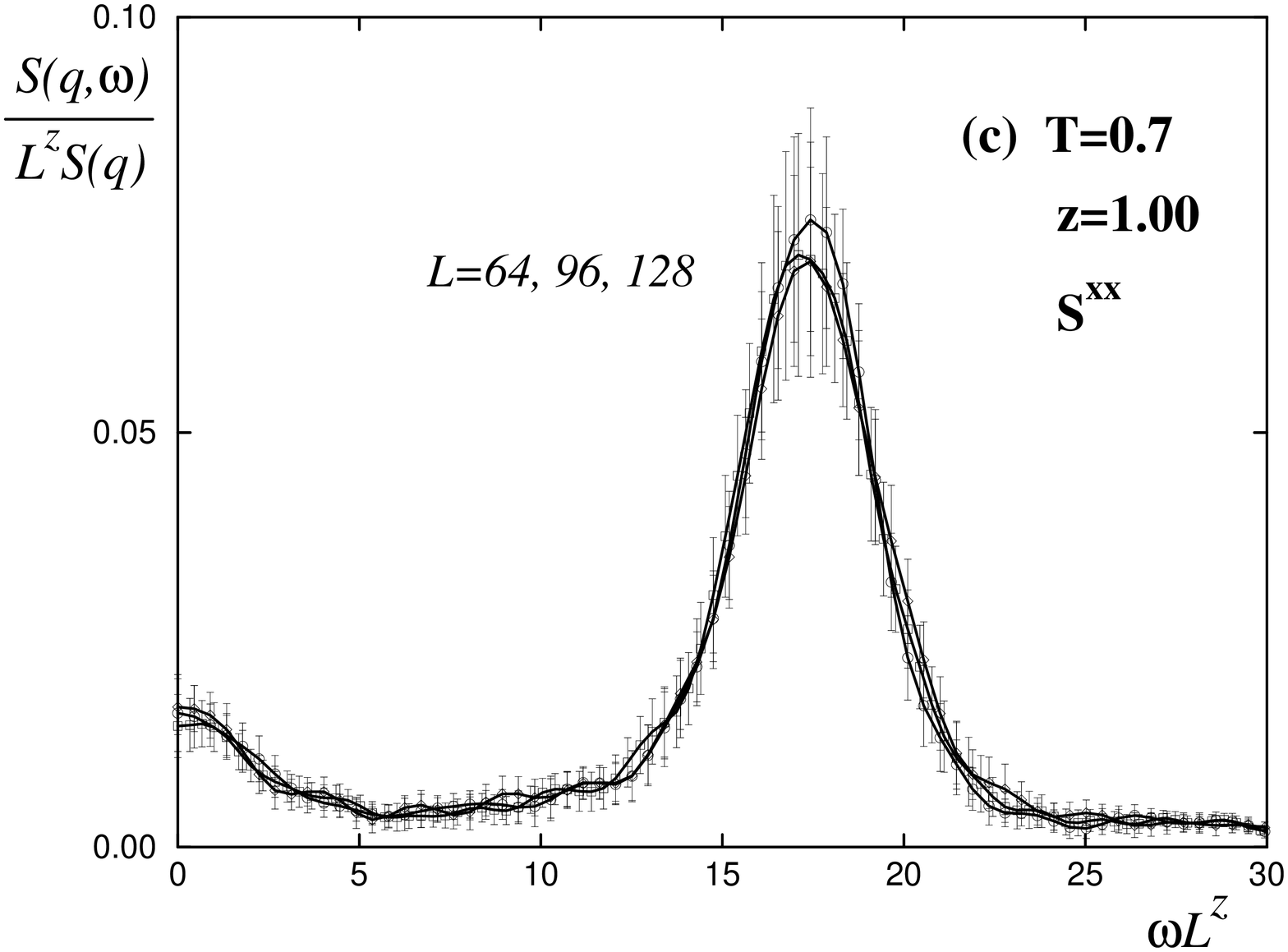,angle=-90,width=\LLL} }
%%      \hskip\SSS
      \parbox[t]{\LLL}{\psfig{file=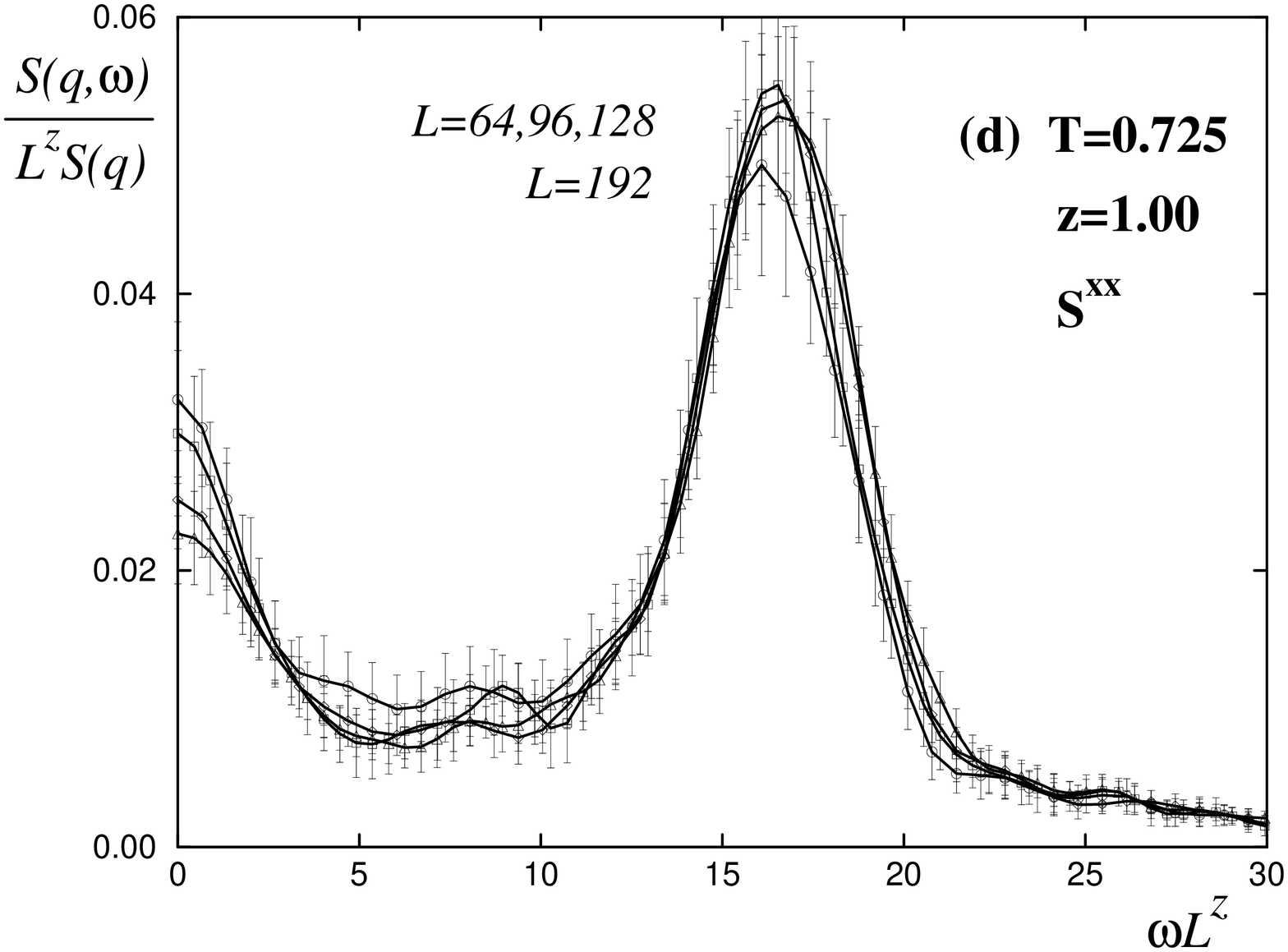,angle=-90,width=\LLL} }
    }%%mbox
    \\
    \mbox{
      \parbox[t]{\LLL}{\psfig{file=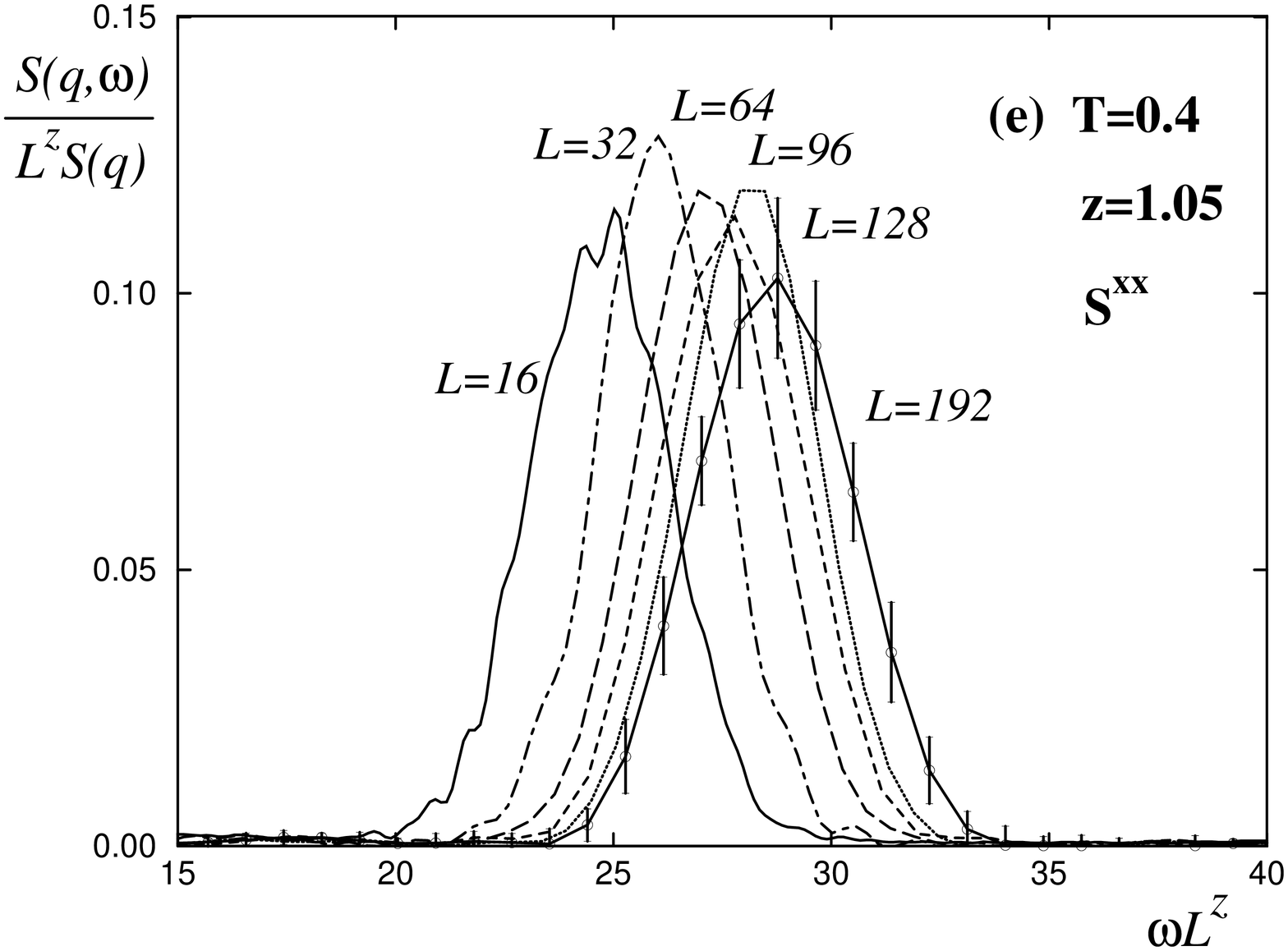,angle=-90,width=\LLL} }
%%      \hskip\SSS
      \parbox[t]{\LLL}{\psfig{file=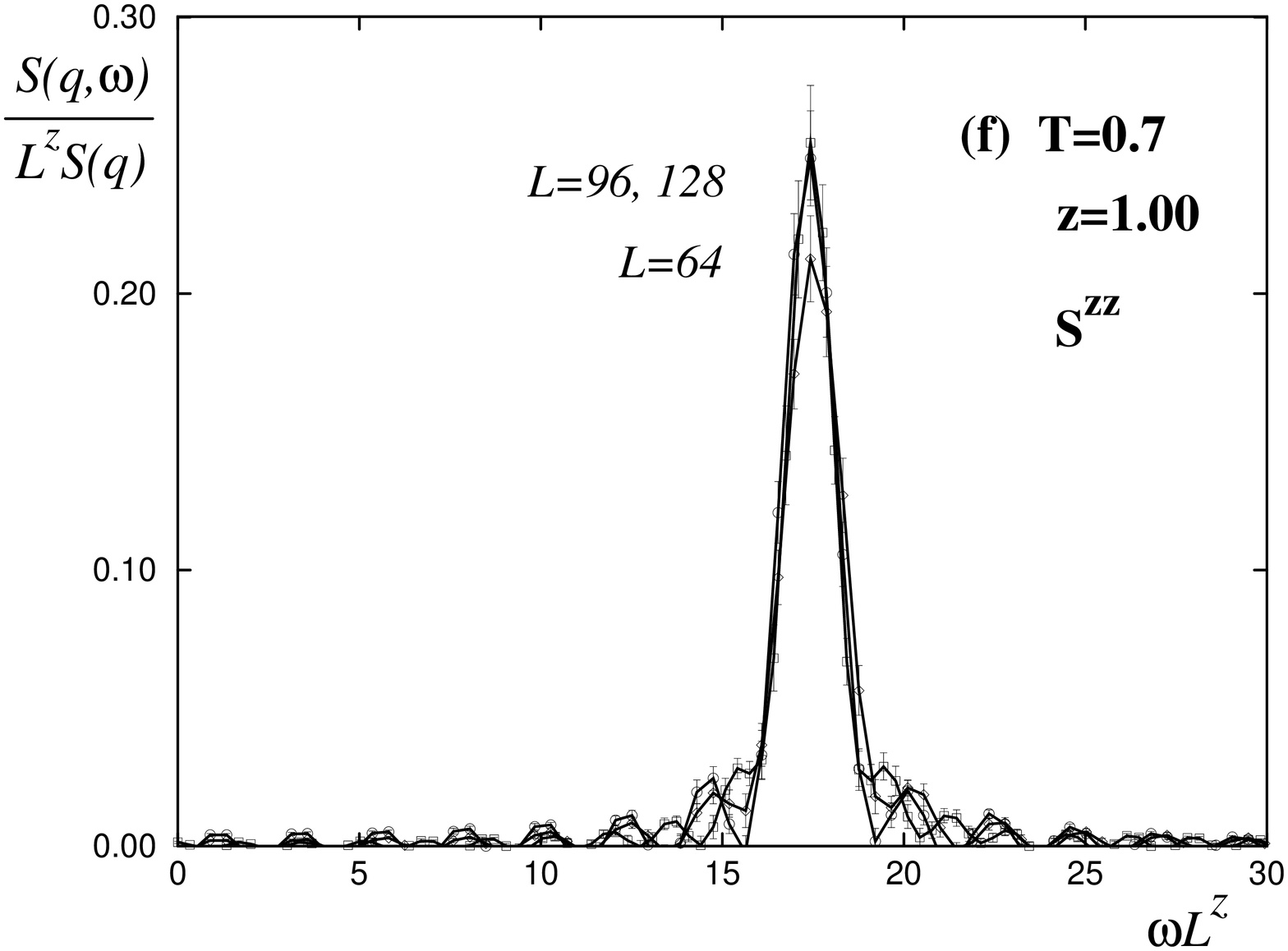,angle=-90,width=\LLL} }
    }%%mbox
    \mycaption{8}{\hfill 
                  Finite size scaling of the neutron scattering function;
                  $S(q,\omega)/(L^z S(q))$ is shown versus $\omega L^z$,
                  with constant $n_q\!\ident\! \frac{q L}{2\pi} =2$.
                  %The  curves correspond
                  %to different lattice sizes, from $L=16$ to $192$.
                  The critical exponent is $z=1.00$ in (a)-(d) and (f).
                  (a)-(d): \Sxx, $z=1.00$; 
                  (e) $z=1.05$ for comparison; 
                  (f) $zz$-component.
                 }%%mycaption
  \end{center}
\end{figure}
%--------------------------------------------------------------

We see that at all  temperatures $T\lsim \TKT$
the data do indeed fall onto a single line within error bars,
when we choose $z=1.00$.
This is not only true for the spin-wave peak itself, but 
for the whole range of $\omega\,L^z$.
Only the data from very small lattices (not shown here)
deviate systematically.
Even at $T=0.725$ the data scale quite well for the values of $L$
for which we have data.
The correlation length at $T=0.725$ is still very large (see appendix);
deviations from scaling 
could presumably be seen if data for much larger lattices were available.

Note that scaling with $\omega\, L^z$ implies that at fixed $qL$ and
for large lattices the spin-wave peak is very narrow in units of $\omega$.
Its width is therefore very sensitive to the time cutoff
in the spin dynamics integration,
and we had to use the very long time integrations 
described in section \ref{Sec:Simulations} 
in order to obtain asymptotic results.

The finite size scaling behavior is very sensitive to variations in $z$.
As an example, figure 8(e) shows that at $T=0.4$ 
the data do not scale when choosing $z=1.05$, even upon excluding 
all lattice sizes $L<96$.
Using similar plots, we obtain
\beq{zresults}
    \begin{array}{ll}
       z = 0.99(2) &\mbox{~~~at~~~} T=0.4\,,\\
       z = 1.00(2) &\mbox{~~~at~~~} T=0.6\,,\\
       z = 1.00(6) &\mbox{~~~at~~~} T=0.7\,,\\
       z = 1.02(3) &\mbox{~~~at~~~} T=0.725\,.
    \end{array}
\eeq
(The relatively large error for $T=0.7$ is a consequence of the limited
amount of data available at this temperature.)
It is  remarkable that the dynamic critical exponent
is the same accross this range of temperatures,
whereas the static exponent $\eta$ varies strongly,
from $\eta=0.082(2)$ at $T=0.4$ to $\eta=0.247(6)$ at $T=0.7$ (see appendix).

The $zz$-component of \Sqw\ is extremely narrow at $T=0.4$ and $T=0.6$,
and cannot show scaling given our maximum integration times.
At $T=0.7$, the spin-wave peak in \Szz\ has become wider,
and we do observe scaling, as shown in figure 8(f). 
%

%%         \setcounter{totalnumber}{2}
%
% FIGURE 9--------------------------------------------------------------
\setlength{\LLL}{\textwidth}
\setlength{\SSS}{2mm}
\divide \LLL by 2
\addtolength{\LLL}{-\SSS}
\begin{figure}[htbp] 
  \begin{center}
  \vskip-5mm
    \mbox{
      \parbox[t]{\LLL}{\psfig{file=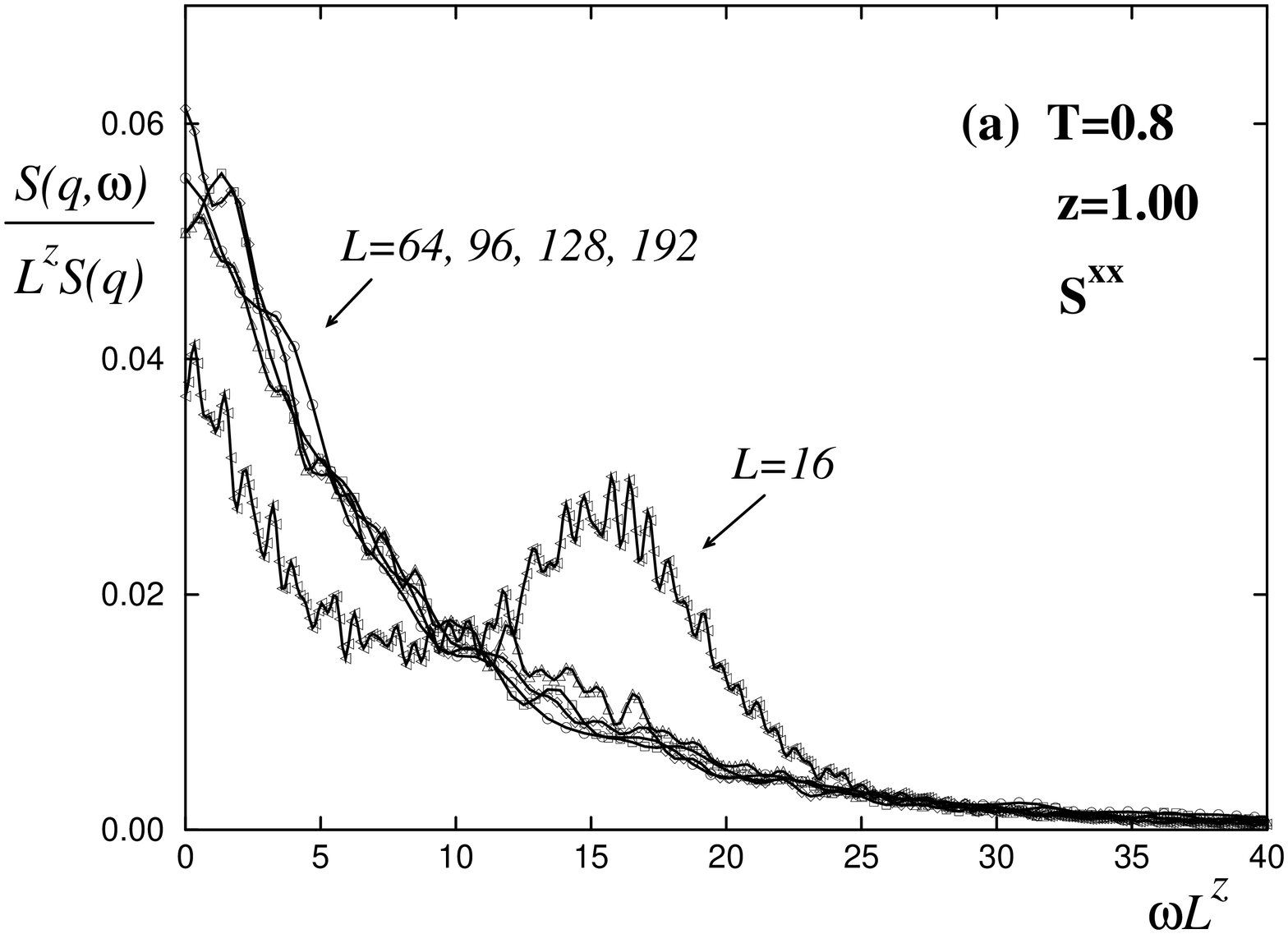,angle=-90,width=\LLL} }
%%      \hskip\SSS
      \parbox[t]{\LLL}{\psfig{file=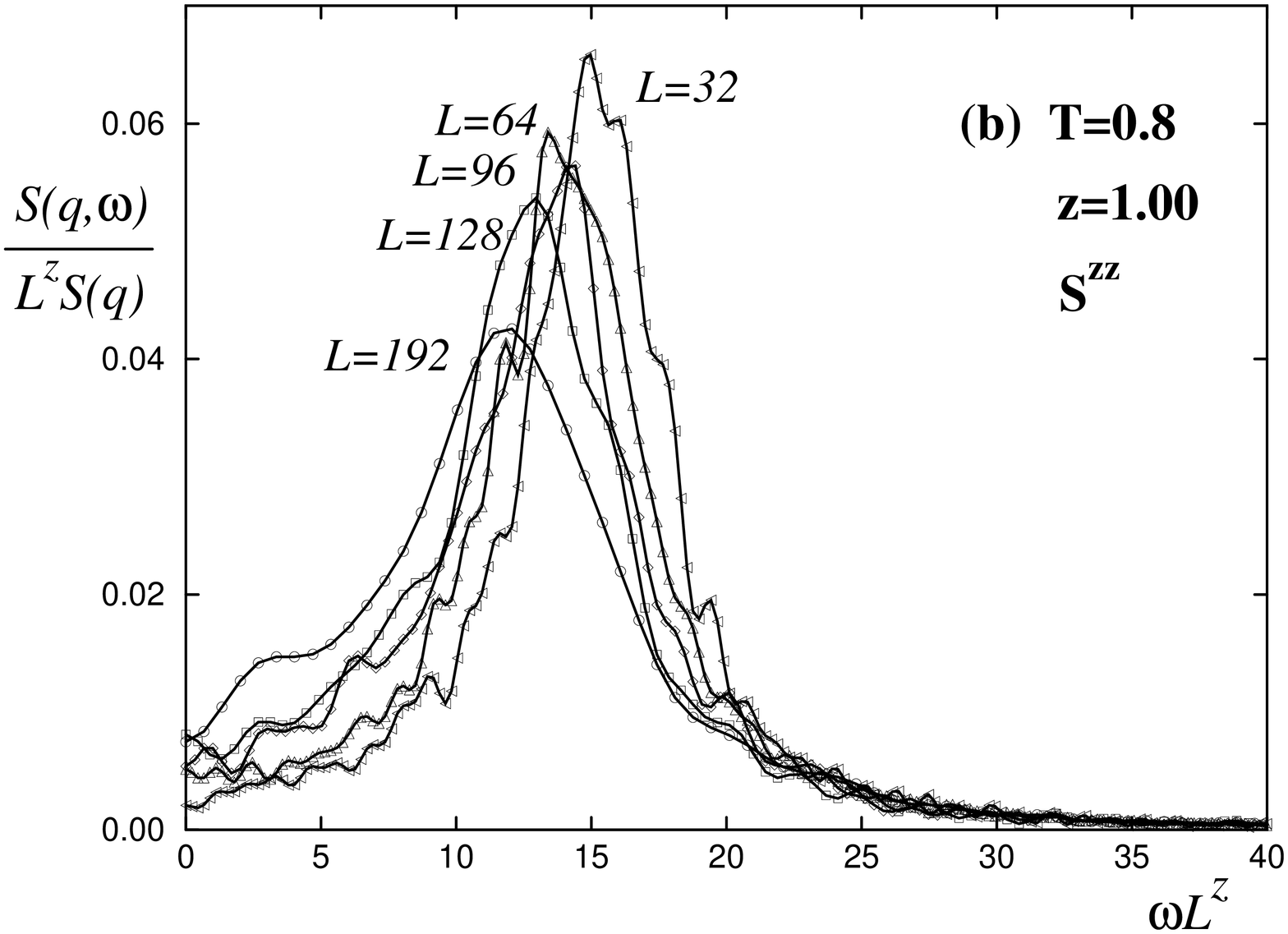,angle=-90,width=\LLL} }
    }%%mbox
    \mycaption{9}{Finite size scaling plot of \Sqw\ for $T=0.8$,
                  above \Tkt,
                  with constant $n_q\ident \frac{q L}{2\pi} \,=2$.
                  (a) \Sxx, (b) \Szz .
                 }%%mycaption
  \end{center}
\end{figure}
%--------------------------------------------------------------
%
Above the phase transition, the relevant length scale is the correlation
length $\xi$, not the lattice size $L$.
We saw earlier (fig.\ 2(d),(e)) that finite size effects are already
small for our lattice sizes.
Yet, surprisingly, there is scaling-like behavior 
for small momenta in \Sxx\ even above the transition, as show in fig.\ 9(a).
Note that, at constant $n_q$, the horizontal scale $\omega L$ is proportional
to $\omega/q$. The data do not scale when different $n_q$ are compared.
At large momenta, a spin-wave peak is visible (see also fig.\ 3(f)).
For the out-of-plane component \Szz\ (fig. 9(b)),
the spin-wave heights do  not obey the scaling equation, \eq{scal}.

%==============================================================================
 \subsection{Tests of Nelson-Fisher scaling form;
             large $\omega$-behavior} \label{Sec:NF_scaling}
%==============================================================================
%
Nelson and Fisher \cite{NelsonFisher}
predicted the scaling form \eq{NelsonS} for $\S^{xx}(q,\omega)$ :
$$
  S^{xx}(q,\omega) \,\sim\, \frac{1}{q^{3-\eta}} \,
               \Psi\left(\frac{\omega}{cq}\right) \;.
$$
This provides an explicit opportunity to compare data 
at different temperatures and at different values of $n_q$.
Equation (\ref{NelsonS}) implies 
the finite size scaling equation (\ref{scal}) used in the previous section, 
with $z=1$:
\beq{NF_FS}
 \frac{S^{xx}(q,\omega)}{L\,S^{xx}(q)} 
 = \frac{1}{cqL} \frac{\Psi\left(\frac{\omega}{cq}\right)}
                    {\int\Psi\left(\frac{\omega}{cq}\right)\,
                          d\frac{\omega}{cq}}
 = f(qL,\, \omega L) \,.
\eeq
It also implies
\beq{NF_scaling}
 cq \frac{S^{xx}(q,\omega)}{S^{xx}(q)} 
 = \frac{\Psi\left(\frac{\omega}{cq}\right)}
                    {\int\Psi\left(\frac{\omega}{cq}\right)\,
                         d\frac{\omega}{cq}}
 = f\left(\frac{\omega}{cq}\right) 
\eeq
for large enough lattice sizes $L$, for which \eq{NelsonS} is valid.
(Note that when $n_q=\frac{qL}{2\pi}$ is constant, 
the arguments $\omega L$ in \eq{NF_FS} and $\frac{\omega}{cq}$ in 
\eq{NF_scaling} are equivalent.)

%
% FIGURE 10--------------------------------------------------------------
\setlength{\LLL}{\textwidth}
\setlength{\SSS}{2mm}
\divide \LLL by 2
\addtolength{\LLL}{-\SSS}
\begin{figure}[htbp] 
  \begin{center}
    \vskip-5mm
    \mbox{
      \parbox[t]{\LLL}{\psfig{file=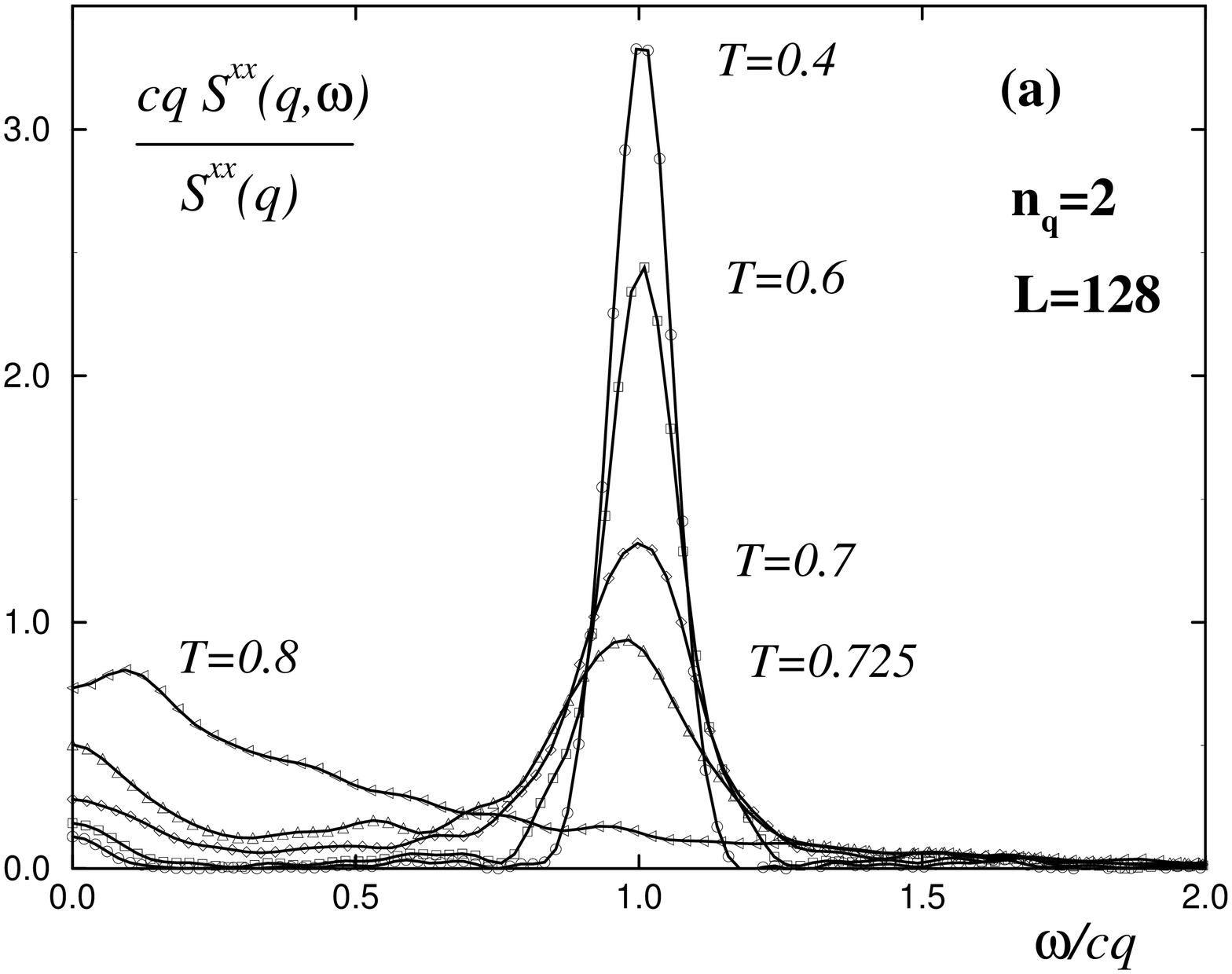,angle=-90,width=\LLL} }
%%      \hskip\SSS
      \parbox[t]{\LLL}{\psfig{file=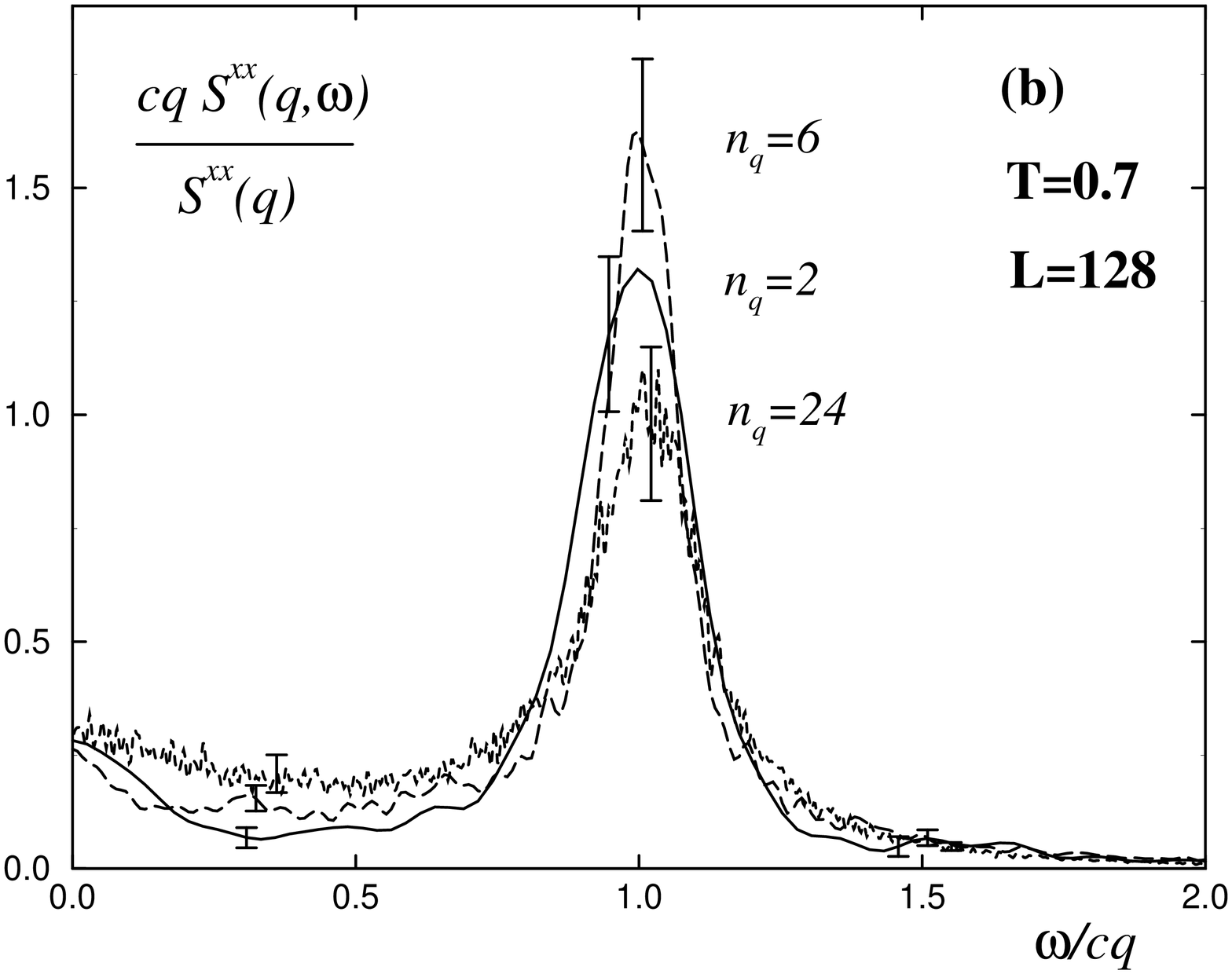,angle=-90,width=\LLL} }
    }%%mbox
\vskip-1mm
    \mycaption{10}{Tests of the Nelson-Fisher scaling form, 
                   \eq{NelsonS}.
                    (a) Different temperatures; 
                    (b) Different values of $n_q$.
                 }%%mycaption
  \end{center}
\end{figure}
%--------------------------------------------------------------
%
In fig.\ 10(a) we use \eq{NF_scaling} to 
compare the in-plane scattering function \Sxx\ 
at different temperatures, for constant $n_q=2$.
The same data appear unscaled in fig.\ 1(a), 
and with dynamic finite size scaling in fig.\ 8.
Obviously, \eq{NF_scaling} is {\em not} satisfied:
the scattering function at different temperatures
within the KT-phase does not scale to the same shape $\Psi(\frac{\omega}{cq})$.
This is also the case at other values of $n_q$.

Different values of $n_q$ are compared in fig.\ 10(b),
at $T=0.7$.
Again, the data do {\em not} scale.
Moreover, the dependence of the spin-wave peak 
on $n_q$ is not monotonous:
for increasing $n_q$ the peak height first grows,
is approximately constant for $n_q=3...8$, and then shrinks.
As $n_q$ becomes large, there is growing intensity below the
spin-wave peak (see also fig.\ 5(d)).
Note that equations (\ref{NelsonS}) and (\ref{NF_scaling})
are at odds with the fact that the additional peaks we observed in \Sxx\ 
(section \ref{Sec:strangepeaks}) have positions which do depend on $n_q$.
The data in fig.\ 10(b) have been obtained with constant time-cutoff
$\tcut=360$. 
The picture is virtually unchanged when 
data with $\tcut \sim\frac{1}{cq}$ (and $\tcut$ sufficiently large) are used.
A similar comparison at other temperatures $T\leq\TKT$ shows still 
stronger deviations from scaling.

%
% FIGURE 11--------------------------------------------------------------
\setlength{\LLL}{\textwidth}
\setlength{\SSS}{2mm}
\divide \LLL by 2
\addtolength{\LLL}{-\SSS}
\begin{figure}[btp] 
  \begin{center}
    \vskip-5mm
    \mbox{
      \parbox[t]{\LLL}{\psfig{file=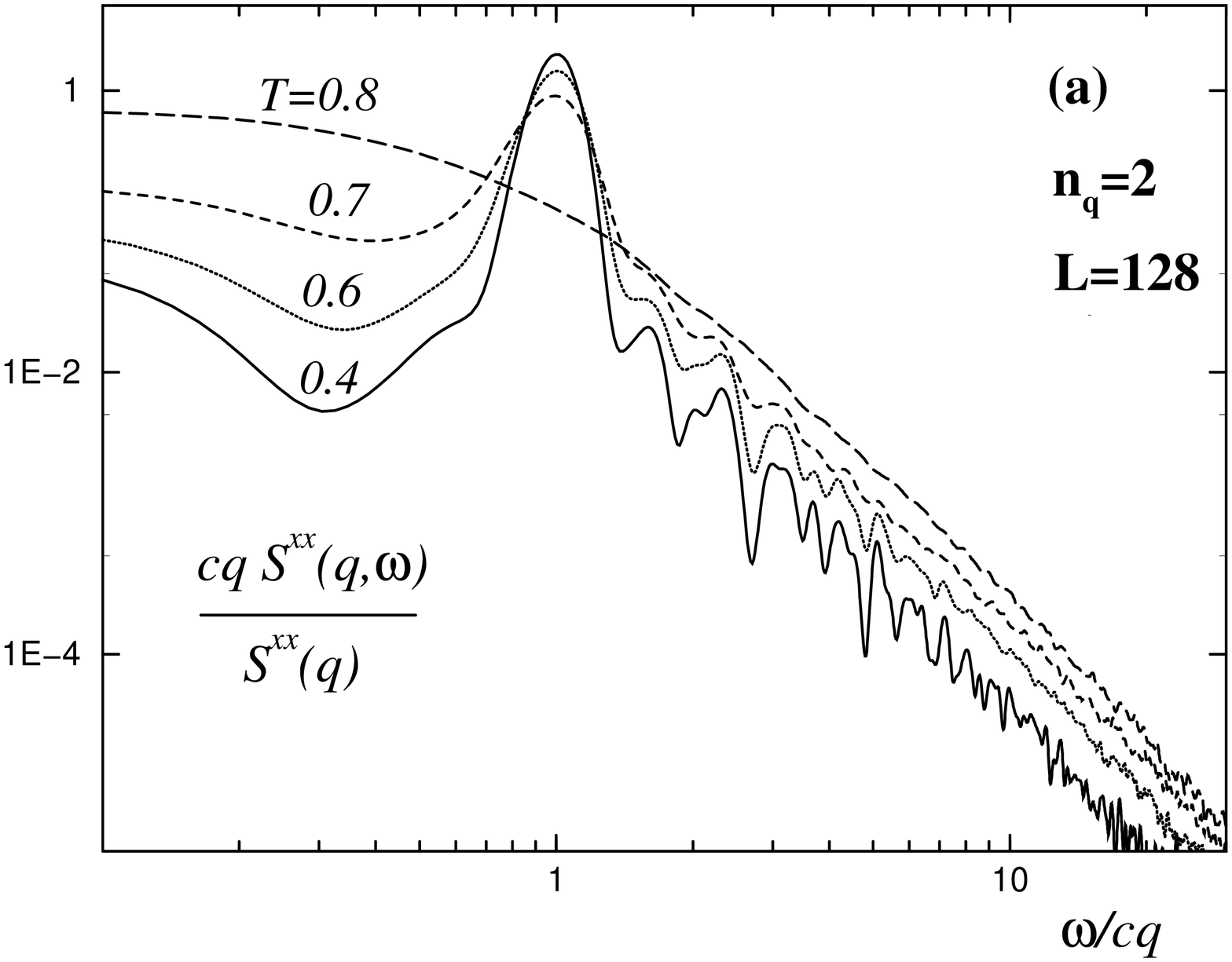,angle=-90,width=\LLL} }
%%      \hskip\SSS
      \parbox[t]{\LLL}{\psfig{file=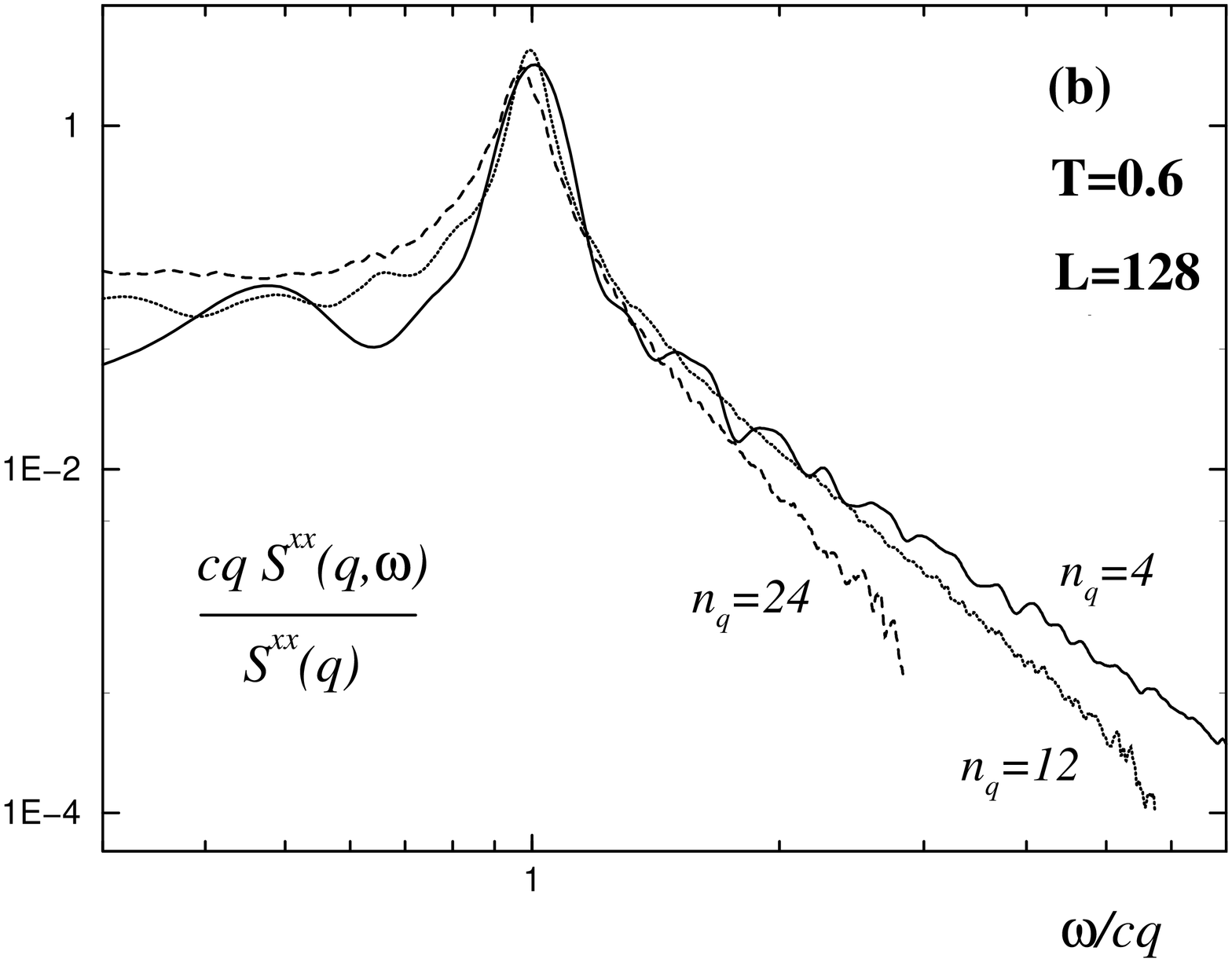,angle=-90,width=\LLL} }
    }%%mbox
    \mycaption{11}{Large frequency behavior of \Sxx, plotted 
                   on a log-log scale in 
                   Nelson-Fisher scaling form, \eq{NelsonS}.
                   The data are  smoothened, with $\delta \omega=0.015$.
                   (a) Different temperatures.
                   (b) Different values of $n_q$.
                 }%%mycaption
  \end{center}
\end{figure}
%--------------------------------------------------------------
%
%%LARGE OMEGA
For {\em large frequencies}, \Sqw\ appears to be independent of lattice size.
We show \Sxx on a log-log scale in fig.\ 11, 
with data scaled similarly as in fig.\ 10.
Nelson and Fisher \cite{NelsonFisher} predict that \Sxx\ is also  independent
of momentum  and  follows a power law $\omega^{-\rho}$, 
with $\rho=3-\eta$ (\eq{largeomega}).
We see somewhat different behavior, which is determined mainly by $n_q$.
For $n_q=2$, fig.\ 11(a), the data can be fitted with $\rho = 3.0(1)$
at all temperatures.
Note the sizeable structure at low $T$.
For larger $n_q$, fig.\ 11(b), $\rho$ increases.
There is also noticeable curvature in $\omega$, with larger $\rho$
at higher $\omega$; $\rho$ also increases slightly with temperature.
The exponents in fig.\ 11(b) range from $3.7(1)$ at $n_q=4$
to $5.4(2)$ at $n_q=24$.
The out-of-plane correlations \Szz\ (not shown) also decay with a power law
with momentum-dependent exponents.

Equation (\ref{NelsonS}) also implies
\beq{Nelson_q}
  q^{3-\eta} \, S^{xx}(q,\omega) \,=\, \Psi\left(\frac{\omega}{cq}\right) \,.
\eeq
In fig.\ 12 we use \eq{Nelson_q} to compare data for different momenta $q$,
at the KT phase transition temperature, using $\eta=0.25$ and constant $n_q$.
Here the data do scale.
This scaling is also implied by dynamic finite size scaling, fig.\ 8(c),
together with a functional dependence $S^{xx}(q)\sim q^{\eta-2}$.
Note that for constant $L$ (instead of constant $n_q$),
we obtain a non-scaling picture indistinguishable from  fig.\ 10(b).

% FIGURE 12--------------------------------------------------------------
\setlength{\LLL}{\textwidth}
\divide\LLL by 2
%%\addtolength{\LLL}{-25mm}
\begin{figure}[htbp] 
  \begin{center}
      \parbox[t]{\LLL}{\psfig{file=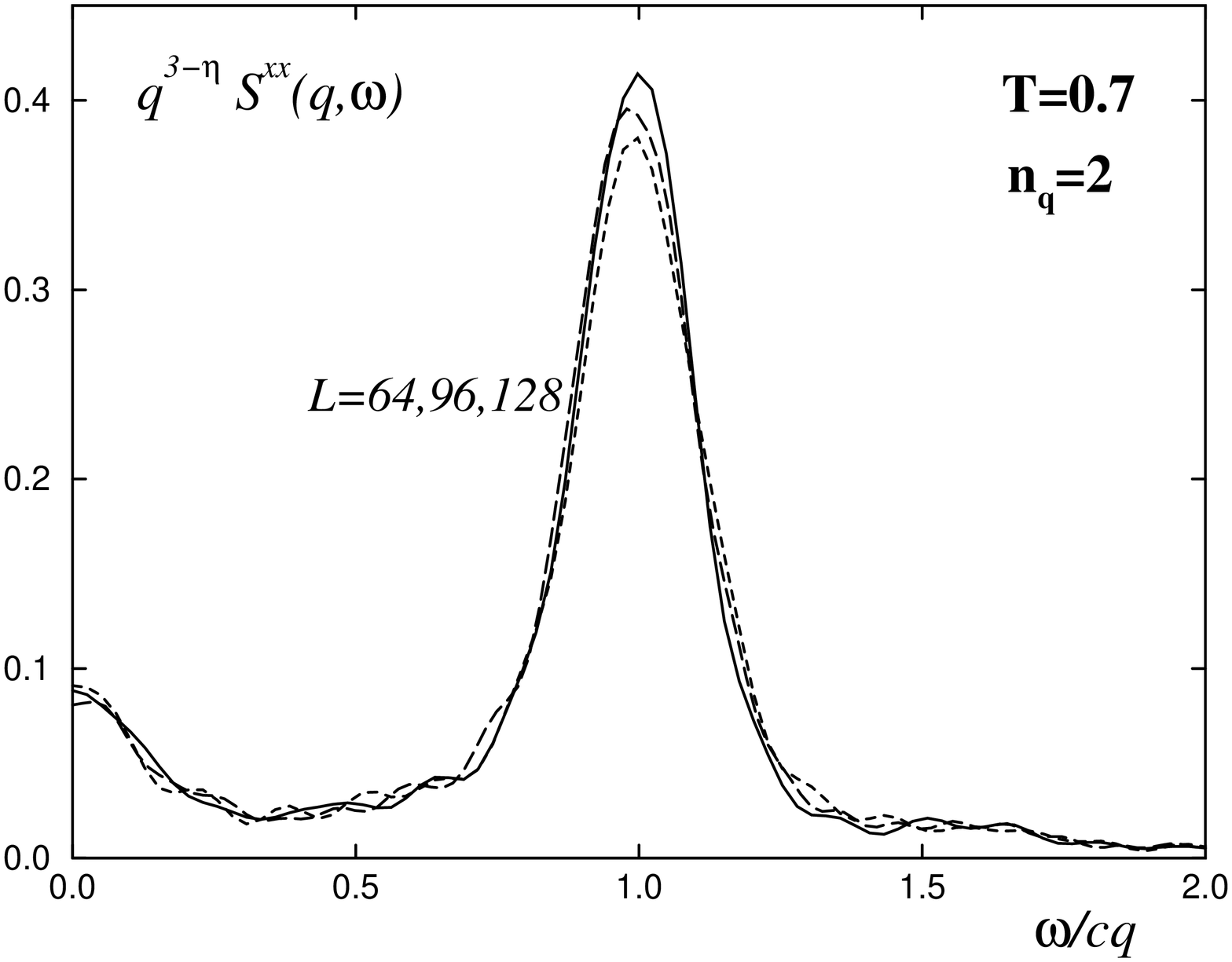,angle=-90,width=\LLL} }
    \mycaption{12}{Test of the Nelson-Fisher scaling form,
                   for different lattice sizes $L$.
                 }%%mycaption
  \end{center}
\end{figure}
%--------------------------------------------------------------
%

%==============================================================================
 \subsection{Lineshapes}
%==============================================================================
%

In figure 13 we compare our results with theoretical predictions
for the shape of \Sxxqw.
We show  data at $T=0.7$, for $L=128$ and $q=\pi/32$ ,
normalized according to \eq{NF_scaling}, 
and we compare with predictions using $\eta=1/4$, 
similarly normalized.
%
% FIGURE 13--------------------------------------------------------------
\setlength{\LLL}{\textwidth}
%%\divide\LLL by 2
\addtolength{\LLL}{-10mm}
\begin{figure}[tbp] 
  \begin{center}
  \vskip-10mm
      \parbox[t]{\LLL}{\psfig{file=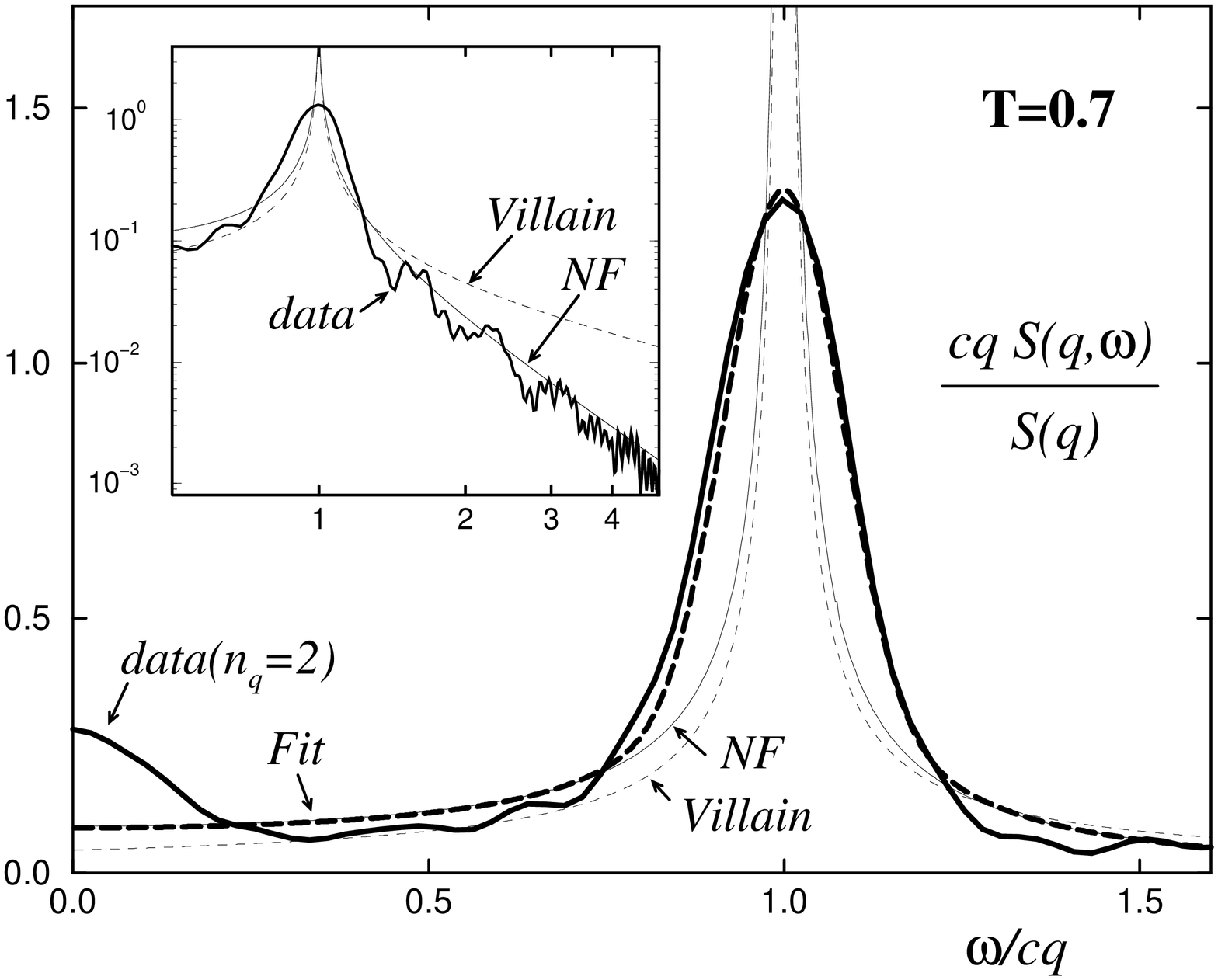,angle=-90,width=\LLL} }
\vskip-6mm
    \mycaption{13}{Comparison of the lineshape of $S^{xx}(q,\omega)$ 
                  with theoretical predictions.
                  Data are at $T=T_{KT}$, $L=128$ and $q=\pi/32$ (thick line),
                  and are normalized according to \eq{NF_scaling}.
                  The two thin lines represent the predictions by 
                  Nelson and Fisher \eq{NelsonS} (continuous line) 
                  and by Villain \eq{Villain} (dashed line), 
                  both with $\eta=0.25$.
                  The thick dashed line is obtained by a fit to the data
                  with an ad-hoc function related to the Nelson-Fisher form
                  (see text).
                  The inset shows the data and predictions
                  on a log-log plot that includes large values of $\omega$.
                 }%%mycaption
  \end{center}
\end{figure}
%--------------------------------------------------------------

The predictions by Nelson and Fisher and by Villain
both have a pole at the spin-wave peak 
(eqs.\ \ref{Villain} and \ref{Psi}), as shown in the figure.
In order to compare better with our data
which have been obtained with a time integration of finite length,
we also tried to convolute the predictions
with the fourier-transform of a finite-time cutoff at $\tcut=360$.
The resulting functions (not shown) exhibit very strong 
oscillations (size$=O(0.5)$), and a spin-wave peak that is much
higher (about 3.7) and more narrow than that of the data.
We conclude that $S^{xx}(q,t)$ decays much faster in time than predicted.

As mentioned before, the rich structure in \Sxxqw\ 
below and above the  spin-wave peak
that was described in section \ref{Sec:strangepeaks}
had not been predicted,
except for a (small) central peak,
\eq{Mcentral}, foreseen by Menezes et al. \cite{Menezes}.

Since the Nelson-Fisher prediction does not agree well with the 
data at the spin-wave peak, 
we tried to find a functional form which does fit these data reasonably well
and should thus be an approximation to the actual form,
so that our data can more easily be compared to the results of
future theoretical calculations.
We found that a modified form of \eq{NF_scaling}
works well, namely a widening of \eq{NF_scaling} 
with a Gaussian resolution function \eq{smoothen}, 
with $\delta\omega$ a free parameter. 
Around the spin-wave peak we obtained fairly good agreement with
our (unconvoluted) data, 
as shown by the thick dashed line in Fig.\ 13, 
which uses $\delta\omega=0.01$.
However, different values of $\delta\omega$ in the modified function
are necessary to describe the data at different $n_q$.
The modified function  can of course not describe the additional structure 
in \Sxxqw, including the central peak.
%(see also section \ref{Sec:NF_scaling}).
%
%The prediction by Villain does not describe the spin-wave peak well,
%even after an artificial widening.
%
%(When the data are also smoothened, the spin-wave peak is still described
%well by the Nelson-Fisher prediction with a correspondingly (in rms-fashion)
%increased $\delta\omega$.)

{\em The large frequency behavior} of the data is shown in the inset 
in figure 13. The prediction by Nelson and Fisher
agrees with the data at large frequencies qualitatively.
However, as described in section \ref{Sec:NF_scaling},
a fit to \Sxxqw\ for large $\omega$ results in a different
power law exponent than predicted.
Not surprisingly, at large frequencies the prediction by Villain,
intended for the spin-wave peak divergence,
does not describe the data correctly.

We conclude that below and above the transition, the
actual lineshape is quite different from the predicted forms,
with a much wider spin-wave peak, and a lot of additional structure.

{\em Above the transition}, the theoretical predictions
 \eq{MertensSxx} and \eq{MertensSzz} do
not describe the data well either.
For the in-plane component \Sxx\ we see two different regimes in $\omega$.
At small $\omega$, it is compatible with a Lorentzian-like peak
$\sim (\omega^2 + a)^{-b}$, but with an exponent $b$ that is momentum-dependent
(e.g.\ $b\approx 1.1(1)$ at $q=\pi/48$, $b\approx 0.43(2)$ at $q=\pi/16$).
At large $\omega$, \Sxx\ decays with a power law $\sim \omega^{-c}$, with
$c=3.2(2)$ for small momenta (see fig.\ 11(a)).
The out-of-plane component \Szz\ does not show the predicted central peak at all;
instead it exhibits a spin-wave peak.

%******************************************************************************
 \section{Conclusions}
%******************************************************************************
%

We have performed the first high precision study of the dynamic critical
behavior of the $XY$-model,
at five different temperatures below, at, and above \Tkt,
on square lattices of size up to $192\times192$.
We have  determined the critical temperature 
to be  $T=0.700(5) \,J/k_B$. 
Starting from about $1000$ equilibrium configurations generated
by an efficient Monte Carlo procedure at each temperature and lattice size, 
we have integrated
the equations of motion of the spins to very large times,
$t_{max}=400 J^{-1}$,
and measured space-displaced, time-displaced correlation functions
to compute the neutron scattering function \Sqw.

At temperatures up to \Tkt, \Sqw\ exhibits very strong and sharp spin-wave peaks
in the in-plane-component \Sxx.
As T increases, they widen slightly and move to lower $\omega$,
but remain pronounced even just above \Tkt.
For increasing momentum they broaden and rapidly lose intensity.
Well above \Tkt, the spin-wave peak disappears in \Sxx, as expected, and we
observe a large central peak instead.

In addition to the spin-wave peak, the in-plane component \Sxx\ 
exhibits a rich structure of small intensity, 
which we interpret to come from two-spin-wave processes. 
Furthermore, \Sxx\ shows a clear central peak, even below \Tkt,
which becomes very pronounced towards the critical temperature.
Neither this strong central peak 
nor the additional structure 
are predicted by existing analytical calculations.

The out-of-plane component \Szz\ is much weaker than \Sxx, 
except for large $q$.
It displays a sharp spin-wave peak at all temperatures, even above \Tkt.
The peak widens with increasing temperature, and only at low T
is it consistent with a delta-function shape.

We measure the dispersion relation, 
i.e.\ the position of the spin-wave peak as a function
of momentum, to be linear to high accuracy.
Its slope, the spin-wave velocity, decreases with increasing temperature
approximately linearly, 
as expected from approximate analytical results at small T.

Examining dynamic finite size scaling,
we show that both the characteristic frequency $\omega_m$
and the neutron scattering function \Sqw\ itself
scale very well for all $T\leq \TKT$,  
with a dynamic critical exponent of $z=1.00(4)$
that does not depend on temperature,
whereas the static exponent $\eta$ varies strongly. 

The shape of the scaling function 
is not well described 
around the spin-wave peak
by the available theoretical predictions,
nor is the shape of the scattering function  above the transition,
and the additional structure had not been predicted at all.
The data which we have presented here are of sufficiently high quality 
that meaningful comparison with theory and experiment is possible. 
We hope that this spin dynamics study will thus serve to stimulate 
further effort in this area.

%%%%%%%%%%%%%%%%%%%%%%%%%%%%%%%%%%%%%%%%%%%%%%%%%%%%%%%%%%%%%%%%%%%%%%%%%%%%%%%
%     ACKNOWLEDGEMENTS
%%%%%%%%%%%%%%%%%%%%%%%%%%%%%%%%%%%%%%%%%%%%%%%%%%%%%%%%%%%%%%%%%%%%%%%%%%%%%%%
\section*{Acknowledgements}

We are indebted to Kun Chen for the initial version of the spin dynamics
program, and to Alex Bunker for helpful discussions.
We would like to thank the Pittsburgh Supercomputer Center for its support;
all of our computer simulations were carried out on the Cray C90 at Pittsburgh.
This research was supported in part by NSF Grant No.\ DMR-9405018.

\pagebreak
%==============================================================================
 \section*{Appendix: Static critical behavior}
%==============================================================================
%
The determination of the transition temperature in $XY$-like systems has been
notoriously difficult \cite{Gerling_Landau_statics,XYsimulations}.
The best previous estimate \cite{Gerling_Landau_statics} for \Tkt\ for the
model considered here was $\TKT=0.725\pm 0.010$, estimated from the
onset of vortex-pair unbinding, which is a procedure that is 
quite difficult to apply with high precision.
The results of our high resolution spin dynamics study 
at $T=0.725$ prompted us to 
perform a new, more accurate determination of \Tkt, using the powerful 
hybrid Monte Carlo algorithm described in section \ref{Sec:Simulations}.

We carried out
a set of static Monte Carlo simulations, 
with lattice sizes $L=64$, $128$, and $256$, 
and $40000$ hybrid Monte Carlo sweeps in each run.
We  then analyzed the static correlations
$C(r) = \frac{1}{2} \langle S^x(0) S^x(r) \,+\, S^y(0) S^y(r)\rangle$
in three different ways:
(i) using finite size scaling, 
(ii) with a fit to a power law decay, and 
(iii) with a fit to the free lattice propagator. 
The results of all three methods are in excellent agreement.

The finite size scaling ansatz for the correlation function 
is \cite{Landau_FSS}
\beq{static_FSS_scaling}
      C(r,L) \,=\, r^{-\eta} \,\,f(\hat{t} L^{1/\nu}, \frac{r}{L}) \,,
\eeq
where $\hat{t}$ is the reduced temperature ($\hat{t}=|1-T/T_c|$), 
and $\nu$ the correlation length
exponent.
Since our model is critical throughout the KT-phase, 
we have $\hat{t}=0$ for all $T\leq\TKT$.
With the correct value of $\eta$, 
the data for different lattice sizes should therefore coalesce
on a plot of $C(r)\,L^\eta$ versus $\frac{r}{L}$.
Figure 14  shows such plots at $T=0.7$,
for $\eta=0.24$, $0.25$, and $0.26$.
%
% FIGURE 14--------------------------------------------------------------
\setlength{\LLL}{\textwidth}
%%\divide\LLL by 2
\addtolength{\LLL}{-25mm}
\begin{figure}[htbp] 
  \begin{center}
      \parbox[t]{\LLL}{\psfig{file=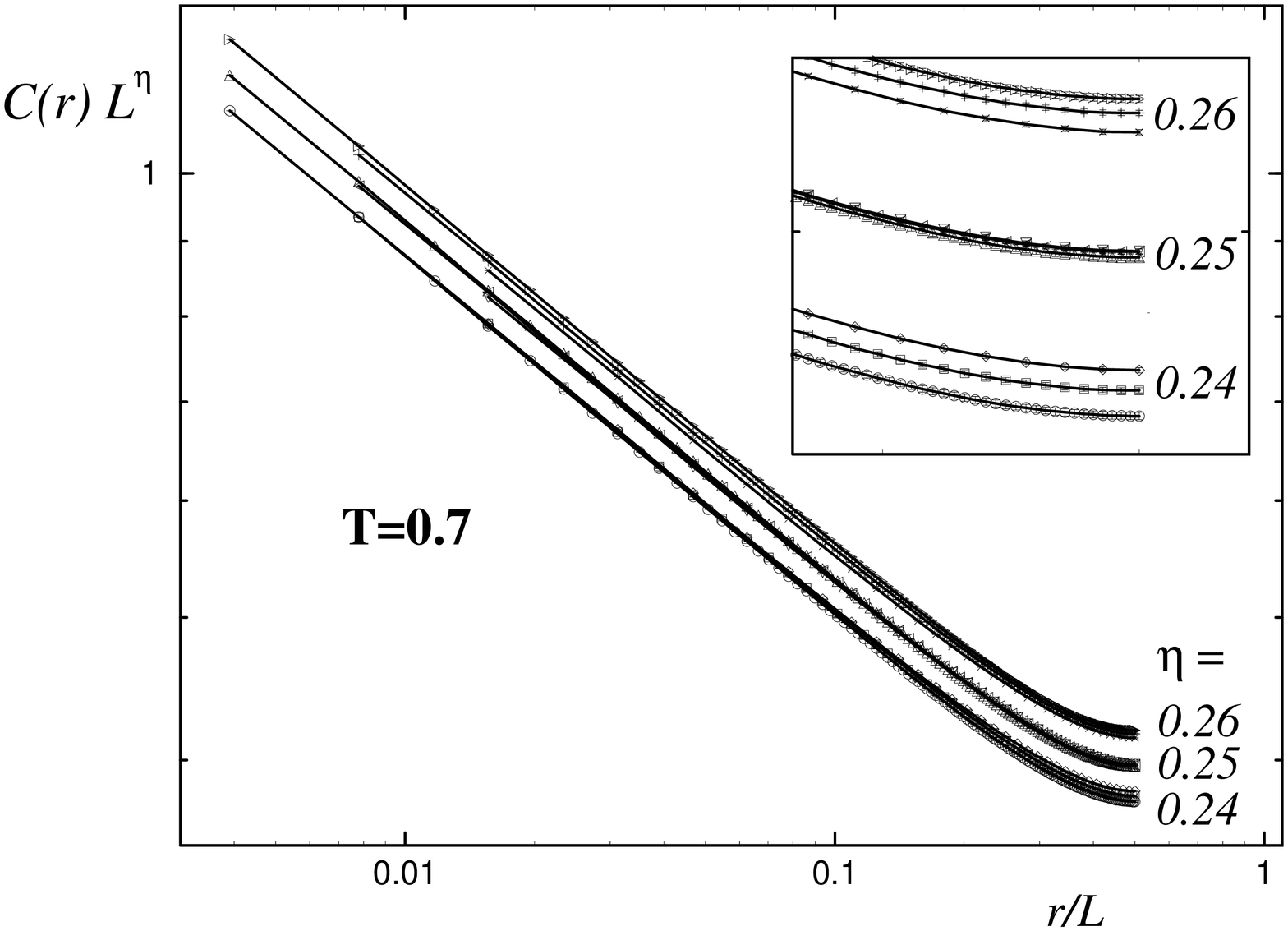,angle=-90,width=\LLL} }
    \mycaption{14}{Finite size scaling plot of the static 
                   correlation function $C^{xx}(r)$ at $T=0.7\approx \TKT $,
                   using three different estimates for $\eta$.
                 }%%mycaption
  \end{center}
\end{figure}
%--------------------------------------------------------------
%
Choosing $\eta=0.25$ results in very good scaling over the whole range of
distances $r$.
At lower temperatures, scaling of similar quality is achieved with
smaller values of $\eta$. 
At a slightly higher temperature of $T=0.71$, however, we observed only 
mediocre scaling, with an effective $\eta$ of $0.29(1)$,
but with small systematic deviations from scaling already visible.
At $T=0.725$, the deviations from scaling are stronger still.
%%the data do not scale at all.

Independent estimates of $\eta$ were obtained from simple 
power law fits $C(r,L) \sim r^{-\eta}$, for lattice size $L=256$,
and $r_{max}\leq 20$.
The results for $\eta$
depend only very little on the fit-range, and agree within error bars
with those from finite size scaling.

A simple power law ansatz ignores the fact 
that $C(r,L)$ is actually periodic in $r$ with period $L$.
The full functional form can be deduced from the fact that within the 
$KT$-phase, the model is thought \cite{KT,Itzykson_Drouffe}
to behave like a free field theory, for which the 
exact finite lattice propagator (correlation function) 
at an effective ``temperature'' $1/\eta$ is \cite{Itzykson_Drouffe} \ 
\def\TPL{\frac{2\pi}{L}}
\begin{eqnarray}
  C(r,L) &=& \exp(-\eta\, \Gamma(r,L)), \mbox{~~~with}\\[.2ex]
  \Gamma(r,L) &=& \frac{2\pi}{L^2} \sum_{q_x,q_y = 0}^{L-1} \;
        \frac{1-\cos(r q_x \TPL)}{4 - 2\cos(q_x \TPL) - 2\cos(q_y \TPL)} \;.
\end{eqnarray}
We also used this functional form to fit $C(r,L)$. 
The results for different lattice sizes and different fit ranges
(up to $r=L/2$ and excluding $r< 5$) agreed with each other
and with the results from finite size scaling.
The quality of fits was very good for all $T\leq 0.70$,
whereas for larger temperatures ($T=0.71$ and above) it deteriorated strongly, 
and the results became lattice size dependent.

We have found no evidence for logarithmic corrections.
It is possible that small corrections are present and introduce a subtle bias.
In this case our results for $\eta$ would need modification.
Assuming that logarithmic corrections are indeed negligible, we obtain 
as our combined results from all three methods
\beq{eta}
    \begin{array}{ll}
       \eta = 0.082(2)  &\mbox{~~~at~~} T=0.4 \\  
       \eta = 0.153(5)  &\mbox{~~~at~~} T=0.6 \\
       \eta = 0.247(6)  &\mbox{~~~at~~} T=0.7 \;.
    \end{array}
\eeq
We note that the expected linear dependence of $\eta$ on $T$
below \Tkt\ does not seem to be satisfied at these temperatures 
in our model.

Both from assuming $\eta=1/4$ at \Tkt\ \cite{KT},
and from the different qualitative behavior of $C(r,L)$ for $T\geq 0.710$,
we conclude that the Kosterlitz Thouless transition temperature is
\beq{Tkt}
          \TKT\,=\,0.700(5) J/k_B\,.
\eeq
This estimate is slightly below the value from \cite{Gerling_Landau_statics}
which we had used at the start of the spin dynamics study, 
and clearly below the estimate of ref.\ \cite{Voelkel}.

Slightly above the KT-transition, at $T=0.725$, the dynamic behavior
of \Sqw\ in our study resembles that at \Tkt\ (whereas at $T=0.8$ it
is very different). This may be explained by looking at the correlation 
lengths:  
The correlation function at $T=0.725$ can be fitted by an
Ornstein-Zernike form
\beq{Ornstein}
  \Gamma(r) \,\sim\, r^{-\frac{1}{2}} \, e^{-r/\xi}
\eeq
with a value of $\xi = O(400)$. 
Since this correlation length is  larger than the lattice sizes we have used
in our study, 
a behavior resembling the KT-phase (where $\xi=\infty$) is not surprising.
A similar (approximate) fit at $T=0.8$, on the other hand,
gives $\xi=O(10)$, much smaller than our lattice sizes.
After our work was completed, we received a paper by 
Cuccoli et al. \cite{Cuccoli} with a Monte Carlo study of the statics of 
our model. Their results are in excellent agreement with ours.

%%%%%%%%%%%%%%%%%%%%%%%%%%%%%%%%%%%%%%%%%%%%%%%%%%%%%%%%%%%%%%%%%%%%%%%%%%%%%%%
%     BIBLIOGRAPHY
%%%%%%%%%%%%%%%%%%%%%%%%%%%%%%%%%%%%%%%%%%%%%%%%%%%%%%%%%%%%%%%%%%%%%%%%%%%%%%%
   \pagebreak%%[3]


\begin{thebibliography}{99}
%
%% \protect\small
%% \addtolength{\baselineskip}{-.1\baselineskip}
   \addtolength{\itemsep}{-\itemsep}
\def\mybf#1{{#1}} %boldface volume numbers look ugly
%-----------------------------------------------------------------------------
%
%
\bibitem{KT}
     J. M. Kosterlitz and D. J. Thouless, J. Phys. C \mybf{6}, 1181 (1973).
%
\bibitem{Gerling_Landau_statics}
     R. W. Gerling and D. P. Landau, in: {\em Magnetic Excitations and
           Fluctuations}, edited by S. W. Lovesey, U. Balucani, F. Borsa,
           and V. Tognetti (Springer, Berlin, 1984).
%
\bibitem{XYsimulations}
     J. Tobochnik and G. V. Chester, Phys. Rev. B \mybf{20}, 3761 (1979); \\
     D. P. Landau and K. Binder, Phys. Rev. B \mybf{24}, 1391 (1981).
%
\bibitem{Rotatorsimulations}
     R. Gupta, J. DeLapp, G. G. Batrouni, G. C. Fox, and C. F. Baillie,
           Phys. Rev. Lett. \mybf{61}, 1996 (1988); \\
     Ph. de Forcrand and P. Samsel, unpublished;\\
%
%
     W. Janke and K. Nather, Phys. Rev. \mybf{B48}, 7419, 1993,
                             Phys. Lett. \mybf{A157}, 11 (1991);\\
     P. Olsson, Phys. Rev. Lett. \mybf{73},3339 (1994);\\
     P. Olsson, preprint {\em Monte Carlo analysis of the 2D XY model: 
       II. Comparison with Kosterlitz' Renormalization Group equations},
       January 1995;\\
     R. Gupta and C. F. Baillie, Phys. Rev. \mybf{B45}, 2883 (1992);\\
     H. Kawamura and M. Kikuchi, Phys. Rev. \mybf{B47}, 1134 (1993);\\
     M. Hasenbusch, M. Marcu, and K. Pinn, Physica \mybf{208A}, 124 (1994);\\
     H. Weber and P. Minnhagen, Phys. Rev. \mybf{B10}, 5986 (1988);\\
     J. E. Van Himbergen and S. Chakravarty, 
                                Phys. Rev. \mybf{B23}, 359 (1981);\\
%%   J. Tobochnik and G. V. Chester, Phys. Rev. B \mybf{20}, 3761 (1979);
     J. F. Fern\'andez, M. F. Ferreira, and J. Stankiewicz,
         Phys. Rev. B \mybf{34}, 292 (1986);\\
     H. Betsuyaku, Physica \mybf{106A}, 311 (1981).
%     
%
\bibitem{HohenbergHalperin}  %% DYNAMIC SCALING
     P. C. Hohenberg and B. I. Halperin , 
           Rev. Mod. Phys. \mybf{49}, 435 (1977).
%
\bibitem{ChenLandau}
   K. Chen and D. P. Landau, Phys. Rev. B \mybf{49}, 3266 (1994).
%
\bibitem{Villain}
     J. Villain, J. Phys. (Paris) \mybf{35}, 27 (1974).     
%
\bibitem{NelsonFisher}
     D. R. Nelson and D. S. Fisher, Phys. Rev. B \mybf{16}, 4945 (1977).
%
\bibitem{Menezes}
     S. L. Menezes, A. S. T. Pires, and M. E. Gouv\^{e}a,
           Phys. Rev. B \mybf{47}, 12280 (1993).
%
\bibitem{Pereira}
     A. R. Pereira, A. S. T. Pires, M. E. Gouv\^{e}a, and B. V. Costa,
           Z. Phys. B. \mybf{89}, 109 (1992).
%
\bibitem{Landau_Gerling_dynamics}
     D. P. Landau and R. W. Gerling, 
           J. Magn. Magn. Mat. \mybf{104-107}, 843 (1992).
%
\bibitem{HalperinHohenberg}
     B. I. Halperin and P. C. Hohenberg, Phys. Rev. \mybf{177}, 952 (1969).
%
\bibitem{MoussaVillain}
     F. Moussa and J. Villain, J. Phys. C \mybf{9}, 4433 (1976).
%
\bibitem{NelsonKosterlitz}
   D. R. Nelson and J.M. Kosterlitz, Phys. Rev. Lett. \mybf{39}, 1201 (1977).
%
\bibitem{MertensOne}
     F. G. Mertens, A. R. Bishop, G. M. Wysin, and C. Kawabata,
           Phys. Rev. Lett. \mybf{59}, 117 (1987); 
           Phys. Rev. B \mybf{39}, 591 (1989).
%
\bibitem{MertensTwo}
     F. G. Mertens, A. R. Bishop, M. E. Gouv\^{e}a, and G. M. Wysin, 
           J. de Physique \mybf{C8}, 1385 (1988);
     M. E. Gouv\^{e}a, G. M. Wysin, A.R. Bishop, and F. G. Mertens,
           Phys. Rev. B \mybf{39}, 11840 (1989).
%
\bibitem{Kawabata}
    C. Kawabata, M. Takeuchi, and A. R. Bishop,
           J. Magn. Magn. Mat. \mybf{54-57}, 871 (1986).
%
\bibitem{Shirakura}
    T. Shirakura, F. Matsubara, and S. Inawashiro,
           J. Phys. Soc. Japan \mybf{59}, 2285 (1990).
%
\bibitem{Wiesler}
    D. G. Wiesler, H. Zabel, and S. M. Shapiro,
           Z. Phys. B \mybf{93}, 277 (1994).
%
\bibitem{experiments}
    S. T. Bramwell, M. T. Hutchings, J. Norman, R. Pynn, and P. Day,
           J. de Physique \mybf{C8}, 1435 (1988);\\
    M. T. Hutchings, P. Day, E. Janke, and R. Pynn, 
           J. Magn. Magn. Mat. \mybf{54-57}, 673 (1986);\\
    L. P. Regnault, J. Rossat-Mignod, J.Y. Henry, and L.J. de Jongh,
           J. Magn. Magn. Mat. \mybf{31-34}, 1205 (1983);\\
    K. Hirakawa, H. Yoshizawa, J. D. Axe, and G. Shirane,
           J. Phys. Soc. Japan \mybf{52}, 4220 (1983);\\
    K. Hirakawa,
           J. Appl. Phys. \mybf{53}, 1893 (1982);\\
    K. Hirakawa and H. Yoshizawa, 
           J. Phys. Soc. Japan \mybf{47}, 368 (1979).
%
\bibitem{Bramwell}
    S. T. Bramwell, P. C. W. Holdsworth, and M. T. Hutchings,
    J. Phys. Soc. Jpn. {\bf 64}, 3066 (1995); \\
    S. T. Bramwell and P. C. W. Holdsworth,
    J. Appl. Phys. {\bf 75}, 5955, (1994).
    
%
\bibitem{SW}
    R. H. Swendsen and J. S. Wang, Phys. Rev. Lett. \mybf{58}, 86 (1987).
%
\bibitem{WolffXY} 
    U. Wolff, Phys. Rev.  Lett. \mybf{62},    361 (1989),
              Nucl. Phys. B     \mybf{322},   759 (1989),
         and  Phys. Lett.       \mybf{228B},  379 (1989).
%
\bibitem{Overrelaxation}
    F. R. Brown and T. J. Woch, Phys. Rev. Lett. \mybf{58}, 2394 (1987);\\
    M. Creutz, Phys. Rev. D \mybf{36}, 515 (1987).
%
\bibitem{SDshort}
   H. G. Evertz, D. P. Landau,
%         Spin dynamics calculations in the two-dimensional classical XY-model
          in Computer Simulations in Condensed Matter Physics VIII,
          ed.\ D.P. Landau et al., 
          Springer Proceedings in Physics, 1995, to appear.
%
\bibitem{Menezes_Szz}
     S. L. Menezes, A. S. T. Pires, and M. E. Gouv\^{e}a,
          Phys. Rev. B \mybf{45}, 10454 (1992).
%
\bibitem{Menezes_spinstiffness}
     S. L. Menezes, M. E. Gouv\^{e}a, and A. S. T. Pires,
           Phys. Lett. A \mybf{166}, 330 (1992).
%
\bibitem{Landau_FSS}
   D. P. Landau, in {\em Finite Size Scaling and Numerical Simulation 
           of Statistical Systems}, ed.\ V. Privman, World Scientific, 
           Singapore 1990.
%
\bibitem{Itzykson_Drouffe} 
   C. Itzykson, J.-M. Drouffe, {\em Statistical Field Theory},
           Cambridge University Press 1989, chapter 4.
%
\bibitem{Voelkel}
   A.R. V\"olkel, G.M. Wysin, A. R. Bishop, and F. G. Mertens,
           Phys. Rev. B \mybf{44}, 10066 (1991).             
%
\bibitem{Cuccoli} A. Cuccoli, V. Tognetti, and R. Vaia,  
           Phys. Rev. \mybf{B52}, 10221 (1995).
%%%%%%%%%%%%%%%%%%%%%%%%%%%%%%%%%%%%%%%%%%
%
%
\end{thebibliography}
\end{document}